\documentclass[%
superscriptaddress,
nofootinbib,
nobibnotes,
pra,
]{revtex4-2}

\usepackage{float,graphicx}
\usepackage{dcolumn}
\usepackage{bm}
\usepackage{cancel,soul,ulem}
\usepackage{mathrsfs}

\usepackage{graphicx}
\usepackage{amssymb,amsmath}
\usepackage[utf8]{inputenc}
\usepackage{multirow}
\usepackage{xcolor}
\usepackage{hyperref}
\hypersetup{colorlinks=true,linkcolor=blue,anchorcolor=blue,citecolor=magenta,filecolor=blue,urlcolor=magenta,bookmarksnumbered=true}
\usepackage[flushleft]{threeparttable}

\usepackage{amsmath}

\usepackage{appendix}

\usepackage{caption}

\usepackage{mathtools}
\usepackage{dsfont}
\usepackage{amsmath}
\usepackage{amssymb}
\usepackage{stackrel}
\usepackage{graphicx}
\usepackage{subcaption}

\usepackage{tikz}
\usetikzlibrary{quantikz}
\usepackage{braket}

\begin{document}

\title{Efficient and precise quantum simulation of ultra-relativistic quark-nucleus scattering}

\author{Sihao Wu}
\affiliation{Department of Modern Physics, University of Science and Technology of China, Hefei 230026, China} 

\author{Weijie Du}
\email[Email:]{\ duweigy@gmail.com}
\affiliation{Department of Physics and Astronomy, Iowa State University, Ames, Iowa 50010, USA}

\author{Xingbo Zhao}
\affiliation{Institute of Modern Physics, Chinese Academy of Sciences, Lanzhou 730000, China}
\affiliation{University of Chinese Academy of Sciences, Beijing 100049, China}

\author{James P. Vary}
\affiliation{Department of Physics and Astronomy, Iowa State University, Ames, Iowa 50010, USA}

\date{\today}

\begin{abstract}

	We present an efficient and precise framework to quantum simulate the dynamics of the ultra-relativistic quark-nucleus scattering.	
	This framework employs the eigenbasis of the asymptotic scattering system and implements a compact scheme for encoding this basis upon lattice discretization. 
	It exploits the operator structure of the light-front Hamiltonian of the scattering system, which enables the Hamiltonian input that utilizes the quantum Fourier transform for efficiency. 
	Our framework simulates the scattering by the efficient and precise algorithm of the truncated Taylor series. 
	The qubit cost of our framework scales logarithmically with the Hilbert space dimension of the scattering system. The gate cost has optimal scaling with the simulation error and near optimal scaling with the simulation time. These scalings make our framework advantageous for large-scale dynamics simulations on future fault-tolerant quantum computers.
	We demonstrate our framework with a simple scattering problem and benchmark the results with those from the Trotter algorithm and the classical calculations, where good agreement between the results is found.
	Our framework can be generalized to simulate the dynamics of various scattering problems in quantum chromodynamics.
	
\end{abstract}

\maketitle

\section{Introduction}
\label{sec:intro}

The quantum chromodynamics (QCD) is a successful theory for the strong force that generates abundant phenomena associated with quarks and gluons. {First-principles} calculations of the quark-gluon systems based on QCD are crucial for understanding the early evolution of the Universe. However, such investigations are in general intractable with classical computers due to the exponential growth of the Hilbert space dimension for the strongly correlated quantum many-body systems.
Quantum computing provides a promising approach for addressing such computational challenges utilizing the synergy between computer science and quantum information theory. Initially proposed by Feynman \cite{feynman2018simulating}, quantum computing technology has undergone rapid development in the last decade. Quantum computing leads to cross-fertilizations among various fields such as quantum chemistry \cite{bauer2020quantum,cao2019quantum,RevModPhys.92.015003}, condensed matter physics \cite{PhysRevB.104.075159, PhysRevLett.121.110504, PRXQuantum.2.030307, PhysRevLett.79.2586, PhysRevA.92.062318, smith2019simulating}, nuclear physics \cite{PhysRevLett.120.210501,PhysRevC.105.064317,PhysRevC.106.034325,Du:2020glq,Turro:2023xgf,Turro:2023dhg,Du:2023bpw,Du:2021ctr,P_rez_Obiol_2023,lv2022qcsh,PhysRevC.105.064308,PhysRevC.105.064318,yang2023shadowbased,PhysRevResearch.1.033176,Wang:2024scd,Du:2024zvr,bhoy2024shellmodel,Liu:2024hmm}, quantum field theories \cite{Jordan_2012,PhysRevA.98.032331, jordan2019quantum, PhysRevD.101.074512,PhysRevD.106.054508,bauer2022quantum,PhysRevLett.131.081901,PRXQuantum.4.030323,PhysRevD.102.016007,PhysRevResearch.2.013272,Kreshchuk:2023btr,Muller:2023nnk,Barata:2021yri,Barata:2022wim,Barata:2023clv,Yao:2022eqm,Kreshchuk:2020dla,Kreshchuk:2020kcz,Kreshchuk:2020aiq,PhysRevA.103.042410,dimeglio2023quantum,su2024coldatom}, and many other areas \cite{dalzell2023quantum}.

Very few works \cite{Barata:2021yri,Yao:2022eqm,Barata:2022wim,Barata:2023clv} have been proposed for quantum simulating the real-time QCD dynamics, where the existing works are mainly based on the Trotter algorithms \cite{lloyd1996universal}. 
While the Trotter algorithms are straightforward and intuitive in implementations, they can be inefficient and less precise in general large-scale dynamics simulations. 
It is then important to develop an efficient and accurate simulation framework to investigate real-time QCD dynamics, where the post-Trotter algorithms such as the truncated Taylor series (TTS) \cite{PhysRevLett.114.090502}, quantum signal processing \cite{PhysRevLett.118.010501}, and quantum singular value transformation \cite{gilyen2019quantum} etc, can be beneficial. 

In this work, we explore the implementation of the TTS algorithm to simulate real-time QCD dynamics, which enables optimal gate scaling with the simulation error \cite{PhysRevLett.114.090502}. While the TTS algorithm only achieves near-optimal gate scaling with the simulation time and requires more ancilla qubits compared to the algorithms such as the quantum signal processing, it provides an intuitive approximation of the time-evolution operator and guarantees the straightforward implementation of the oblivious amplitude amplification \cite{PhysRevLett.114.090502,berry2017exponential} (OAA) scheme to ensure the success probability of simulations.
To the best of our knowledge, the application of the TTS algorithm in real-time QCD simulations has not yet been investigated in the literature. 

Our focus in this work is to present our TTS-based simulation framework for solving the problem of the scattering between an ultra-relativistic quark and a heavy nucleus, where the quark is scattered by the SU(3) color field generated by the heavy nucleus. 
We expect that this framework can be generalized to quantum simulate the dynamics of a broad range of QCD systems with good efficiency and precision.
We follow the time-dependent basis light-front quantization approach \cite{PhysRevD.88.065014,PhysRevD.101.076016,PhysRevD.104.056014}, and work with the eigenbases of the asymptotic scattering system where the external interaction is off.
These eigenbases are discretized and mapped to the multidimensional lattice bases, which are then encoded as binaries in the quantum registers utilizing the straightforward compact encoding scheme. 
We devise an input scheme that encodes the full light-front (LF) Hamiltonian of the scattering system employing the linear combination of unitaries \cite{childs2012hamiltonian}. Exploiting the operator structures of the LF Hamiltonian of the SU(3) gauge field, this Hamiltonian input scheme takes advantage of the quantum Fourier transformation (qFT) \cite{nielsen_chuang_2010} for efficiency.
We employ the TTS algorithm to simulate the scattering dynamics based on our qFT-assisted Hamiltonian input scheme.
We show that the transition probabilities (which in turn determine the elastic and inelastic cross section) and various quantities of the scattering system can be evaluated from the quantum simulations.
We present detailed analysis of the qubit and gate costs for TTS-based simulation framework, which are compared with the cost of the Trotter-based approach \cite{lloyd1996universal}.

We illustrate our TTS-based simulation framework with a model problem, where we employ the classical color-glass-condensation (CGC) theory \cite{gelis2010color} to model the SU(3) color field generated by the nucleus. Within a restricted model space, we design the quantum circuit for implementing the TTS algorithm based on the qFT-assisted Hamiltonian input scheme. We perform our simulations with the IBM QASM simulator of the Qiskit package \cite{Qiskit}. Based on the simulation results, we also compute a set of quantities of the quark under scattering. We benchmark our results with those from the Trotter algorithm \cite{lloyd1996universal} and from the classical calculations.

The arrangement of this paper is as follows. In Sec.~\ref{sec:theory}, we briefly introduce the LF formalism for an ultra-relativistic quark-nucleus scattering and establish our notation. In Sec.~\ref{sec:basis_encoding_scheme}, we set up the encoding scheme, which maps the physical degrees of freedom to the computational basis of quantum computers. In Sec.~\ref{sec:H_input_model_sec}, we introduce the qFT-assisted Hamiltonian input scheme that takes advantage of the operator structure of the LF Hamiltonian. Section~\ref{sec:algorithms} provides the details of the TTS and Trotter algorithms for the dynamics simulations. We introduce the model problem for our numerical demonstration in Sec.~\ref{sec:Model_Problem}. 
We present the results and discussions for the model problem in Sec.~\ref{sec:results_discussion}.  We conclude in Sec. \ref{sec:conclusions_outlook}, where we also provide an outlook.

\section{Theory}
\label{sec:theory}

To make our discussion self-contained, we present the necessary details of the LF Hamiltonian of the ultra-relativistic quark-nucleus scattering system and the formalism of the time-dependent basis light-front quantization approach for solving scattering dynamics. 
Interested readers are referred to Refs. \cite{PhysRevD.88.065014,PhysRevD.101.076016} for more details.
Our special foci in this section are on the structure of the Hamiltonian operator, the choice of the basis, and the lattice discretization of the elected basis; the properties of these theoretical ingredients are exploited in our algorithmic designs (Secs. \ref{sec:basis_encoding_scheme}, \ref{sec:H_input_model_sec}, and \ref{sec:algorithms}).

\subsection{LF Hamiltonian of the scattering system}
\label{sec:LF_Hamiltonian_scattering_Sys}

We consider the problem of an ultra-relativistic quark scattered by the SU(3) color field generated by a nucleus, which was described in Ref.~\cite{PhysRevD.101.076016}. We employ the LF field theory to solve the dynamics of the scattering process. 
For exploratory purpose, we start with a simple setup, in which we only retain the Fock state of a single quark denoted as $| q \rangle $. The QCD Lagrangian with this Fock sector truncation reads~\cite{PhysRevD.101.076016}
\begin{equation}
	\mathcal{L}_{q} = \bar{\Psi}\left(i\gamma^{\mu}\boldsymbol{D}_{\mu}-\boldsymbol{m} \right)\Psi,
	\label{eq:LF_Lagrangian}
\end{equation}
where $ \Psi $ denotes the quark field operator. The coupling with the gauge field is defined with the covariant derivative $\boldsymbol{D}^{\mu} \equiv \partial_{\mu}\boldsymbol{I}+ig \boldsymbol{ \mathcal{A} }^{\mu} $, where $\boldsymbol{I}$ is the 3 by 3 identity matrix in the color space, $g$ denotes the coupling constant,  $  \boldsymbol{ \mathcal{A} }^{\mu} = \mathcal{A}^{a \mu }\boldsymbol{T}^{a} $ is the local gauge field generated by the nucleus with $\boldsymbol{T}^a$ ($a=1,\ 2,\cdots,8$) being the ${\rm SU}(3)$ generator and $\mathcal{A}^{a}_{\mu} $ being the color vector potential. $ \boldsymbol{m} = m \boldsymbol{I} $ denotes the mass matrix that is diagonal in the color space, and $m $ is taken to be the quark mass (we assume the  quarks of different species to be of the same mass, which is taken to be $0.02$ GeV in this work).

We obtain the LF Hamiltonian from the LF Lagrangian via the standard Legrendre transformation as \cite{Brodsky:1997de}
\begin{equation}
	\boldsymbol{P}^{-}=\int dx^{-}d^{2}\boldsymbol{x}_{\perp} \Big[ \frac{1}{2}\bar{\Psi}\gamma^{+}\frac{ \boldsymbol{ m }^{2} - \boldsymbol{\nabla}_{\perp}^{2}}{i\partial^{+}}\Psi+g\bar{\Psi}\gamma^{\mu}\boldsymbol{T}^{a}\Psi\mathcal{A}_{\mu}^{a} + \frac{g^{2}}{2}\bar{\Psi}\gamma^{\mu}\boldsymbol{T}^{a}\mathcal{A}_{\mu}^{a}\frac{\gamma^{+}}{i\partial^{+}}\gamma^{\nu}\boldsymbol{T}^{b}\mathcal{A}_{\nu}^{b}\Psi  \Big] ,
	\label{eq:LF_Hamiltonian}            
\end{equation}
where $ \gamma ^{\mu} $ denotes the gamma matrix, with $ \mu = +, - , \perp $. 
The conventions and notations of the LF field theory are shown in Appendix \ref{sec:APP_conventions}, which can also be found in Ref.~\cite{PhysRevD.101.076016}.
When higher Fock sectors of the physical quark are included, one expects additional terms (e.g., those corresponding to the gluon emission and absorption processes, and the kinetic energy of the gluon) in the resulting LF Hamiltonian. Such terms can be systematically obtained following the prescriptions presented in Refs. \cite{BRODSKY1998299,Harindranath2005}.

To simplify our discussion below, we rewrite $\boldsymbol{P}^{-}$ in terms of the kinetic energy $\boldsymbol{P}^{-} _{\rm QCD} $ and the interaction $ \boldsymbol{V} $ as
\begin{equation}
	\boldsymbol{P}^{-} = \boldsymbol{P}^{-} _{\rm QCD} + \boldsymbol{V} ,
	\label{eq:LF_Hamiltonian_1} 
\end{equation}
with 
\begin{align}
	\boldsymbol{P}^{-} _{\rm QCD} & \equiv \int dx^{-}d^{2}\boldsymbol{x}_{\perp} \ \frac{1}{2}\bar{\Psi}\gamma^{+}\frac{ \boldsymbol{ m }^{2}- \boldsymbol{ \nabla }_{\perp}^{2}}{i\partial^{+}}\Psi , \label{eq:kinetic_part} \\
	\boldsymbol{V}  & \equiv \int dx^{-}d^{2}\boldsymbol{x}_{\perp}\ \Big[ g \bar{\Psi}\gamma^{\mu} \boldsymbol{T}^{a} \Psi \mathcal{A}_{\mu}^{a} + \frac{g^{2}}{2}\bar{\Psi}\gamma^{\mu}\boldsymbol{T}^{a} \mathcal{A}_{\mu}^{a} \frac{\gamma^{+}}{i\partial^{+}}\gamma^{\nu}\boldsymbol{T}^{b}\mathcal{A}_{\nu}^{b}\Psi \Big] . \label{eq:interaction_part}
\end{align}
We remark that the light-cone gauge $ \boldsymbol{ \mathcal{A} }^{+} = 0 $ is adopted in obtaining $ \boldsymbol{P}^{-} $. As we retain only the leading sector $|q\rangle $, the kinetic term $ \boldsymbol{P}^{-} _{\rm QCD} $ contains only the quark kinetic energy. By the division in Eq. \eqref{eq:LF_Hamiltonian_1}, we note that 1) $ \boldsymbol{P}^{-} _{\rm QCD} $ (referred to as the reference Hamiltonian) describes the kinetics of the system when the external interaction is turned off (e.g., before and after the scattering); and 2) $ \boldsymbol{V} $ (referred to as the interaction Hamiltonian) determines the dynamics of the system by coupling the system with the external gauge field.

We note that $ \boldsymbol{P}^{-} _{\rm QCD}  $ is local in the longitudinal and transverse momentum spaces. $ \boldsymbol{V} $ is local in the transverse coordinate space. In general, the background field generated by the heavy nucleus is time varying and presents explicit dependence on the longitudinal coordinate. Hence, $\boldsymbol{V} $ varies with the LF time $x^+$ and longitudinal coordinate $x^-$. For demonstration purposes, however, we assume that gauge field depends on neither $x^+$ nor $x^-$ in this work. In other words, we take $\boldsymbol{V} $ to be constant of LF time, and it does not carry the longitudinal momentum. We comment that it is straightforward to generalize the approach in this work to simulate the scattering with the time-dependent external gauge field \cite{PhysRevD.101.076016,PhysRevD.104.056014}.

\subsection{Dynamics simulation}

The dynamics of the scattering system can be solved based on the equation of motion (EOM). We proceed with the Schr\"odinger picture in this work. The EOM is defined according to the LF Hamiltonian $ \boldsymbol{P}^{-} $ of the scattering system as
\begin{equation}
	i\frac{\partial}{\partial x^{+}}\ket{\psi;x^{+}}= \frac{1}{2} \boldsymbol{P}^{-} \ket{\psi;x^{+}} ,
	\label{eq:Schrodinger_eq}
\end{equation}
where $ \ket{\psi;x^{+}} $ denotes the state of the scattering system at the moment $x^+$. The EOM admits the solution
\begin{equation}
	\ket{\psi;x^{+}}=\boldsymbol{\mathcal{U}}\left(x^{+}\right) \ket{\psi;0},
	\label{eq:time_evo}
\end{equation}
where $ \ket{\psi;0} $ denotes the initial state (we take the scattering to start at $x^+=0$). The time-evolution operator is
\begin{equation}
	\boldsymbol{\mathcal{U}}\left(x^{+}\right) = {\mathcal{T}}_+ \exp\left\{ - \frac{i}{2} \int_{0}^{x^{+}}ds^+ \big[  \boldsymbol{P}^{-} _{\rm QCD} + \boldsymbol{V}  \big] \right\} ,
	\label{eq:time_evo_op}
\end{equation}
with $ {\mathcal{T}}_+ $ being the time-ordering operator towards the positive LF time $x^+$. Once the initial state $ \ket{\psi;0} $ and the explicit form of the LF Hamiltonian $\boldsymbol{P}^- $ are known, we can, in principle, solve the state $ \ket{\psi;x^{+}} $ at arbitrary LF time $x^+ \geq 0 $ during the scattering. 

Based on $ \ket{\psi;x^{+}} $, we can evaluate the expectation of the quantity defined by the operator $ \boldsymbol{O} $ during the scattering as
\begin{equation}
	\braket{\boldsymbol{O}(x^+)}= \braket{\psi;x^+| \boldsymbol{O} |\psi;x^+} . \label{eq:observable_expectation}
\end{equation}
By projective measurements, we can also probe the instantaneous transition probability of the scattering system. In particular, by setting $ \boldsymbol{O} = | \phi _f \rangle \langle \phi _f | $ (with $ \ket{\phi_f}$ being the state of interest), the transition probability from $  \ket{\psi;0} $ to  $ \ket{\phi_f}$ can be evaluated as 
\begin{equation}
	\big| c_{\phi _f}(x^+) \big|^2 \equiv  \braket{\psi;x^+| \phi _f \rangle \langle \phi _f |\psi;x^+} .
\end{equation} 

\subsection{Basis representation}

To solve the EOM, one elects a set of bases and constructs the corresponding basis representation. Within the basis representation, the EOM can be solved numerically as an initial value problem. Convenient choice of the basis set should preserve the symmetries of the scattering system and should be advantageous in reducing the complexity in numerical calculations.

We adopt the idea of the time-dependent basis light-front quantization \cite{PhysRevD.88.065014,PhysRevD.101.076016,PhysRevD.104.056014} and elect our basis to be the eigenbasis of the reference Hamiltonian $ \boldsymbol{P}^{-} _{\rm QCD} $ solved from the eigenequation 
\begin{equation}
	\boldsymbol{P}^{-} _{\rm QCD} \ket{\beta} = K_{\beta} \ket{\beta} ,
	\label{eq:eigenEquation}
\end{equation}
where $ K_{\beta} $ and $ \ket{\beta} $ are the eigenenergies and the corresponding eigenstates, respectively. We can define $\ket{\beta}$ as
\begin{equation}
	\ket{\beta} \equiv \ket{p^+, p^1, p^2, \lambda,  c} = \ket{p^+}\otimes\ket{p^1}\otimes\ket{p^2} \otimes \ket{\lambda} \otimes\ket{c} , 
	\label{eq:eigenBasis}
\end{equation}
where $p^1$ and $p^2$ are the components of the transverse momentum. The transverse momentum is denoted as $ \boldsymbol{p}^{\perp} = (p^1, p^2) $. $ \lambda = \pm {1}/{2}$ denotes the helicity of the quark, and $c$ denotes the color of the quark. In the SU(3) gauge theory, the color $c$ takes the value of `Red', `Green', and `Blue'. $p^+$ denotes the longitudinal momentum. 
We comment that this basis is similar to that of the momentum discretization in Refs. \cite{Pauli:1985pv,PhysRevD.35.1493}.

We construct the basis representation with the eigenbases $\{ \ket{\beta} \}$. In the basis representation, the state vector of the scattering system $ \ket{\psi;x^{+}} $ admits the decomposition
\begin{equation}
	\ket{\psi;x^{+}} = \sum  _{\beta} c_{\beta }(x^+) | \beta \rangle ,
	\label{eq:amplitude_basis_expansion}
\end{equation}
where $ c_{\beta}(x^+) \equiv \braket{\beta|\psi;x^+} $ denotes the coefficient. The initial state $ \ket{\psi; 0} $ is represented by a vector that consists a set of coefficients $\{ c_{\beta}(0) \} $ according to the above expansion. The time-evolution operator $ \boldsymbol{\mathcal{U}}\left(x^{+}\right) $ [Eq. \eqref{eq:time_evo_op}] becomes a matrix. Being a vector consisting of the coefficients $\{ c_{\beta}(x^+) \}$, the state $\ket{\psi;x^{+}} $ can be solved according to Eq. \eqref{eq:time_evo} in terms of matrix-vector multiplication. 

We comment that our choice of the basis $ \{ \ket{\beta} \}$ in Eq. \eqref{eq:amplitude_basis_expansion} enables the direct projection of the state vector of the scattered system to the eigenstates of the system after the scattering (when the interaction is off); this facilitates the computation of scattering observables such as the elastic and inelastic cross sections. With $ \ket{\psi;x^{+}} $ and $ \boldsymbol{O} $ in the basis representation, one can also evaluate the quantity $ \braket{\boldsymbol{O}(x^+)} $ according to Eq. \eqref{eq:observable_expectation} in terms of matrix-vector multiplication.

It is worth noting that the above mentioned matrix-vector multiplication operations can become intractable with classical computers for complex applications, e.g., in which the contributions from higher Fock sectors of the quark system, and/or multiple quarks and gluons are involved. In such cases, the required Hilbert space dimension grows exponentially. We anticipate that our approach developed from the simplified quark-nucleus scattering problem could lay the foundation to address such computationally challenging problems.

\subsection{Structure of the LF Hamiltonian matrix}
\label{sec:Structure_analysis_Hamiltonian}

The LF Hamiltonian $\boldsymbol{P}^{-} $ [Eq. \eqref{eq:LF_Hamiltonian_1}] possesses a hybrid matrix structure within the elected basis set $\{ \ket{\beta }\} = \{  \ket{p^+,p^1,p^2,\lambda , c} \} $. The matrix element of the LF Hamiltonian $ \boldsymbol{P}^- $ in the basis representation receives contributions from the reference Hamiltonian and the interaction part as
\begin{align}
	\langle \beta | \boldsymbol{P}^{-} | \beta ' \rangle = \langle \beta | \boldsymbol{P}^{-}_{\rm QCD} | \beta ' \rangle + \langle \beta | \boldsymbol{V} | \beta ' \rangle .
	\label{eq:LF_Hamiltonian_with_QFT}
\end{align}
We note that the matrix  of the reference Hamiltonian $\boldsymbol{P}^{-} _{\rm QCD} $ is diagonal according to our choice of the basis representation [Eq. \eqref{eq:eigenEquation}], i.e., $ \langle \beta | \boldsymbol{P}^{-}_{\rm QCD} | \beta ' \rangle  = K_{\beta} \delta _{\beta \beta '}$.
The matrix representation of the interaction Hamiltonian $ \boldsymbol{V} $ is more complex in the basis set $\{  \ket{p^1} \otimes \ket{p^2} \}$ because $ \boldsymbol{V} $ is diagonal in the basis set $\{  \ket{x^1} \otimes \ket{x^2} \}$. Note that $ \boldsymbol{V} $ does not carry the longitudinal momentum according to our assumption that the gauge field does not depend on $x^-$.

One can implement the Fourier transform to switch between the transverse momentum basis and the transverse coordinate basis, in which $ \boldsymbol{V} $ admits a convenient expression. In particular, we can define $ \mathcal{F} \equiv \mathcal{F}_{ 1} \otimes \mathcal{F}_{ 2} $ to denote a two-dimensional Fourier transform, where $\mathcal{F}_i $ ($i=1,2$) transforms from the basis set $ \{ \ket{p^i} \} $ to the basis set $ \{ \ket{x^i} \} $. With the Fourier transform, we transform from the basis set $\{ \ket{\beta} \} $ to $\{ \ket{\alpha} \} \equiv  \{  \ket{p^+,x^1,x^2,\lambda , c} \}$. The matrix element $ \langle \beta | \boldsymbol{V} | \beta ' \rangle  $ can then be written as
\begin{equation}
	\langle \beta | \boldsymbol{V} | \beta ' \rangle = \sum _{\alpha} \sum _{\alpha '}  \langle \beta | \alpha \rangle \langle \alpha |  \boldsymbol{V} | \alpha' \rangle \langle \alpha' | \beta ' \rangle , \
	\label{eq:interaction_FT}
\end{equation}
where it is, in general, more convenient to calculate the matrix element $ \langle \alpha |  \boldsymbol{V} | \alpha' \rangle $. 
Combining Eqs. \eqref{eq:LF_Hamiltonian_with_QFT} and \eqref{eq:interaction_FT}, we denote the matrix form of the LF Hamiltonian as
\begin{equation}
	P^- = P^-_{\rm QCD} + \mathcal{F} V \mathcal{F}^{\dag} ,
	\label{eq:closed_form_LF_Hamiltonian}
\end{equation}
where $  P^- $ denotes the matrix of $ \boldsymbol{P}^{-} $ in the basis set $\{ \ket{\beta} \}$. 
$ V $ denotes the matrix of $ \boldsymbol{V}$ in the basis set $\{ \ket{\alpha} \}$. $\mathcal{F}$ denotes the matrix of the Fourier transform with elements $ \langle \beta | \alpha \rangle  $, while $ \mathcal{F} ^{\dag} = \mathcal{F}^{-1}$ is the inverse Fourier transform. It is worth noting that, while $V$ (the matrix of $ \boldsymbol{V}$ in the basis set $\{ \ket{\beta} \}$) is not a sparse matrix and is of complex structure, it can be expressed in terms of the sparse matrix $ \mathcal{F} V \mathcal{F}^{\dag} $  with the actions of the two-dimensional Fourier transform.

\subsection{Lattice discretization}
\label{sec:lattice_discretization_XXL}

In the following, we first introduce the lattice discretization of the transverse coordinate and momentum bases, and show the discretized Fourier transform between the two sets of bases (in terms of lattice states). Then, we discuss the lattice discretization of the coordinate and momentum in the longitudinal direction where special care is required due to the positiveness of the longitudinal momentum.
Finally, we present the scheme to map the helicity and color degrees of freedom to respective lattice states for consistency. 

As a brief summary of this section, we map the elected basis set to the multidimensional lattice space with proper regularizations and discretizations, where we take advantage of the efficient Fourier transform and its inverse to switch between the lattice states corresponding to $ \{ \ket{p^+, p^1, p^2, \lambda,  c} \}$ [Eq. \eqref{eq:eigenBasis}] and those corresponding to $ \{ \ket{p^+, x^1, x^2, \lambda,  c} \}$.

\subsubsection{Transverse degrees of freedom}

We start by discretizing the basis $\ket{  \boldsymbol {x}^{\perp } } = \ket{ x^1 } \otimes \ket{ x^2 }$ in two-dimensional transverse coordinate space. In each transverse direction, we employ a lattice that spans from $-L^{\perp}$ to $L^{\perp}$ with the periodic boundary condition imposed. We adopt $ 2 N^{\perp} $ sites for the discretization, and the lattice spacing is $a^{\perp}_r = L^{\perp}/N^{\perp}$.
With this, the components of the transverse coordinates $  \boldsymbol{x}^{\perp } $ are
\begin{equation}
	x^i = n^{i} a^{\perp}_r , \label{eq:lattice_R}
\end{equation}
with $i = 1,\ 2$ and $ n^i = -N^{\perp} ,\ -N^{\perp} +1,\ \dots,\ N^{\perp} -1 $. Accordingly, we map the state $ \ket{ \boldsymbol{x}^{\perp }  } =  \ket{ x^1,\ x^2 }$ to the lattice state $ \ket{ \boldsymbol{n}^{\perp } }  = \ket{ n^1,\ n^2 }$. 

We also discretize the transverse momentum $ \ket{ \boldsymbol{p }^{\perp } } = \ket{ p^1,\ p^2 } $ utilizing a two-dimensional lattice that is reciprocal to that for discretizing the transverse coordinates. We map the discretized transverse momentum $ \ket{ \boldsymbol{p}^{\perp } } = \ket{ p^1,\ p^2 } $ to the lattice state $ \ket{ \boldsymbol{q}^{\perp } } = \ket{ q^1, \ q^2 } $ as
\begin{equation}
	p^i = q^i a^{\perp}_p , \label{eq:lattice_P}
\end{equation}
with the lattice sites being $ q^i = -N^{\perp} ,\ -N^{\perp} +1,\ \dots,\ N^{\perp} -1 $ and $i=1,\ 2$. The spacing of the transverse lattice in each direction is $ a^{\perp}_p = 2\pi /(2L^{\perp})  $. As such, $p^i  $ spans from $- \pi N^{\perp}  /L^{\perp}$ to $ \pi N^{\perp}  /L^{\perp} $. Based on the lattice spacing and extension, one can also introduce the infrared cutoff, $\Lambda _{\rm IR} =  \pi  /L^{\perp}$, and the ultraviolet cutoff $ \Lambda _{\rm UV} = \pi N^{\perp}  /L^{\perp}$ for the transverse momentum. We find that $ \Lambda _{\rm UV} $ and $\Lambda _{\rm IR}$ are related to $ N^{\perp} $ as $ N^{\perp} = \Lambda _{\rm UV}/ \Lambda _{\rm IR} $ \cite{PhysRevD.101.076016}.

One can define the discrete Fourier transform $\mathcal{F}$ and its inverse $\mathcal{F}^{-1}$ between the lattice states  $ \ket{ \boldsymbol{n}^{\perp }  } $ and $ \ket{ \boldsymbol{q}^{\perp }  } $, which reads \cite{somma2015quantum}
\begin{align}
	\ket{ \boldsymbol{q}^{\perp }  }  & =\frac{1}{2 N^{\perp}} \sum_{n^1=-N^{\perp}}^{N^{\perp}-1} \sum_{n^2=-N^{\perp}}^{N^{\perp}-1} e^{+i2\pi\frac{ \boldsymbol{q}^{\perp }  \cdot \boldsymbol{n}^{\perp } }{2N^{\perp }}}  \ket{ \boldsymbol{n}^{\perp }  }  \equiv\mathcal{F}  \ket{ \boldsymbol{n}^{\perp }  }   , \label{eq:FT_n_to_q} \\
	\ket{ \boldsymbol{n}^{\perp }  }  & =\frac{1}{2N^{\perp}} \sum_{q^1=-N^{\perp}}^{N^{\perp}-1} \sum_{q^2=-N^{\perp}}^{N^{\perp}-1}  e^{- i2\pi\frac{ \boldsymbol{q}^{\perp }  \cdot \boldsymbol{n}^{\perp } }{2N^{\perp }}}   \ket{ \boldsymbol{q}^{\perp } } \equiv\mathcal{F}^{-1}  \ket{ \boldsymbol{q}^{\perp }  }  \label{eq:FT_q_to_n}.
\end{align}
One can implement the above transformation via the efficient qFT \cite{nielsen_chuang_2010,somma2015quantum} in quantum computing.

\subsubsection{Longitudinal degree of freedom}

We discretize the longitudinal momentum $p^+$ of the quark by confining it to a longitudinal lattice of extension $x^- = 2 L^{\parallel }$, where the antiperiodic boundary condition is applied for the fermionic system. The longitudinal momentum of the quark is discretized as \cite{PhysRevD.104.056014}
\begin{align}
	p^+ = q^+ a^{\parallel}_p ,
\end{align}
with the longitudinal lattice sites being $q^+ = \frac{1}{2}, \ \frac{3}{2}, \ \frac{5}{2}, \ \cdots $. The resolution $ a^{\parallel}_p \equiv \pi/L^{\parallel} $ of the discretized longitudinal momentum depends on the lattice extension $ L^{\parallel } $. In this way, we map the state $\ket{p^+}$ to the corresponding lattice state $\ket{q^+}$. With the cutoff  $p^+_{\rm max} $ to the longitudinal momentum of the scattering system, we can relate $p^+_{\rm max} $ to the number of sites (modes) as $ N^{\parallel}  = \lceil q^+_{\rm max} \rceil = \lceil p^+_{\rm max} / a^{\parallel}_p \rceil $.

We note that the Fourier transform between the longitudinal momentum basis set $\{ \ket{p^+} \}$ and the longitudinal coordinate basis set $\{  \ket{x^-}  \}$ takes additional cautions in general numerical treatments. To avoid discussions on this involving topic, we restrict our focus to the set of problems where the Hamiltonian of the scattering system is local in the longitudinal momentum space. Or, equivalently, we assume that the gauge field is independent of the variable $x^-$. Even with this restriction, one can still treat a wide array of scattering problems in which the longitudinal momentum conserves (there can be various modes that are specified by distinct $p^+$ values in the longitudinal direction). We expect that this assumption can be relaxed in future generalizations.

\subsubsection{Helicity and color degrees of freedom}

The helicity and color of the quark takes discrete values. In particular, the helicity  state is $ \ket{\lambda} $ with $\lambda =\pm 1/2$. We can map these two basis states to a binary lattice state $\ket{n_{\lambda}}$ ($n_{\lambda}=0,\ 1$), with $\ket{-{1}/{2}} \mapsto \ket{0} $ and $\ket{ {1}/{2} } \mapsto \ket{1} $. Similarly, the color states $\ket{ c }$ (with $c=$ Red, Green, and Blue) can be mapped to the lattice states as
 \begin{equation}
 	\ket{\text{Red}} \mapsto \ket{0}, \ \ket{\text{Green}} \mapsto \ket{1}, \ \ket{\text{Blue}} \mapsto \ket{2}.
 \end{equation}

\section{Basis encoding scheme}
\label{sec:basis_encoding_scheme}

We employ the compact encoding scheme to map the lattice-discretized basis $\ket{ p^+, p^1,p^2,\lambda ,c } $ [Eq. \eqref{eq:eigenBasis}] to the register state on the quantum computer. The details of the scheme are as follows.

\paragraph{Longitudinal basis $ \ket{p^+}$.}
We map the longitudinal momentum $ \ket{p^+}$ to the lattice state $\ket{ q^+ } $, where we have $q^+ = \frac{1}{2}, \ \frac{3}{2}, \ \frac{5}{2}, \ \cdots $ for the quark. In our scheme, we encode $\lceil q^+ \rceil $ in binaries. Provided the total number of sites being $N^{\parallel}  = \lceil q^+_{\rm max} \rceil = \lceil  p^+_{\rm max} / a^{\parallel}_p \rceil $, it takes $ \lceil  \log _2 N^{\parallel}  \rceil $ qubits to encode all the possible values of $q^+$ (we assume $ N^{\parallel} \geq 1$). 

\paragraph{Transverse bases $ \ket{p^1,p^2} $ and $ \ket{x^1,x^2} $.}
The transverse momentum $ \ket{p^1,p^2} $ is mapped to the lattice state $\ket{q^1,q^2}$, where the transverse lattice site $ q^i  $ (with $i=1,\ 2$) takes the values of $ -N^{\perp} ,\ -N^{\perp} +1,\ \dots,\ N^{\perp} -1 $. We define the encoding scheme for $ q^i $ ($i=1,\ 2$) as
\begin{equation}
	{-N^{\perp} } \mapsto {0\cdots 0 0},\ 	(-N^{\perp} +1) \mapsto {0\cdots 0 1}, \ (-N^{\perp} +2 ) \mapsto {0\cdots 10}, \ \cdots .
	\label{eq:encoding_scheme_trans_momentum}
\end{equation} 
As such, it takes $ \lceil \log_2 2N^{\perp} \rceil $ qubits to encode the $2N^{\perp}$ lattice sites on each transverse lattice. One then needs $ 2 \lceil \log_2 2N^{\perp} \rceil $ qubits to encode all the possible lattice sites for the transverse momentum.

Meanwhile, we map the transverse coordinate state $ \ket{ \boldsymbol{x}^{\perp }  } =  \ket{ x^1,\ x^2 }$ to the lattice state $ \ket{ \boldsymbol{n}^{\perp }  }  = \ket{n^1, n^2} $, where $ n^i = -N^{\perp} ,\ -N^{\perp} +1,\ \dots,\ N^{\perp} -1 $. We implement the same scheme described in Eq. \eqref{eq:encoding_scheme_trans_momentum} to encode $n^{i}$ ($i=1,\ 2$) as binary strings. 

We recall that $n^i$ and $q^i$ denote the components of states in reciprocal lattice spaces. According to Eqs. \eqref{eq:FT_n_to_q} and \eqref{eq:FT_q_to_n}, one can apply the qFT between the register states that correspond to $ \ket{q^i} $ and $ \ket{n^i}$, respectively. However, it is worth noting that the lattice states, $\ket{n^i}$ and $\ket{q^i}$, both count from $-N^{\perp}$ to $N^{\perp}-1$, while the corresponding register states count from $0$ to $2N^{\perp} - 1$. In view of this shift, we adopt a modified version of the qFT \cite{somma2015quantum} in our simulations.

\paragraph{Helicity basis $\ket{\lambda}$ and color basis $\ket{c}$.}

We also encode the helicity and color states in terms of binaries on the quantum computer. The helicity is mapped to the lattice state $ \ket{n_{\lambda}} $ with $ n_{\lambda } =0,\ 1$; we encode the states $ \ket{n_{\lambda }  = 0} $ and $ \ket{n_{\lambda} =1}$ as the binaries 0 and 1, respectively. Similar to helicity, the color basis states $\ket{c}$ are mapped to the lattice state $\ket{n_c}$ with $n_c=$ 0, 1, and 2. In a straightforward approach, we encode $\ket{n_c} = \ket{0}, \ \ket{1} $ and $\ket{2}$ as the binary strings ${00}$, ${01}$, and ${10}$, respectively.

With the compact encoding scheme discussed above, the total number of qubits required to encode the lattice-discretized basis set of dimension $ 6 \cdot (2N^{\perp})^2 \cdot N^{\parallel} $ is
\begin{equation}
	N_{\rm sys} = 2 \lceil \log_2 (2N^{\perp}) \rceil   +  \lceil  \log _2 N^{\parallel}  \rceil + 3 .
	\label{eq:number_sys_bits} 
\end{equation} 
These are the number of qubits necessary to encode the state of the scattering system (hence the subscript ``sys" denotes). With the compact encoding scheme, $ N_{\rm sys} $ is the logarithm of the basis space dimension of the scattering system. 

For more complete Fock sector expansion of the quark system, one includes more constituent particles for the physical quark. It is straightforward to generalize the eigenbasis set $\{ \ket{ \beta } \}$ of the single quark to include the bases of these additional particles (in terms of tensor products) applying the same techniques illustrated above via the single quark sector. Meanwhile, analogous lattice-discretization and basis encoding schemes can be applied for the bases of these particles. As such, one again expects that the qubit cost for the calculations including higher Fock sectors of the quark system scales as the logarithm of the corresponding basis space dimension.

\section{Hamiltonian input schemes}
\label{sec:H_input_model_sec}

We divide the total LF Hamiltonian into the reference part and interaction part. In the chosen basis $\{ \ket{\beta } \}$, the reference Hamiltonian is diagonal. However, the interaction Hamiltonian is non-sparse in this basis. The major complexity\footnote{We recall that even though our interaction matrix is local in the spatial degrees of freedom (i.e., the longitudinal momentum space and the transverse coordinate space), it induces interaction between different helicity states $\{\ket{\lambda} \}$ and between different color states $\{\ket{c}\}$. However, we can treat the interaction Hamiltonian in the helicity and color bases $\{\ket{\lambda}\otimes \ket{c} \}$ directly as these bases are of limited dimensions.} is due to the fact that the interaction Hamiltonian is non-local in transverse momentum basis (it is local in transverse coordinate basis instead). This non-sparsity hampers the efficiency of the Hamiltonian input and hence the dynamics simulations. To address this difficulty, we utilize the Fourier transform for the efficient basis transformation and express the interaction Hamiltonian matrix in terms of a sparse matrix in the coordinate basis. Assisted by the Fourier transform, the efficiency of the input scheme of the total LF Hamiltonian can be improved.

In this work, we encode the sparse matrices $P^-_{\rm QCD}$ and $V$ in the LF Hamiltonian $P^-$ [Eq. \eqref{eq:closed_form_LF_Hamiltonian}] in terms of the linear combination of unitaries. It is also possible to implement other efficient sparse-matrix based input schemes \cite{Low2019hamiltonian,PhysRevLett.118.010501}. For the convenience of the following discussions, we scale $  P^-  $ by a factor of $1/2$ and denote $  H =  {P}^{-} /2 $. We can decompose the LF Hamiltonian as a linear combination of unitaries\footnote{Readers are also referred to Sec. \ref{sec:VB_model_space_and_Hamiltonian} for an illustration of the scheme.}
\begin{equation}
	 H = \frac{1}{2}  {P}_{\rm QCD}^{-}  + \frac{1}{2} \mathcal{F} ^{\dag} {V} \mathcal{F}  
	= \sum_{\ell _P=0}^{L_1-1} \alpha_{P, \ell _P} {h}_{P,\ell _P}  + \mathcal{F}^{\dag} \left( \sum_{\ell _V=0}^{L_2-1} \alpha_{V, \ell _V} {h}_{V,\ell _V} \right) \mathcal{F},
	\label{eq:Hamiltonian_decomposition}
\end{equation}
where ${h}_{P,\ell _P } $ and $ {h}_{V,\ell _V }  $ denote the unitaries for the expansions of $  {P}_{\rm QCD}^{-} /2$ and $ {V} /2$, respectively. $\alpha_{P,\ell _P} $ (with ${\ell _P} = 0,\ 1, \ \cdots , \ L_1-1 $) denotes the expansion coefficients of the matrix $  {P}_{\rm QCD}^{-} /2$, while $\alpha_{V,\ell _V} $ (with ${\ell _V} = 0,\ 1, \ \cdots , \ L_2 -1 $) is the expansion coefficients of the matrix $ {V} /2$. The total number of the unitaries in Eq. \eqref{eq:Hamiltonian_decomposition} is $L=L_1+L_2$. Without loss of generality, we assume $ \alpha_{P,\ell _P }, \ \alpha_{V,\ell _V} >0 $ (if needed, one can absorb a phase within the corresponding unitary).
The $\mathbb{L}^1$ norm of the expansion coefficients is obtained by summing up all these coefficients $\{ \alpha_{P, \ell _P} \}$ and $ \{ \alpha_{V, \ell _V} \}$, which is 
\begin{equation}
	\Lambda = \sum _{\ell _P =0}^{L_1-1}  \alpha _{P,\ell _P} + \sum _{\ell _V= 0}^{L_2-1} \alpha _{V, \ell _V} . 
\end{equation}
It is preferable to have the decomposition of $H$ in Eq. \eqref{eq:Hamiltonian_decomposition} with smaller $\Lambda$ and $L$ for more efficient dynamics simulations.

In this work, we adopt a straightforward approach to decompose $ P^-_{\rm QCD}$ and $ V $ in terms of the tensor product of Pauli matrices, i.e., both $ {h}_{P,\ell _P} $ and $ {h}_{V,\ell _V} $ are taken as Pauli strings. We can estimate the upper bound of the total number of terms $L$ as\footnote{We adopt the typical convention in computer science in this work. For any function $f$ and $g$, $f\in O(g)$ denotes that $f$ is asymptotically upper bounded by $g$, while $f\in \Theta (g)$ indicates that $f$ is asymptotically both upper and lower bounded by multiples of $g$.}
\begin{equation}
	L=L_1 + L_2 \in {O} (2 ^{ 2 \lceil \log 2 N^{\perp} \rceil  + \lceil \log N^{\parallel} \rceil }  ) .
	\label{eq:L_scaling_Pauli_product}
\end{equation}
This upper bound can be understood from the following analysis.

\paragraph{Scaling of $L_1$.}
We note that $ L_1 $ is dominated by the dimension of the discretized momentum basis (recall that the basis dimensions in the spin and color spaces are limited for the quark). The matrix $ {P}_{\rm QCD}^{-} $ is diagonal in the discretized bases $\{  \ket{p^+}\otimes\ket{p^1}\otimes\ket{p^2}  \} $, where ${p^i}$ ($i=1,2$) takes $2N^{\perp}$ values and $p^+$ takes $N^{\parallel}$ values according to our discussion in Sec. \ref{sec:basis_encoding_scheme}.  On the other hand, a general diagonal matrix of dimension $2^{D}$ ($D \in \mathbb{Z}^{\ast}$) can be expressed in terms of the Pauli strings, where each string contains $D$ Pauli matrices. While each constituent Pauli matrix can only be either the Pauli-$I$ or Pauli-$Z$ matrix, there exist $2^{D}$ distinct Pauli strings. As such, we have $L_1 \in {O} (2 ^{ 2 \lceil \log 2N^{\perp} \rceil  + \lceil \log N^{\parallel} \rceil }  )$. 

\paragraph{Scaling of $L_2$.}

Though the interaction Hamiltonian $ \boldsymbol{V} $ is not local in the helicity and color spaces, the bases $\{\ket{ \lambda } \otimes \ket{c} \}$ of these two spaces are limited. Hence, we see that $ L_2 $ is dominated by the dimension of discretized bases $\{  \ket{p^+}\otimes\ket{x^1}\otimes\ket{x^2}  \} $. According to the analysis in Sec. \ref{sec:Structure_analysis_Hamiltonian}, the matrix $ V $ is diagonal in the discretized bases $\{  \ket{p^+}\otimes\ket{x^1}\otimes\ket{x^2}  \} $. As $x^i$ ($i=1,2$) takes $2N^{\perp}$ values and $p^+$ takes $N^{\parallel}$ values, we follow the analysis in $L_1$ to estimate the upper bound of $L_2 $ to be $L_2 \in {O} (2 ^{ 2 \lceil \log 2N^{\perp} \rceil  + \lceil \log N^{\parallel} \rceil }  )$.

\paragraph{Total scaling of $L$.}

The sum of $ L_1 $ and $L_2 $ then gives Eq. \eqref{eq:L_scaling_Pauli_product}. We find that $L$ asymptotically scales as 
\begin{equation}
	L \in {O} \big( (N^{\perp}  )^2 N^{\parallel} \big) ,
	\label{eq:afjss}
\end{equation}
which is also the scaling of the dimension of $ \{ \ket{\beta} \} $ as the restricted spin and color basis dimensions contribute only a multiplicative factor. We recall $N^{\perp} = \Lambda _{\rm UV}/ \Lambda _{\rm IR} $ and $ N^{\parallel}  = \lceil p^+_{\rm max} / a^{\parallel}_p \rceil  $ from Sec. \ref{sec:lattice_discretization_XXL}. Therefore, we can rewrite the upper bound of $L $ as
\begin{equation}
	L \in {O} \left(   \left( \Lambda _{\rm UV}/ \Lambda _{\rm IR} \right)^2 \cdot \left( p^+_{\rm max} / a^{\parallel}_p \right)  \right) .
\end{equation}

We remark that, though being straightforward, the decomposition scheme in terms of Pauli strings yields the number of unitaries that scales proportionally to the Hilbert space dimension. Better choices of the unitaries for decomposing the diagonal matrices are preferable to improve the scaling of $L$ and hence the efficiency of simulation, which is an open question to address in future research efforts (see an alternative choice of unitaries in Appendix \ref{sec:decompo_to_one_sparse}). 

In addition to the Hamiltonian input scheme that is based on the unitaries discussed above, recent works \cite{Du:2024zvr,Du:2023bpw,Liu:2024hmm,Du:2024ixj,PhysRevA.104.042607} provide new approaches to input the Hamiltonian with efficiency, where these new input schemes do not depend on any specific choice of unitaries. We expect that they can improve the efficiency and practicability of our simulation framework discussed in this work.

\section{Algorithms for the quantum simulations}
\label{sec:algorithms}

An essential step for simulating the dynamics of the LF Hamiltonian $\boldsymbol{P}^{-} $ [Eq. \eqref{eq:LF_Hamiltonian_1}] is to construct the quantum circuit for the time-evolution unitary $ \boldsymbol{\mathcal{U}} (x^{+} )  = e^{-i \frac{1}{2} \boldsymbol{P}^{-} x^+ } $ [Eq. \eqref{eq:time_evo_op}]. 
By taking $ \boldsymbol{H} =  \boldsymbol{P}^{-} /2 $ and dividing the total simulation time $x^+$ into $r$ segments of equal step size $\tau = x^+ / r$, we have 
\begin{equation}
	\boldsymbol{\mathcal{U}} (x^{+} ) = \big[ \boldsymbol{\mathcal{U}} (\tau ) \big]^r = \big[ e^{-i  \boldsymbol{H} \tau } \big]^r .
	\label{eq:time_ev_unitary_Trotterized}
\end{equation}
Here, $ \boldsymbol{\mathcal{U}} (\tau ) $ denotes the time-evolution operator on a single time step of size $ \tau $. 

In this section, we present the details of the TTS algorithm and its adaption to solve Eq. \eqref{eq:time_ev_unitary_Trotterized} with our qFT-assisted Hamiltonian input scheme.
The details of the Trotter algorithm are also shown for comparison. We illustrate our circuit designs for both simulation algorithms, where we also present the analysis of the gate and qubit cost. 
The TTS algorithm enables dynamics simulations with optimal and near optimal gate scaling with the simulation error and time, respectively. Compared to the Trotter algorithm, the TTS algorithm is more efficient and precise for large-scale quantum simulations on future fault-tolerant quantum computers.

\subsection{The TTS algorithm}

We first discuss the TTS algorithm and its implementation with our qFT-assisted Hamiltonian input scheme [Sec. \ref{sec:H_input_model_sec}], where we also present the necessary details for the practical application of the OAA scheme. 
As its name suggests, the TTS algorithm approximates the time evolution unitary of each time step by its Taylor series up to certain truncation order [e.g., Eq. \eqref{eq:Taylor_expansion_model_1}]. The retained terms in the Taylor series can be expressed in terms of the linear combination of unitaries [e.g., Eq. \eqref{eq:V_j_label}]. One can then block encode the approximate (truncated) evolution unitary with the standard protocol of the linear combination of unitaries [e.g., Eq. \eqref{eq:TTS}]. 
In order to enhance the success probability of the time evolution, it is important to implement the OAA scheme \cite{PhysRevLett.114.090502,berry2017exponential}. In particular, by selecting the step size of the time evolution and by defining the amplification operator, one can amplify the success probability of implementing the desired time evolution unitary to almost 100$\%$ for each time step [e.g., Eqs. \eqref{eq:OAA_total_eq} and \eqref{eq:error_TTS_OAA_SingleStep}].

In particular, we have the LF Hamiltonian matrix $P^-$ given in Eq. \eqref{eq:closed_form_LF_Hamiltonian} in the basis representation. 
Within our input scheme, $ H =P^-/2$ is decomposed in terms of the linear combination of unitaries [Eq. \eqref{eq:Hamiltonian_decomposition}]. Following the TTS approach, we have the time-evolution operator $\boldsymbol{\mathcal{U}}_{K}(\tau )$ in the basis representation up to the truncation order $K$ as
\begin{equation}
	\mathcal{U}_K(\tau ) = \sum_{k=0}^{K}\frac{(-i\tau)^{k}}{k!} {H}^{k}  = \mathds{1} + \sum_{k=1}^{K}\sum_{\ell_{1},\ldots,\ell_{k}=0}^{L-1}\frac{(-i\tau)^{k}}{k!} \widetilde{\alpha} _{\ell_{1}} \cdots \widetilde{\alpha}_{\ell_{k}} \widetilde{h}_{\ell_{1}} \cdots \widetilde{h }_{\ell_{k}} ,
	\label{eq:Taylor_expansion_model_1} 
\end{equation}
where $\mathcal{U}_{K}(\tau )$ approximates the exact evolution unitary $ \mathcal{U} (\tau ) $ (with $K \rightarrow \infty $) up to the truncation error $ {O}(\epsilon)$ for a small evolution step $\tau$ of which the size is to be determined later. We take $ \widetilde{\alpha} _{\ell_{k}} \in \{  \alpha _{P,0},  \cdots , \alpha _{P,L_1 -1}, \alpha _{V,0}, \ \cdots , \ \alpha _{V,L_2-1} \} $, and $ \widetilde{h}_{\ell_{k}} \in \{ h_{P,0}, \cdots ,  h_{P,L_1-1} ,  \mathcal{F}^{\dag}  {h}_{V,0}  \mathcal{F}, \cdots , \ \mathcal{F}^{\dag}  {h}_{V,L_2-1}  \mathcal{F}    \} $ based on Eq. \eqref{eq:Hamiltonian_decomposition}. Without loss of generality, we sort the elements such that the first $L_1$ elements correspond to $P^-_{\rm QCD}/2$ and the next $L_2$ elements correspond to $V/2$ in each set, where one notes that $\{ h_{P,0}, \cdots ,  h_{P,L_1-1} \}$ are composed of the diagonal Pauli-$Z$ and Pauli-$I$. This sorting is also important to facilitate the implementation of the qFT in the circuit design, which will be clear below.

Following Refs. \citep{PhysRevLett.114.090502,Babbush_2016}, we rewrite Eq. \eqref{eq:Taylor_expansion_model_1} as
\begin{equation}
	\mathcal{U}_{K} (\tau ) = \sum_{j}\eta_{j} (\tau) {V}_{j},
	\label{eq:V_j_label}
\end{equation}
with $ \eta_j (\tau) \equiv \frac{\tau^{k}}{k!} \widetilde{\alpha}_{\ell_{1}}\cdots \widetilde{\alpha}_{\ell_{k}} $, $ {V}_{j}\equiv (-i)^{k} \widetilde{h}_{\ell_{1}}\cdots \widetilde{h}_{\ell_{k}} $, and $j $ being the multi-index $j \equiv (k,\ell_{1},\ldots,\ell_{k} ) $. Note that $ \mathcal{U}_{K} (\tau ) $ approaches full unitarity as $K \rightarrow \infty $, while ${V}_{j} $ is unitary. 

Following Ref. \cite{Babbush_2016}, we prepare the state $\ket{j}^{\otimes J}$ in the ancilla registers that are initialized in the state $ \ket{0} ^{\otimes J} $:
\begin{equation}
	\mathcal{P} (\tau) \ket{0} ^{\otimes J} = \frac{1}{\sqrt{\mathcal{N}_{K}}} \sum_{j} \sqrt{\eta_{j} (\tau) }\ket{j}^{\otimes J},
	\label{eq:prepare_oracle}
\end{equation}
where $J$ denotes the total number of qubits in the ancilla registers. $\mathcal{P} (\tau) $ is the ``preparation" operator. $ \mathcal{N}_{K} $ denotes the normalization factor. 

We also introduce the ``selection" operator to apply ${V}_{j}$ to the state $\ket{\psi} $ of the scattering system, which is encoded in the system register, controlled by the ancilla state $\ket{j}^{\otimes J} $. The selection operator $ \mathcal{S} $ acts as \cite{PhysRevLett.114.090502}
\begin{equation}
	\mathcal{S} \ket{j}^{\otimes J} \ket{\psi} \equiv \ket{j}^{\otimes J} {V}_{j} \ket{\psi}.
	\label{eq:select_Operator}
\end{equation}
Based on the operators $\mathcal{P} $ and $ \mathcal{S} $, one defines the TTS operator as \cite{PhysRevLett.114.090502}
\begin{equation}
	\mathcal{W} (\tau) \equiv \left(\mathcal{P}^{\dagger} (\tau) \otimes\mathds{1}\right) \mathcal{S} \left(\mathcal{P} (\tau) \otimes \mathds{1}\right) ,
	\label{eq:W_operator}
\end{equation}
where $ \mathcal{W} (\tau) $ block encodes $\mathcal{U}_{K}(\tau)$ as
\begin{equation}
	\mathcal{W} (\tau) \ket{0}^{\otimes J} \ket{\psi}=\frac{1}{\mathcal{N}_{K}} \ket{0}^{\otimes J} \mathcal{U}_{K}(\tau) \ket{\psi} + \sqrt{1- \frac{1}{\mathcal{N}^2_{K}} } \ket{\Phi^{\perp}},
	\label{eq:TTS}
\end{equation}
where $ \ket{\Phi^{\perp}} $ is orthogonal to $ \ket{0}^{\otimes J} \mathcal{U}_{K}(\tau) \ket{\psi} $.

\subsubsection{The OAA scheme}

According to Eq. \eqref{eq:TTS}, one can postselect those simulations that measure the ancilla registers being in the state $\ket{0}^{\otimes J}$, in which cases $ \mathcal{U}_{K} (\tau) $ is expected to be implemented to the system state $ \ket{\psi} $. The probability of such successful implementations is $1/ \mathcal{N}^2_{K} $, where
\begin{equation}
	\mathcal{N}_{K} \equiv\sum_{j} \lvert \eta_{j} \rvert = \sum _{k=0}^K \frac{\tau ^k}{k!} \Big( \sum _{\ell =0}^{L-1} \widetilde{\alpha} _{\ell} \Big)^k = \sum_{k=0}^{K} \frac{\left(\Lambda \tau\right)^{k}}{k!} ,
	\label{eq:N_factor}
\end{equation}
and $ \Lambda = \sum _{\ell =0}^{L-1} \widetilde{ \alpha }_{\ell} $ with $\widetilde{ \alpha }_{\ell} $ being the expansion coefficients in Eq. \eqref{eq:Hamiltonian_decomposition} 
(recall we take $\widetilde{ \alpha } _{\ell} >0 $). In the limit of $K \rightarrow \infty$, we have $ \mathcal{N}_{K} \rightarrow  e^{\Lambda \tau} $. That is, $	\mathcal{N}_{K} $ increases exponentially with $\tau $ and $ \Lambda $. The success probability of implementing $ \mathcal{U}_{K}(\tau) $ can be suppressed in the simulations with large $\tau $, and the total success probability of multiple implementations of $ \mathcal{U}_{K} (\tau) $ can be significantly suppressed.

One way to address the problem of the suppressed success probability is via the OAA scheme \cite{PhysRevLett.114.090502,berry2017exponential}. The OAA scheme helps to improve the success probability to almost $100\%$ for implementing $ \mathcal{U}_{K} (\tau)$ that is a nonunitary approximation of the unitary $ \mathcal{U} (\tau)$. 

To implement the OAA scheme, we take\footnote{A subtlety arises when $r= x^+ \Lambda / \ln 2 $ is not an integer. This subtlety can be treated with the techniques proposed in Ref. \cite{PhysRevLett.114.090502}.} 
\begin{equation}
	\tau = x^+/r = \ln 2 / \Lambda .
\end{equation}
We then have $ \mathcal{N}_{K} \approx 2 $ for the truncation order $K$ with $|| \mathcal{U}_{K}(\tau) - \mathcal{U} (\tau) || \in {O}(\epsilon )$. The amplification operator can be defined as \cite{PhysRevLett.114.090502,Babbush_2016}
\begin{equation}
      \mathcal{Q} (\tau) \equiv - \mathcal{W} (\tau) \mathcal{R}  \mathcal{W}^{\dagger} (\tau) \mathcal{R} ,
      \label{eq:amplification_operator_Q}
\end{equation}
where $ \mathcal{R} \equiv \big[ \mathds{1}-2 ( \ket{0} \bra{0} )^{\otimes J} \big] \otimes \mathds{1}$ denotes the reflection operator acting on the ancilla registers. With the projection operator ${\Pi} \equiv ( \ket{0} \bra{0} )^{\otimes J} \otimes \mathds{1} $ onto the ancilla state $\ket{0}^{\otimes J}$, it can be proved that \citep{PhysRevLett.114.090502,Babbush_2016}
\begin{equation}
	\Pi \mathcal{Q} (\tau) \mathcal{W}(\tau) \ket{0}^{\otimes J} \ket{\psi } = \ket{0}^{\otimes J} \Big( \frac{3}{\mathcal{N}_{K}} - \frac{4}{{\mathcal{N}_{K}}^{3}}  \Big) \mathcal{U} _{K} \ket{\psi} \approx \ket{0}^{\otimes J} \mathcal{U}_{K} (\tau) \ket{\psi} .
	\label{eq:OAA_total_eq}
\end{equation}

Following Ref. \cite{PhysRevLett.114.090502}, one can verify the error bound of the simulation via the OAA-assisted TTS implementation. Provided that $  (\mathcal{N}_{K} - 2) \in {O}( \epsilon )$ and $ || \mathcal{U}_{K} (\tau) - \mathcal{U} (\tau ) || \in {O} (\epsilon) $, one has $ ||  \mathcal{U} _{K} (\tau) \mathcal{U} _{K} ^{\dag} (\tau) - \mathds{1} || \in {O}(\epsilon)$ and Eq. \eqref{eq:OAA_total_eq} implies
\begin{equation}
	\Big| \Big|  \Pi \mathcal{Q}(\tau) \mathcal{W}(\tau) \ket{0}^{\otimes J} \ket{\psi }  - \ket{0}^{\otimes J}  \mathcal{U} (\tau ) \ket{\psi} \Big| \Big| \in {O} (\epsilon).
	\label{eq:error_TTS_OAA_SingleStep}
\end{equation}
That is, the OAA amplifies the success probability of the desired evolution from $1/\mathcal{N}_{K}^{2}$ to almost $1$.

\subsubsection{Circuit construction}

We present the circuit design to implement the TTS algorithm with our qFT-assisted Hamiltonian input scheme [Eq. \eqref{eq:Hamiltonian_decomposition}] in this section. 

\paragraph{Preparation operator $\mathcal{P} (\tau) $.}

The TTS algorithm implements the ancilla registers to encode the state $\ket{j} ^{\otimes J} = \ket{k}_{y_0} \ket{\ell_1}_{y_1}  \ket{\ell_2}_{y_2} \cdots  \ket{\ell_K}_{y_K}$, where  $0 \leq k \leq K$, $ 0 \leq \ell _v \leq L-1 $ for $v = 1,\ 2, \ \cdots , K$, and the subscript $y_i$ with $i = 0, \ 1, \cdots K$ specifies the registers. We can rewrite the operator $\mathcal{P} (\tau) $ in Eq. \eqref{eq:prepare_oracle} for the convenience of discussion as
\begin{equation}
	\mathcal{P} (\tau) \equiv \mathcal{P}_{y_0} (\tau) \otimes \mathcal{P}_{y_1} \otimes \mathcal{P}_{y_2} \cdots \otimes \mathcal{P}_{y_K} ,   
	\label{eq:total_prepare_operator}
\end{equation}
where the operator $ \mathcal{P}_{y_i}$ acts on the ancilla register $y_i$. 

The register $y_0$ encodes the order $k \in [0,K] $ of the Taylor series in unaries as the state $\ket{k} _{y_0} \equiv \ket{1^{k}0^{K-k}} _{y_0}$; $K$ qubits are required for this unary encoding. The action of $ \mathcal{P}_{y_0} (\tau) $ on the register $y_0$ is
\begin{equation}
	\mathcal{P}_{y_0} (\tau) \ket{0}_{y_0}^{\otimes K} = \frac{1}{\sqrt{\mathcal{N}_K}} \sum _{k=0}^{K} \sqrt{\frac{(\Lambda \tau)^k}{k!}} \ket{k} .
	\label{eq:prepare_oracle_y0}
\end{equation}

The gate construction follows the designs in Ref. \citep{Babbush_2016}. In particular,
utilizing the rotational gate $R_y(\theta) \equiv \exp [-i \theta Y /2] $ ($Y$ denoting the Pauli-Y gate), we can apply a single $ R_y(\theta _1 (\tau) )$ to the first qubit for the unary encoding. Then, we implement a sequence of controlled rotations, where the gate $R_y(\theta _k (\tau) ) $ is applied to the $k^{\rm th}$ qubit controlled by the $(k-1)^{\rm th}$ qubit with $k = 2, \ 3, \ \cdots , \ K$. The angle $\theta _k (\tau) $ with $ k \in [1, K]$ is defined as
\begin{equation}
	\theta_{k} (\tau) =2\arcsin\left(\sqrt{1-\frac{\left(\Lambda \tau\right)^{k-1}}{\left(k-1\right)!}\left(\sum_{q=k-1}^{K}\frac{\left(\Lambda \tau\right)^{q}}{q!}\right)^{-1}}\right).
	\label{eq:prepare_oracle_rotation_angle}
\end{equation} 

The register $y_k$ (with $k=1,\ \cdots , \ K $) encodes the expansion index $\ell _k \in [0, L-1]$ as the register state $\ket{\ell _k}_{y_k}$, where such state is weighted by $\sqrt{\widetilde{ \alpha }_{\ell _k }}$ [recall $ \widetilde{ \alpha }_{\ell _k}$ is the expansion coefficient indexed by $\ell _k$ in Eq. \eqref{eq:Hamiltonian_decomposition}]. With compact binary encoding scheme, it takes $ n_L = \lceil \log _2 L \rceil $ qubits in the register. The action of $\mathcal{P}_{y_k}$ on the register $y_k$ initialized as $\ket{0}^{\otimes n_L}_{y_k}$ is defined as
\begin{equation}
	\mathcal{P}_{y_k} \ket{0}^{\otimes n_L}_{y_k} = \frac{1}{\Lambda} \sum _{\ell _k =0}^{L-1} \sqrt{ \widetilde{ \alpha } _{\ell _k} } \ket{\ell _k} .
	\label{eq:prepare_oracle_Pyk}
\end{equation}

\paragraph{Selection operator $\mathcal{S}$.}

We now explain how the selection operator $\mathcal{S}$ [Eq. \eqref{eq:select_Operator}] functions with the state $\ket{j}^{\otimes J}$ of the ancilla registers, which are prepared by $\mathcal{P}$ [Eq. \eqref{eq:prepare_oracle}]. Recall that the expansion order $k\in [0, \ K]$ is encoded as the state $\ket{k} _{y_0} $ with the unary encoding in the register $y_0$. The $k^{\rm th}$ qubit in the $y_0$ register connects to the qubits of the register $y_k$, which in turn connect to the gates acting on the system register. Controlled by the state $\ket{1}$ of $k^{\rm th}$ qubit in the $y_0$ register and the state $\ket{\ell _{k}}_{y_k}$ in the register $y_k$ , the desired unitary $\widetilde{ {h}}_{\ell_{k}} $ functions on the state of the system register.

For example, we note that 1) for $\ket{k}_{y_0} = \ket{0} = \ket{00 \cdots 0} $ ($K$ bits in total), no operation (i.e., a unit operator) functions on the system register, which corresponds to the $0^{\rm th}$ order term of the Taylor series; 2) for $\ket{k}_{y_0} = \ket{1} = \ket{10 \cdots 0} $, only the register $y_1$ is activated by the state $\ket{1}$ of the first qubit in the register $y_0$, and the unitary $ \widetilde{h}_{\ell_{1}} $ is selected to act on the system register controlled on the state $\ket{l_1}_{y_1}$ of the register $y_1$; and 3) when $\ket{k}_{y_0} = \ket{2} = \ket{11 \cdots 0} $, the registers $y_1$ and $y_2$ are activated in sequence by the states $\ket{1}$ of the first two qubits in the register $y_0$, and the unitary $ \widetilde{h}_{\ell_1}  \widetilde{h}_{\ell_2} $ are selected to act on the system register controlled by the states $\ket{\ell_1}_{y_1}$ and $\ket{\ell_2}_{y_2}$ of the registers $y_1$ and $y_2$, respectively. Cases with larger $k$ values follow suit.

\paragraph{Circuit of single-step evolution.}

\begin{figure}[!ht]
	\centering
	\includegraphics[scale=0.55]{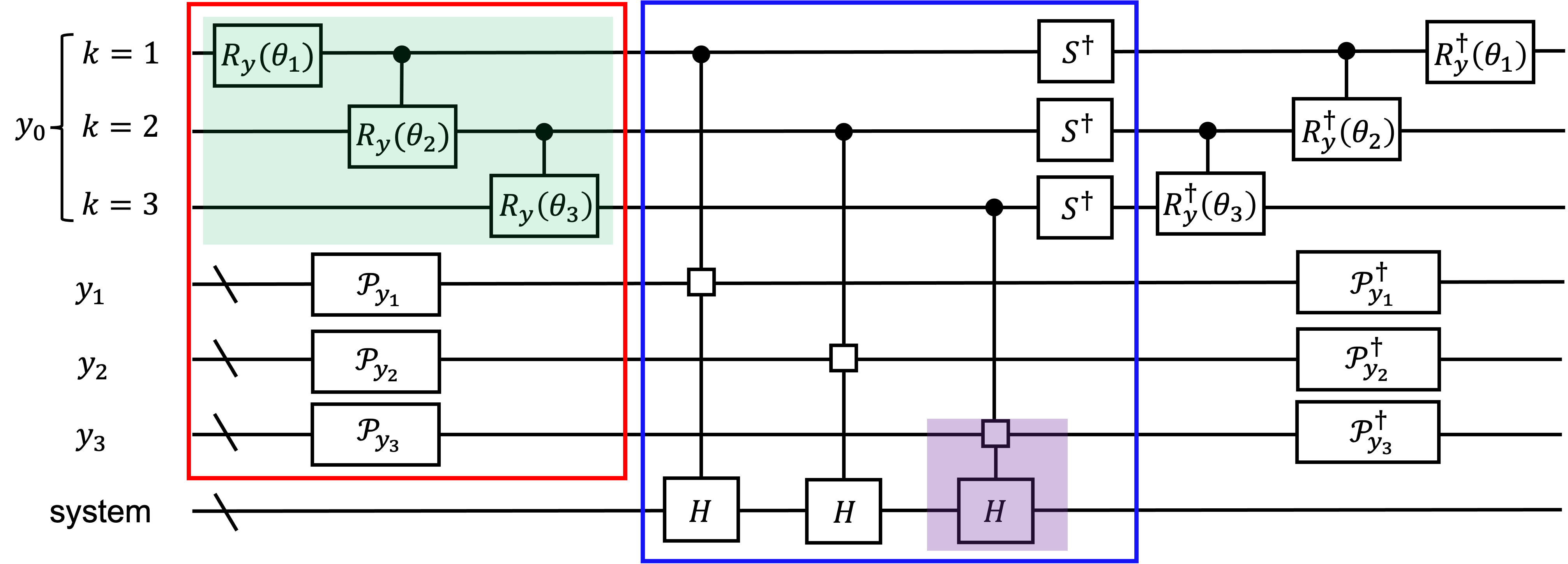}
	\caption{(color online)
		Schematic circuit design of the TTS operator $\mathcal{W}(\tau)$ [Eq. \eqref{eq:TTS}] with $K=3$. $y_0$, $y_1$, $y_2$, and $y_3$ denote the ancilla registers. The bottom line denotes the multi-qubit system register to encode the state of the scattering system. The red box encloses the circuit for the preparation operator $\mathcal{P}(\tau)$ [Eq. \eqref{eq:prepare_oracle}]. The circuit shaded in green is for Eq. \eqref{eq:prepare_oracle_y0} with $K=3$, with the angles $\theta _k(\tau)$ ($k=1,2,3$) given in Eq. \eqref{eq:prepare_oracle_rotation_angle}. $\mathcal{P}_{y_k} $ [Eq. \eqref{eq:prepare_oracle_Pyk}] is implemented with the precomputed expansion coefficients $ \{\alpha_{P,\ell _P}, \ \alpha_{V,\ell _V} \} $ in Eq. \eqref{eq:Hamiltonian_decomposition}. The blue box encloses the circuit for the select operator $\mathcal{S}$ [Eq. \eqref{eq:select_Operator}]. $S^{\dag}$ denotes the conjugate of the $S $ gate \cite{nielsen_chuang_2010} that treats the factor of $(-i)$ in $ V_j $ [Eq. \eqref{eq:V_j_label}]. The multi-controlled $H$ gate (e.g., the part shaded in purple) is defined according to $H$ [Eq. \eqref{eq:Hamiltonian_decomposition}], of which the circuit design is shown in  Fig. \ref{fig:TTS_circuit_detailed}. The slashed line denotes multiple qubits. The $\tau $ dependence is suppressed in the circuit for clarity. The circuit design of $\mathcal{W} (\tau )$ can be generalized to general truncation orders $ K $.
	}
	\label{fig:TTS_circuit}
\end{figure}

\begin{figure}[!ht]
	\centering
	\includegraphics[scale=0.35]{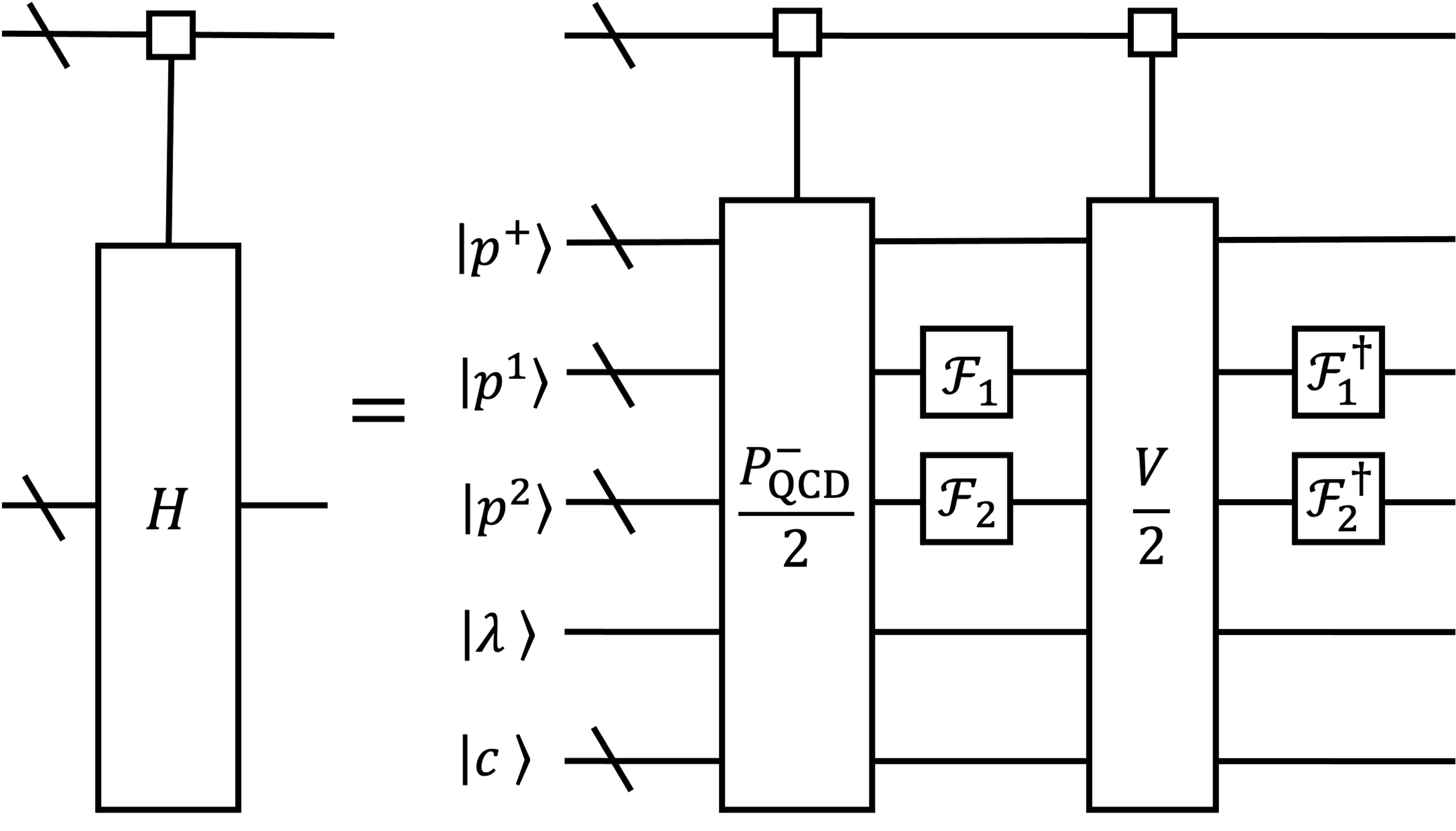}
	\caption{
		Circuit design for the multi-controlled $H$ gate in Fig. \ref{fig:TTS_circuit}. $H$ is decomposed in terms of the linear combination of unitaries according to Eq. \eqref{eq:Hamiltonian_decomposition}. These unitaries are implemented to the register that encodes the state of the scattering system in the lattice-discretized bases $\{ \ket{\beta} \}$. The qFT $ \mathcal{F} = \mathcal{F}_{ 1} \otimes \mathcal{F}_{ 2} $ and its inverse $\mathcal{F}^{\dag}$ are implemented for efficient basis transformations in the transverse degrees of freedom.
		The controlling qubits in the register $y_k$ are prepared in the state shown as the right hand side of Eq. \eqref{eq:prepare_oracle_Pyk}.
	}
	\label{fig:TTS_circuit_detailed}
\end{figure}

To implement $ \mathcal{U} _K(\tau )$ to the state $\ket{\psi }$ of the scattering system encoded in the system register, we implement the TTS operator $\mathcal{W} (\tau) $ [Eq. \eqref{eq:TTS}]. The circuit of $\mathcal{W} (\tau) $ is composed of  $\mathcal{P} (\tau) $, its inverse $\mathcal{P}^{\dagger} (\tau) $, and $\mathcal{S}$. The schematic circuit design of $\mathcal{W} (\tau) $ is shown in Fig. \ref{fig:TTS_circuit}, while the characteristic circuit design of our qFT-assisted Hamiltonian input scheme is in Fig. \ref{fig:TTS_circuit_detailed}. 

In order to improve the success probability of implementing $ \mathcal{U}_K(\tau ) $, it is crucial to implement the OAA scheme, where the amplification operator $\mathcal{Q} (\tau) $ [Eq. \eqref{eq:amplification_operator_Q}] is applied in combination with $\mathcal{W} (\tau) $. The circuit of $\mathcal{Q} (\tau) $ is constructed based on the circuits of $\mathcal{W} (\tau) $ and the reflection operator $\mathcal{R}$. Via the implementation of the OAA scheme, the success probability of implementing $\mathcal{U}_K(\tau )$ reaches to almost $100\% $, where the probability of measuring the ancilla registers to be in the state $\ket{0}^{\otimes J}$ is approximately unit.

\paragraph{Circuit for multiple steps.}

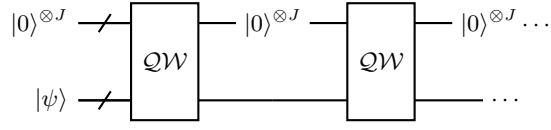
\begin{figure}
	\centering
	\begin{quantikz} 
		\lstick{$\ket{0}^{\otimes J} $} & [2mm] \gate[wires = 2]{\mathcal{QW}} \qwbundle{} & \ \push{ \ket{0} ^{\otimes J } } \  & \gate[wires = 2]{\mathcal{QW}}  &  \ \push{\ket{0}^{ \otimes J}}   \cdots \\ 
		\lstick{$\ket{\psi}$} & [2mm] \qwbundle{} & \qw & & \qw \ \cdots   
	\end{quantikz}
	\caption{Schematic circuit design for the multi-step evolution via the TTS algorithm. The top line denotes the qubits of the ancilla registers, which is initialized in the state $\ket{0}^{\otimes J}$. The bottom line denotes the system register that encodes the state of the scattering system. For each step, the TTS operator $\mathcal{W}(\tau)$ [Eq. \eqref{eq:TTS}] is implemented with the amplification operator $\mathcal{Q}(\tau)$ [Eq. \eqref{eq:amplification_operator_Q}], which effectively uncomputes the ancilla registers. After each single-step evolution, the ancilla registers are measured. At the beginning of each evolution step, the ancilla register is initialized as $\ket{0}^{\otimes J}$. The $\tau $ dependence is suppressed in the circuit for clarity. 
	}
	\label{fig:multiple_step}
\end{figure}

We construct the circuit for multiple implementations of $ \mathcal{U} _K(\tau )$ based on the circuit of the single-step evolution. This circuit is illustrated in Fig. \ref{fig:multiple_step}. At the end of each OAA-assisted evolution step, we measure the ancilla register. We then reset the ancilla state to be $\ket{0}^{\otimes J}$ at the beginning of the following evolution step. We remark that it is important to ensure the successful implementation of the OAA scheme in each single-step evolution, which effectively uncomputes the ancilla registers back to the state $\ket{0}^{\otimes J}$.

\subsubsection{Qubit and gate cost}

\paragraph{Single-step evolution.}

In order to achieve the simulation within error $\epsilon $ for a single step, the required truncation order $K$ in the TTS algorithm [Eq. \eqref{eq:Taylor_expansion_model_1}]  can be evaluated based on the remainder of the Taylor series
\begin{equation}
	\left\Vert \sum_{k=K+1}^{\infty}\frac{1}{k!}\left(-i {H} \tau \right)^{k}\right\Vert \leq \sum_{k=K+1}^{\infty}\frac{1}{k!}\left(\Lambda \tau\right)^{k} = e^{\xi}\frac{\left(\ln2\right)^{K+1}}{\left(K+1\right)!} < \epsilon ,
\end{equation}
with $ \xi\in\left(0,\ln2\right)  $ and $ \Vert  {H}   \Vert  \leq \Lambda  $. By solving the above equation, we obtain the upper bound for $K$ as  \cite{PhysRevLett.114.090502}
\begin{equation}
	K\in {O}\left(\frac{\log\left(1/\epsilon\right)}{\log\log\left(1/\epsilon\right)}\right) ,
	\label{eq:truncation_order_K_value}
\end{equation}
such that $  \Vert \mathcal{U} (\tau ) - \mathcal{U} _K (\tau ) \Vert < \epsilon $. 

We can evaluate the total qubit cost for the implementing one TTS operator $\mathcal{W}(\tau)$ with the truncation order $K$. In particular, the unary encoding of $k \in [0,K]$ takes $K $ qubits. Meanwhile, we need $ \lceil \log_2 L \rceil  $ qubits to encode $l_v =0, \ 1, \ \cdots, \ L-1$ [i.e., the expansion indices in Eq. \eqref{eq:Hamiltonian_decomposition}] as binaries in the register $ {y_v} $ ($v=1, \ 2, \ \cdots , K$). In addition, $ N_{\rm sys} $ qubits [Eq. \eqref{eq:number_sys_bits}] are required for the system register to encode the state of the scattering system. Therefore, the total number of the qubits required for the circuit of $\mathcal{W}(\tau) $ is\footnote{Note that we recycle the ancilla qubits during the evolution.} 
\begin{equation}
	N_{\rm q,tot} = K + K \lceil \log_2 L \rceil  + N_{\rm sys} . 
	\label{eq:qubit_cost_single_step}
\end{equation}
This is also the qubit cost for implementing $\mathcal{Q}(\tau) \mathcal{W}(\tau)$.

The total gate cost for the circuit of $\mathcal{W}(\tau)$ can be estimated following Ref. \cite{PhysRevLett.114.090502}. 
The implementation of $\mathcal{P}_{y_0}$ takes $ {O}(K)$ gates. Meanwhile, each $\mathcal{P}_{y_i}$ can be implemented with $ {O}(L)$ gates with a classically precomputed database of the expansion coefficients $ \{ \alpha _{P, \ell_P} \} $ and  $ \{ \alpha _{V, \ell_V } \}  $ (with $\ell _P \in [0, L_1 -1]$, $\ell _V \in [0, L_2 -1]$, and $L_1 + L_2 =L$) in Eq. \eqref{eq:Hamiltonian_decomposition} following the prescriptions in Ref. \cite{Shende_2006}. Controlled by the state $\ket{1}$ of the $k^{\rm th}$ qubit in the register $y_0$ and the state $\ket{\ell }_{y_k}$ (with $\ell \in [0,L-1]$) in the register $y_k$, a unitary is selected from $ \{ h_{P, \ell _P} \} $ or $ \{ h_{V, \ell _V} \} $ to operate on the system register. 
The implementation of each of these $( \lceil  \log_2 L  \rceil +1 )$-fold controlled operators takes $\Theta (\log L)$ gates.
We assume these unitaries in Eq. \eqref{eq:Hamiltonian_decomposition} are 1-sparse matrices of dimension $2^{N_{\rm sys }}$ (e.g., the $N_{\rm sys} $-fold Pauli strings), each of which can be implemented with $ {O}(N_{\rm sys })$ gates to the system register containing $N_{\rm sys }$ qubits.
Therefore, the gate cost for the $\mathcal{P}$, and $\mathcal{P}^{\dag}$ and $\mathcal{S}$ can be estimated as
\begin{equation}
	 {O}\left[  K L ( N_{\rm sys} + \log L ) \right] .
\end{equation}

Moreover, we need to include the gate cost for the qFT in each evolution step, which is $ {O} (N^2_{\rm sys})$ \cite{nielsen_chuang_2010}. Therefore, the gate cost to implement $\mathcal{W}(\tau)$ with our qFT-assisted Hamiltonian input scheme can be estimated as
\begin{equation}
	 {O}\left[  K L ( N_{\rm sys} + \log L ) +  N^2_{\rm sys} \right] .
	\label{eq:asymptotic_TTS}
\end{equation}

The above equation is also the asymptotic gate cost for implementing $\mathcal{Q}(\tau) \mathcal{W}(\tau)$ [Eq. \eqref{eq:OAA_total_eq}]. This gate scaling can be understood from the facts that $\mathcal{Q}(\tau)$ contains several copies of $\mathcal{W}(\tau)$ and one can omit the minor gate cost to implement the reflection operator $\mathcal{R}$.

\paragraph{Multi-step evolution.}

Based on the cost of the single-step evolution with the step size $\tau = (\ln 2) / \Lambda $, we can evaluate the gate cost of the simulation for the entire duration $x^+$. As stated above, we assume $x^+$ to be an integer number of $\tau = \ln 2 / \Lambda $ for simplicity. Therefore, the total number of segments $ r $ is
\begin{equation}
	r= x^+ / \tau = \Lambda x^+ /\ln 2 .
	\label{eq:segment_number}
\end{equation}
If we still require the simulation error to be bounded by $\epsilon $ for the $r$-step evolution, then we should restrict the error for each single-step evolution to be $\epsilon / r$. Therefore, the truncation order $ K $ for each single-step evolution should be modified from Eq. \eqref{eq:truncation_order_K_value} to
\begin{equation}
	K_r \in {O}\left(\frac{\log\left( \Lambda x^+ /\epsilon\right)}{\log\log\left( \Lambda x^+ /\epsilon\right)}\right). 
	\label{eq:K_r_expression}
\end{equation}

The total qubit cost for the $r$-step evolution is  
\begin{equation}
  N_{\rm q,tot} = K_r + K_r \lceil \log_2 L \rceil  + N_{\rm sys},
  \label{eq:total_Q_cost}
\end{equation}
where $K$ in Eq. \eqref{eq:qubit_cost_single_step} is substituted by $K_r$. 
We comment that $K_r$ is expected to be finite in practical numerical simulations. In such cases, $N_{\rm q,tot} $ scales as the logarithm of the basis space dimension of the scattering system; this is also the scaling of the number of the ancilla qubits, $ K_r + K_r \lceil \log_2 L \rceil $, according to Eq. \eqref{eq:afjss}.

The total gate cost for implementing the $r$-step evolution based on the circuit of $\mathcal{Q}(\tau)\mathcal{W}(\tau)$ for each single-step evolution is 
\begin{equation}
	{O}\left[ r K_r L ( N_{\rm sys} + \log L ) + r N^2_{\rm sys} \right] .
	\label{eq:multiStepIntermediate}
\end{equation}
Combining Eq. \eqref{eq:K_r_expression}, we rewrite Eq. \eqref{eq:multiStepIntermediate} to finalize the gate cost for all the $r$-step evolution as
\begin{equation}
	 {O}\left[    
	\Lambda x^+ \frac{\log\left( \Lambda x^+ /\epsilon\right)}{\log\log\left( \Lambda x^+ /\epsilon\right)} L ( N_{\rm sys} + \log L ) + \Lambda x^+  N^2_{\rm sys} 
	\right] ,
	\label{eq:gate_cost_by_Pauli_strings_overall}
\end{equation}
where $N_{\rm sys} $ is given in Eq. \eqref{eq:number_sys_bits}, which scales logarithmically with the basis space dimension. 

This gate cost [Eq. \eqref{eq:gate_cost_by_Pauli_strings_overall}] is optimal (near optimal) in its scaling with the simulation error $\epsilon $ (simulation time $x^+$): 1) it depends linearly on $x^+$ with the scaling factor being the $\mathbb{L}^1$ norm $\Lambda $ of the expansion coefficients; and 2) it depends logarithmically on the inverse of the desired precision $\epsilon$. In addition, we find that the gate cost of the TTS algorithm scales linearly with the number of unitaries $L$ [see dominant term in Eq. \eqref{eq:gate_cost_by_Pauli_strings_overall}]. As for efficient implementations of the TTS algorithm, it is advantageous to have smaller $L $ and $\Lambda $. A well designed decomposition scheme [Eq. \eqref{eq:Hamiltonian_decomposition}] with good choice of the unitaries is expected to further improve the simulation efficiency.

\subsubsection{Implementation of the TTS algorithm}

We summarize the procedures to implement of the TTS algorithm as follows.
\begin{itemize}

    \item \textbf{Step} 1: We decompose the Hamiltonian $H$ in terms of the linear combination of unitaries according to Eq. \eqref{eq:Hamiltonian_decomposition}, where we compute the  coefficients and unitaries. We also compute the $\mathds{L}^{1}$ norm $ \Lambda $ of the coefficients. 
     
    \item \textbf{Step} 2: Based on the total simulation time $x^+$ and precision $\epsilon$, we specify the total number of the evolution steps, $r$, and the truncation order $K_r$ according to Eqs. \eqref{eq:segment_number} and \eqref{eq:K_r_expression}.
    
    \item \textbf{Step} 3: We construct the circuit of the preparation operator $ \mathcal{P} (\tau) $ according to Eqs. \eqref{eq:total_prepare_operator}, \eqref{eq:prepare_oracle_y0}, and \eqref{eq:prepare_oracle_Pyk}. We also construct circuit of the selection operator $\mathcal{S}$ according to Eq. \eqref{eq:select_Operator}. 
    
    \item \textbf{Step} 4: We construct the circuit of $\mathcal{W} (\tau) $ [Eq. \eqref{eq:W_operator}] based on the those of $ \mathcal{P} (\tau) $ and $\mathcal{S}$.
    
    \item \textbf{Step} 5: We then construct the circuit of the amplification operator $\mathcal{Q}(\tau) $ according to Eq. \eqref{eq:amplification_operator_Q}. The complete TTS circuit with the OAA implementation is constructed as $\mathcal{Q}(\tau) \mathcal{W} (\tau)$, which block encodes the time evolution unitary for a single step $\tau$ [Eq. \eqref{eq:OAA_total_eq}].
\end{itemize}

The above steps present the circuit of the TTS algorithm for a single time-evolution step, where the OAA scheme is also implemented. 
The multi-step simulation for the duration $x^+ = r \tau $ [Eq. \eqref{eq:segment_number}] consists of $r$ sequential implementations of the single-step evolution. We perform the measurement at the end of each single-step evolution, and reset the ancilla qubits to be in the state $\ket{0}^{\otimes J}$ at the beginning of the next step. Here, we recall that the successful OAA implementation effectively uncomputes the ancilla qubits to be in the state $\ket{0}^{\otimes J}$ at the end of each single-step evolution.

\subsection{The Trotter algorithm}

In this work, we also perform dynamics simulations via the Trotter algorithm \cite{lloyd1996universal} to benchmark the TTS results. Though being of less favorable gate scaling with the simulation time and error and hence inefficient for complex large-scale problems, the Trotter algorithm admits more straightforward circuit design that necessitates no ancilla qubits and fewer multi-controlled gates when compared to the TTS algorithm.

\begin{figure}[!ht]
	\centering
	\includegraphics[scale=0.50]{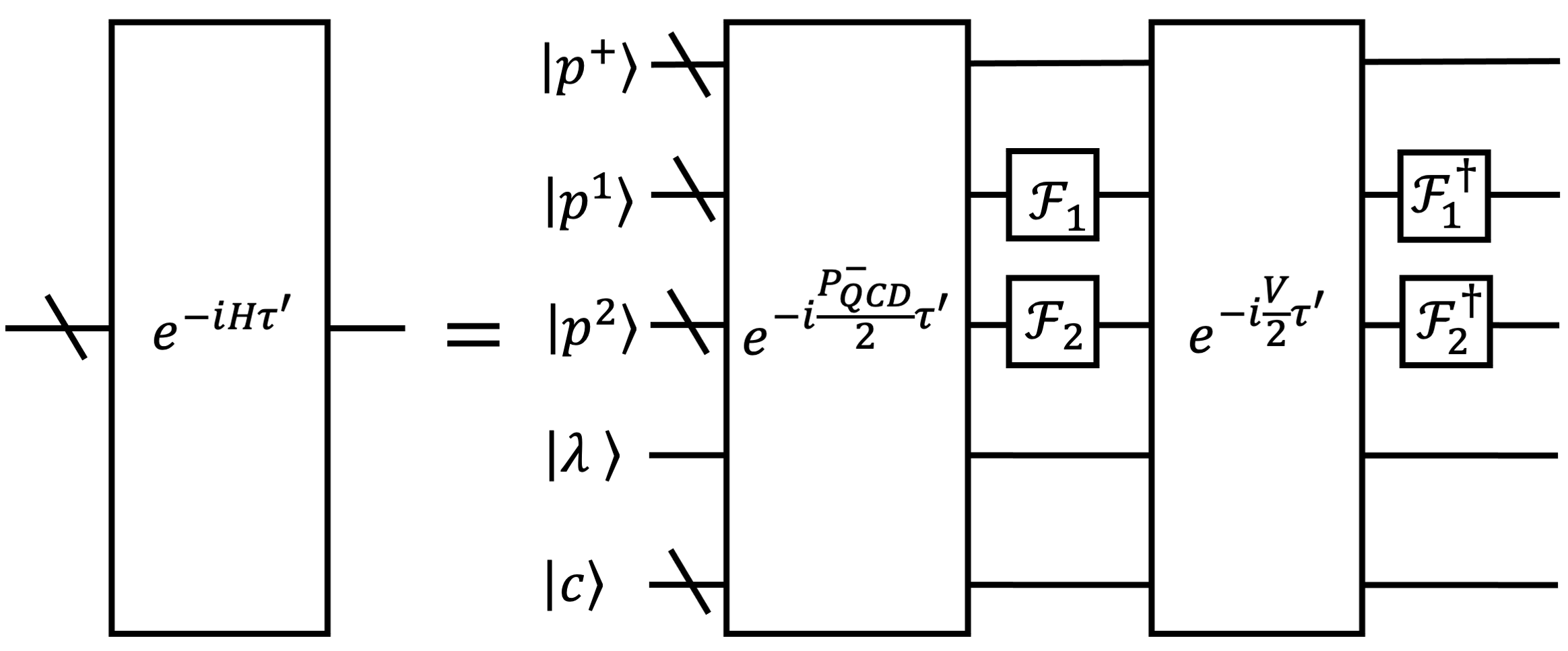}
	\caption{Schematic circuit design for $ \mathcal{U} (\tau ')$ via the Trotter algorithm. The circuit is designed according to Eq. \eqref{eq:Trotter_first_order_line1}. The register encodes the state of the scattering system in the lattice-discretized bases $\{ \ket{\beta} \}$. The qFT and its inverse are implemented for efficient basis transformations in the transverse degrees of freedom. The modules $\exp (  -i \tau ' P^-_{\rm QCD}/2 )$ and $\exp \left(-i \tau ' V/2 \right)$ are designed according to the product formula shown in Eq. \eqref{eq:Trotter_first_order}.}
	\label{fig:trotter_one_step_circuit}
\end{figure}

According to the first-order Trotter formalism \cite{doi:10.1063/1.526596,Huyghebaert_1990,nielsen_chuang_2010}, we can approximate the evolution operator ${\mathcal{U}} (\tau ') $ for a single time step of size $\tau '$ as 
\begin{align}
	 e^{-i  {H} \tau ' } =&  e^{-i \tau '  {P^-_{\rm QCD}}/2 } \cdot \mathcal{F}^{\dag} e^{-i \tau ' V/2}  \mathcal{F}  \label{eq:Trotter_first_order_line1} \\
	 \approx &  \left[ \prod _{\ell=0}^{L_1-1} \exp \big[ -i \tau ' \alpha_{P, \ell} {h}_{P,\ell} \big] \right] \cdot \mathcal{F} ^{\dag} \left[ \prod _{\ell=0}^{L_2-1} \exp \big[ -i \tau ' \alpha_{V, \ell} {h}_{V,\ell} \big] \right] \mathcal{F} , 
	 \label{eq:Trotter_first_order}
\end{align}
where we have applied the qFT-assisted Hamiltonian input scheme [Eq. \eqref{eq:Hamiltonian_decomposition}]. One takes $\tau '$ with $ || {H} || \tau ' \ll 1 $ for the dynamics simulations, where $ || {H} ||  $ denotes the spectral norm of $ {H}  $. 
The error $\epsilon$ of the first-order Trotter formalism scales as $O \left((\tau ')^2 \right)$ for a single step.  
Higher-order Trotter formalisms achieve better precision at the cost of more complex expansions, resulting in more expensive gate cost. In this work, we only apply the first-order Trotter formalism [Eq. \eqref{eq:Trotter_first_order}] in the simulations. The quantum circuit for a single Trotter step is designed according to Eq. \eqref{eq:Trotter_first_order}, and is illustrated in Fig.~\ref{fig:trotter_one_step_circuit}. 

\begin{figure}[!ht]
	\centering
	\includegraphics[scale=0.6]{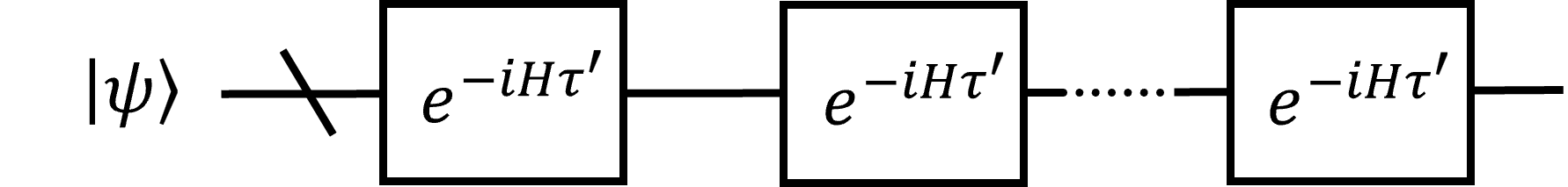}
	\caption{
		Quantum circuit for the multi-step evolution via the Trotter approach. The total circuit is composed of sequential implementations of the single-step circuit (Fig. \ref{fig:trotter_one_step_circuit}) according to Eq. \eqref{eq:trotter_alg_multi_step}.
	}
	\label{fig:trotter_r_step}
\end{figure}

According to Eq. \eqref{eq:time_ev_unitary_Trotterized}, the multi-step Trotter evolution is constructed from sequential implementations of the single-step evolution
\begin{equation}
	\mathcal{U} (x^+ ) = \big[ \mathcal{U} (\tau ' ) \big]^{r'} = \big[ e^{-i H \tau ' } \big]^{r'} , 
	\label{eq:trotter_alg_multi_step}
\end{equation}
with $x^{+} = r' \tau '$. To simulate within error $\epsilon $, it suffices to choose 
$ r' \in O \big( ( \Lambda x^+ )^2 / \epsilon \big) $  \citep{PhysRevX.11.011020}, where we note $ ||H|| \leq \Lambda$.
Correspondingly, the circuit of $ \mathcal{U} (x^+ ) $ is constructed with consecutive $r'$ implementations of the single-step evolution circuit, which is shown as Fig. \ref{fig:trotter_r_step}.
Since we adopt the same scheme to encode the basis of the scattering system, the qubit cost takes the form of Eq. \eqref{eq:number_sys_bits}, which has the same asymptotic qubit scaling of that of the TTS approach with finite truncation orders.
The gate cost for the dynamics simulation from $0$ to $x^+$ ($r'$ steps in total) according to the first-order Trotter formalism scales as 
\begin{equation}
	O \left( \frac{( \Lambda x^{+})^{2}}{\epsilon}\left(LN_{\rm sys}+N_{\rm sys}^{2}\right) \right) ,
	\label{eq:complexity_Trotter}
\end{equation}
with $L=L_1 + L_2$. Though being more intuitive, the Trotter algorithm presents a worse gate scaling with the simulation time and error when compared to the TTS algorithm.

\subsubsection{Implementation of the Trotter algorithm}

The procedures to implement the Trotter algorithm are as follows.

\begin{itemize}

    \item \textbf{Step} 1: We decompose the Hamiltonian in terms of the linear combination of unitaries according to Eq. \eqref{eq:Hamiltonian_decomposition}. We also compute the $\mathds{L}^{1}$ norm $ \Lambda $ of the coefficients. (This step is the same as the first step of the TTS implementation.)
       
    \item \textbf{Step} 2: We specify the simulation time $x^+$ and precision $\epsilon$. The total number of evolution steps can be set as $r^{\prime} \equiv (\Lambda x^{+})^{2}/\epsilon$ with the time step size being $\tau ' = \epsilon /( \Lambda ^2 x^+) $. 
    
    \item \textbf{Step} 3: We construct the circuit of the single-step evolution $\mathcal{U}(\tau ')$ according to Eq. \eqref{eq:Trotter_first_order}.

\end{itemize}

The above steps complete the circuit design for a single-step evolution via the Trotter algorithm.
The circuit of the complete evolution $\mathcal{U}(x^+)$ [Eq. \eqref{eq:trotter_alg_multi_step}] is made up by $r'$ sequential applications of the single-step evolution.

\section{Demonstration}
\label{sec:Model_Problem}

We demonstrate our framework by a simple model problem. We first present the setup of the model problem. Then, we show our choice of the basis set for the numerical demonstration. With the elected basis set, we illustrate our basis encoding scheme and the qFT-assisted Hamiltonian input scheme. Finally, we present the parameter setup for our circuit construction, together with the details of the simulation conditions and measurement.

\subsection{Model problem setup}

\begin{figure}[!ht]
	\centering
	\includegraphics[scale=0.50]{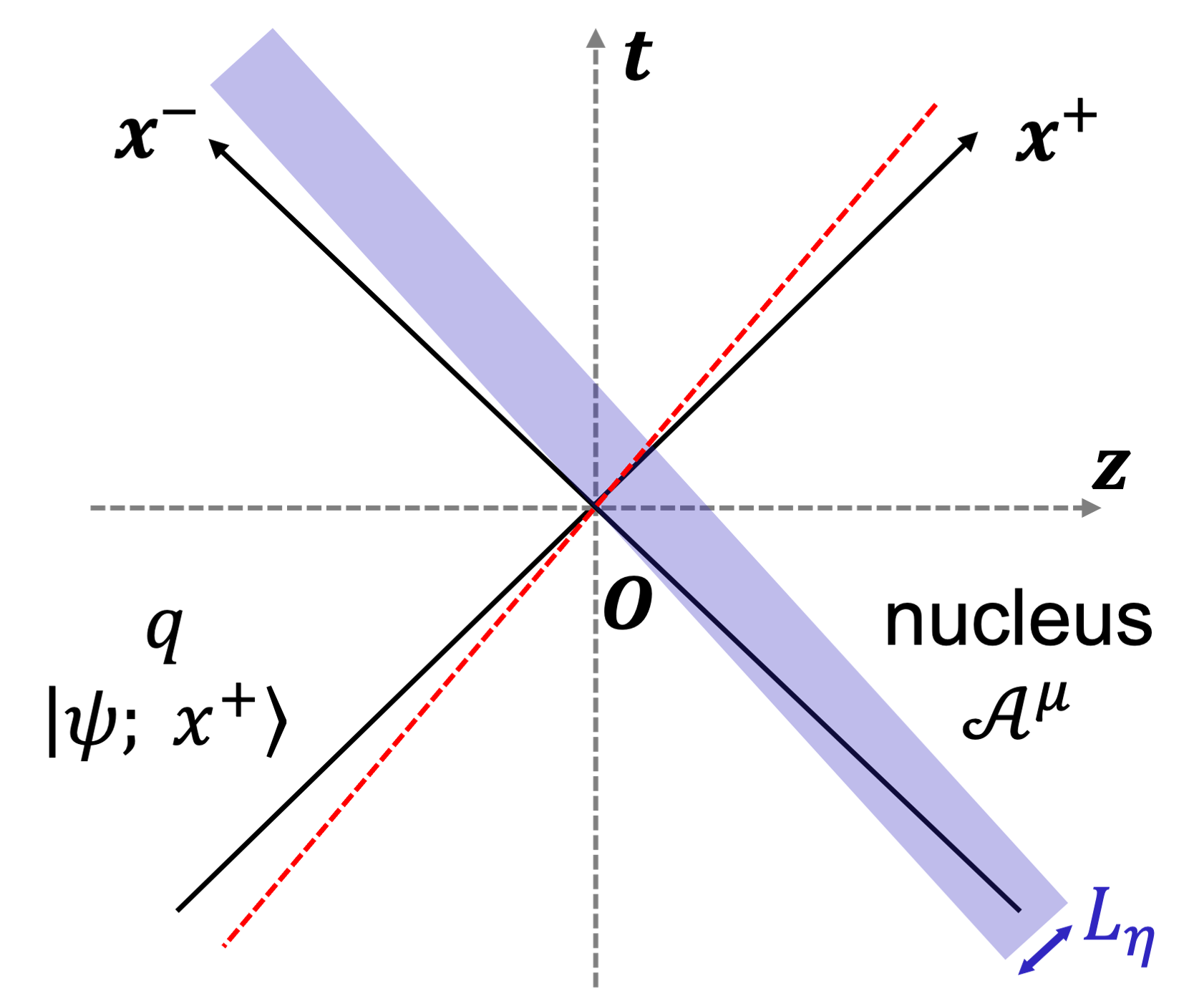}
	\caption{(color online)
		Setup for the ultra-relativistic quark-nucleus scattering (figure adapted from Ref. \cite{PhysRevD.101.076016}). See the text for more details.   
	}
	\label{fig:scattering_setup}
\end{figure}

To illustration our approach of dynamics simulation, we revisit the model problem of ultra-relativistic quark-nucleus scattering in Ref. \cite{PhysRevD.101.076016}. As shown in Fig. \ref{fig:scattering_setup}, an ultra-relativistic quark moves along the positive-$\hat{z}$ direction, while a high-energy heavy nucleus moves along the negative-$\hat{z}$ direction. At high energies, we have $p^+ \gg p^{\perp}$. During the scattering, the quark interacts with the SU(3) color field $ \mathcal{A}^{\mu }$ generated by the nucleus. The nuclear medium is assumed to be distributed within the range $0 \leq x^+ \leq L_{\eta}$ along the direction of the light-front time.

Following Ref. \cite{PhysRevD.101.076016}, we model the $ \mathcal{A}^{\mu } $ field as a classical stochastic field approximated by the CGC theory \cite{PhysRevD.96.074020}. The CGC theory provides an effective description of the gluon dynamics in the small-$x$ regime, which is applicable to the high-energy scattering processes. In the CGC framework, the background field is treated classically; it can be solved from the Yang-Mills equation
\begin{equation}
	\boldsymbol{D}_{\mu}  \mathcal{F}^{\mu \nu} = J^{\nu}.
\end{equation}
In the above equation, $\mathcal{F}^{\mu \nu} $ denotes the strength tensor of the color field, while $J^{\nu} = J^{\nu} _a T_a $ ($a=1,\ 2, \cdots , \ 8$) denotes the color current, with $T_a$ being the color generator and the dummy index $a$ implying the summation over it. As for the color current generated by the high-energy nucleus moving in the negative $\hat{z}$ direction, it only contains one non-vanishing component, i.e., $J^{\nu} _a = \delta ^{\nu -} \rho _a $, whereas this component is independent of $x^-$ of the nucleus \cite{kovchegov_levin_2012}.

In the covariant gauge $\partial _{\mu} \mathcal{A} ^{\mu} =0 $, the color field has only one non-vanishing component $\mathcal{A}^{-}$. It can be solved from the regularized Poisson equation
\begin{equation}
	\left(m_{g}^{2}-\nabla_{ \perp }^{2}\right ) \mathcal{A}^{-}_{a} \left(\boldsymbol{x}_{ \perp } ,x^+\right) = \rho_{a} \left(\boldsymbol{x}_{ \perp },x^+\right),
	\label{eq:reg_Poisson_Eq}
\end{equation}
where the color field and valence charge density are explicit functions of $x^+$ and $\boldsymbol{x}_{ \perp } $ of the nucleus. The gluon mass $m_g$ (taken to be $0.1$ GeV in this work) is introduced to regularize the infrared divergence of the color field in simulating the color neutrality of the source distribution \cite{KRASNITZ2003268}. 
As in Ref.~\cite{PhysRevD.101.076016}, we apply the McLerran-Venugopalan (MV) model \cite{PhysRevD.49.2233,PhysRevD.49.3352} to solve for the SU(3) color field. 
With the MV model, the valence charge density $\rho _a (\boldsymbol{x}_{\perp},x^+)$ of the nucleus is treated as stochastic random variables. These random variables follow the Gaussian distribution 
\begin{equation}
	f\Big[ \rho ^2 _a (\boldsymbol{x}_{ \perp },x^+)  \Big] = \exp \Bigg[  - \frac{\delta x^+  \delta ^2 \boldsymbol{x}_{ \perp }	  }{ g^2 \mu ^2 }   \rho ^2 _a (\boldsymbol{x}_{ \perp },x^+)   \Bigg],
\end{equation}
where $ \delta x^+ $ and $ \delta ^2 {\boldsymbol{x} }_ {\perp} $ denote the unit length in the $x^+ $ and $ {\boldsymbol{x} }_{\perp} $ directions, respectively. $\mu$ serves as a measure of the density of scattering centers in the nuclear medium. The factor $g^2 \mu$ denotes the color charge density parameter. In the ultra-relativistic limit, the distribution function of the color charge density peaks at the vicinity of $x^+=0$ due to the Lorentz contraction. In the MV model, the so-defined valence charge density satisfies the correlation relation
\begin{equation}
	\braket{\rho_{a} (\boldsymbol{x}_{\perp},x^+) \rho_{b}(\boldsymbol{y}_{\perp},y^+)} = g^{2}\mu^{2}(\boldsymbol{x}_{\perp})\delta_{ab}\delta^{(2)}(\boldsymbol{x}_{\perp}-\boldsymbol{y}_{\perp})\delta(x^{+}-y^{+}).
	\label{eq:rho}
\end{equation}
One can solve the color field component $\mathcal{A}^{-}_{a} (\boldsymbol{x}_{ \perp } ,x^+ ) $ via the Green's function approach based on Eq. \eqref{eq:reg_Poisson_Eq}, where it is necessary to regularize the ultraviolet divergence due to the unphysical large momentum modes in the nuclear wave function. In this model problem, the regularization is achieved via applying the ultraviolet cutoff $\Lambda_{\rm UV}$ to the transverse momentum of the basis states.\footnote{Interested readers are referred to Ref. \cite{PhysRevD.101.076016} and references therein for more details.}

The saturation scale of the MV model is defined as
\begin{equation}
	Q_s^2 = \frac{(g^2 \mu )^2  L_{\eta } }{2\pi ^2	} ,
	\label{eq:saturation_scale}
\end{equation}
which is a constant for fixed $g^2 \mu $ and the extension of the nucleus in the longitudinal direction, $L_{\eta}$. For the numerical demonstration in this work, we take  $g=1$, $ g^2 \mu = 0.407294 $ GeV$^{3/2}$, and $L_{\eta} = 50 $ GeV$^{-1}$. The corresponding saturation scale is fixed as $Q_{s}^{2} \approx 0.420 $ GeV$^{2}$. It is noted that the above definition for $Q_s$ does not include any quantum corrections.

The only non-vanishing component of the background color field in this model is the minus component $ \mathcal{A} ^-  $. In this case, the instantaneous interaction between the quark and the color field [i.e., the second term in the integrand of Eq. \eqref{eq:interaction_part}] hence vanishes, and the interaction Hamiltonian $	\boldsymbol{V} $ becomes
\begin{equation}
	\boldsymbol{V} = \int dx^{-}d^{2} \boldsymbol{x}_{\perp} \ \Big[ g \bar{\Psi}\gamma^{\mu}\boldsymbol{T}^{a}\Psi \mathcal{A}_{\mu}^{a}  \Big] ,
	\label{eq:statistical_LF_pot}
\end{equation}
with $ \mathcal{A}^{a-} =2 \mathcal{A}_{+}^{a}  $ according to the convention of the LF formalism.

\subsection{Hamiltonian input scheme}
\label{sec:VB_model_space_and_Hamiltonian}

We demonstrate our approach for the dynamics simulations implementing the IBM QASM simulator of the Qiskit package \cite{Qiskit}. The simulations are performed on our local computer with Intel 8-core CPU and 32GB RAM, which is capable of handling 32-qubit simulations at most. 
We work with the straightforward Hamiltonian input scheme, where the Hamiltonian is expressed in terms of the linear combination of Pauli strings. 
In view of the limited computational resources, we elect to retain a restricted Hilbert space, and keep only a limited number of the Pauli strings in the illustrations of the TTS approach.

As for the restricted basis space, we take $p^i = -2 a_p^{\perp}, \ -a_p^{\perp},\ 0$, and $a_p^{\perp}$ (with $i=1,\ 2$) for the transverse momenta, with $ a^{\perp}_p = 2\pi /(2L^{\perp})  $ and $L^{\perp } =5 $ GeV$^{-1}$. Accordingly, the transverse coordinates are taken as $x^i = -2 a_r^{\perp}, \ -a_r^{\perp},\ 0,$ and $a_r^{\perp}$, with $ a_r^{\perp} = L^{\perp} /2 $. Meanwhile,  we fix the value of $ p^+= 850$ GeV and $\lambda =1/2$. In addition, we take the values of color as $c=$ Red, Green, and Blue. As such, we have $  N^{\perp} =2 $, $ N^{\parallel}$ = $N_{\lambda} = 1$, and $N_c = 3 $  to encode the all the retained bases $\{ \ket{p^+, p^1, p^2, \lambda , c} \}$ and $\{ \ket{p^+, x^1, x^2, \lambda , c} \}$. 

With the above bases, we obtain the LF Hamiltonian ${P}^{-} $ matrix as a summation of the reference part and the interaction part. The reference Hamiltonian matrix in the bases $\{ \ket{p^+, p^1, p^2, \lambda , c} \}$ can be expressed as 
\begin{equation}
	{P}^{-}_{\rm QCD}  \equiv \mathscr{K} (p^+ , p^1, p^2, \lambda ) \otimes  T^{0}(c) .
\end{equation}
In terms of the linear combination of Pauli strings, we write $  \mathscr{K} (p^+ , p^1, p^2, \lambda ) $ as
\begin{eqnarray}
		\mathscr{K} = & 1.39383 * {I \otimes II  \otimes II \otimes I  } + 0.232226 *{I \otimes II  \otimes  IZ \otimes I  }+0.464452 *{I \otimes II  \otimes ZI \otimes I  } \nonumber \\
	 & + 0.464452 *{I \otimes II  \otimes ZZ \otimes I  } + 0.232226 * {I \otimes IZ  \otimes II \otimes I  } +0.464452 * {I \otimes ZI  \otimes II \otimes I  } \nonumber \\
	 & + 0.464452 *{ I \otimes ZZ  \otimes II \otimes I  } , \label{eq:Pauli_string_of_K}
\end{eqnarray}
where $\mathscr{K} $ is in the unit of $ 10^{-3}$ GeV. ${ X }$ and ${ Z }$ denote the Pauli-X and Pauli-Z matrices, respectively. ${ I }$ denotes the two-by-two identity matrix. The decomposition coefficients are solved as
\begin{equation}
    \kappa  = \frac{1}{2^{n}} \textrm{Tr} (\mathscr{K} {{\sigma}}_{n}),
\end{equation}
where ${{\sigma}}_{n}$ denote the tensor products of $n$ Pauli matrices. For the tensor product $ {\zeta}_1 \otimes  {\zeta}_2  {\zeta}_3 \otimes  {\zeta}_4  {\zeta}_5 \otimes  {\zeta}_6 $ in each term, the first Pauli $ {\zeta}_1 $ acts on the discretized longitudinal momentum state $\ket{p^+}$. $ {\zeta}_2  {\zeta}_3 $ and $ {\zeta}_4  {\zeta}_5 $ act on the discretized transverse momentum states $\ket{p^1}$ and $\ket{p^2}$, respectively. ${\zeta}_6 $ acts on the helicity state $\ket{\lambda }$. $ {T}^{0}(c) = { I I }$ denotes the identity acting on the color state $\ket{c}$. 

As discussed in Sec. \ref{sec:H_input_model_sec}, we compute the interaction Hamiltonian $ {V} $ [Eq. \eqref{eq:statistical_LF_pot}] in the bases $\{ \ket{p^+, x^1, x^2, \lambda , c} \}$. The resulting interaction matrix is also expressed in terms of the linear combination of Pauli strings. However, due to the restriction of our computing resources, we can only perform the dynamics simulations via the TTS approach for problems with only a limited number of Pauli strings on our platform (see the discussion of qubit cost below). Therefore, we select to retain only the terms with $a=1$ in ${V} $. The simplified interaction matrix can be written as
\begin{equation}
	{V}' \equiv {W} ^{1-} (p^+ , x^1, x^2, \lambda ) \otimes {T}^1 (c) .
	\label{eq:vPrime_part}
\end{equation}
In terms of Pauli strings, $ {W} ^{1-} (p^+ , x^1, x^2, \lambda) $ can be decomposed as
\begin{eqnarray}
	{W} ^{1-} = & \ \ 346.525 * {I \otimes II \otimes II \otimes I } - 0.709063*{I \otimes II \otimes IZ \otimes I } - 2.73394*{I \otimes II \otimes ZI \otimes I } \nonumber \\
	&  - 6.04144*{I \otimes II \otimes ZZ \otimes I } + 3.56781*{I \otimes IZ \otimes II \otimes I } - 1.50894*{I \otimes IZ \otimes IZ \otimes I } \nonumber \\
	&  - 1.98356*{I \otimes IZ \otimes ZI \otimes I } - 0.209813*{I \otimes IZ \otimes ZZ \otimes I } + 10.7856*{I \otimes ZI \otimes II \otimes I } \nonumber \\
	&  + 1.18556*{I \otimes ZI \otimes IZ \otimes I } - 4.93806*{I \otimes ZI \otimes ZI \otimes I } + 8.65394*{I \otimes ZI \otimes ZZ \otimes I } \nonumber \\
	&  - 28.8128*{I \otimes ZZ \otimes II \otimes I } + 2.46544*{I \otimes ZZ \otimes IZ \otimes I } - 8.74144*{I \otimes ZZ \otimes ZI \otimes I } \nonumber \\ 
	& -3.80169*{I \otimes ZZ \otimes ZZ \otimes I },
\end{eqnarray}
in the unit of $ 10^{-3} $ GeV. The arrangement of the Pauli strings follows that in $ \mathscr{K} $ [Eq. \eqref{eq:Pauli_string_of_K}], except that the $2^{\rm nd}$ and $3^{\rm rd}$ ($4^{\rm th}$ and $5^{\rm th}$) Pauli matrices in each term now operate on the transverse coordinate state $\ket{x^1}$ ($\ket{x^2}$). 
${T}^{1}(c)=({IX}+{ZX})/4$ operates on the color state $ \ket{c} $.

To sum, we employ the simplified LF Hamiltonian matrix 
\begin{equation}
  P^- = \mathscr{K} (p^+ , p^1, p^2, \lambda ) \otimes  {T}^{0}(c) + \mathcal{F} ^{\dag} {W} ^{1-} (p^+ , x^1, x^2, \lambda) \mathcal{F} \otimes {T}^{1}(c) , 
  \label{eq:given_H}
\end{equation}
for our numerical demonstrations with $H=P^-/2$. The LF Hamiltonian is encoded in terms of the linear combination of Pauli strings. 
The qFT and its inverse are utilized to assist our Hamiltonian input scheme, which performs efficient basis transforms between the sets $\{ \ket{x^1 } \}$ and $ \{ \ket{p^1} \} $, and between the sets $ \{  \ket{x^2} \} $ and $\{ \ket{p^2} \} $.

\subsection{Circuit parameters}

We implement the TTS algorithm for numerical simulations based on the Hamiltonian $H$ and the qFT-assisted input scheme discussed above. Although higher truncation order improves the simulation precision, more qubits become necessary in the corresponding implementation [see Eq. \eqref{eq:total_Q_cost}]. In this work, we select $K_r=3$ with our restricted computational resources.

In our circuit design, we employ $ 6 $ qubits to encode the retained basis set ($\{ \ket{p^+, p^1, p^2, \lambda , c} \}$): 1) $4$ qubits to encode the $16$ possible basis states of the discretized transverse momentum states; 2) $2$ qubits for the color basis states. We save the two qubits necessary for encoding the longitudinal momentum and helicity bases as the longitudinal momentum and the helicity do not change in this simplified model problem.
This can also been seen from the identity operations for these two basis states (i.e., the rightmost and leftmost operators are identities in each term of $ \mathscr{K}  $ and $ {W} ^{1-} $).

We evaluate the total number of the ancilla qubits in order to implement our TTS simulations with $K_r=3$. In particular, the number of the ancilla qubits can be evaluated as $K_r+K_r \lceil \log _2 L \rceil $. According to Eq. \eqref{eq:Hamiltonian_decomposition}, we have $L_1=7$ and $L_2=32$ for the expressions of ${P}^{-}_{\rm QCD} $ and ${V}' $ in Sec. \ref{sec:VB_model_space_and_Hamiltonian}, and $L=L_1 +L_2=39$. Therefore, the total number of ancilla qubits in our simulation is $ 21 $.

The total number of qubits in our simulation via the TTS approach is 27 according to Eq. \eqref{eq:total_Q_cost}.
This qubit cost is within the capability of our simulation platform. We note that if we had retained all the components of the color background field in $ 	{V}  $ [Eq. \eqref{eq:statistical_LF_pot}] (i.e., for all the components indexed by $a=1,\ 2,\ \cdots, \ 8$), the total number of the expansion coefficients would reach $L=288$. The corresponding qubit count for TTS simulations with $K_r=3$ would require our simulations to be run on high-performance supercomputers.

We estimate the step size $\tau $ for the implementation of the TTS algorithm together with the OAA scheme. In particular, the $\mathds{L}^{1}$ norm of the LF Hamiltonian in Eq. (\ref{eq:given_H}) is $\Lambda = 0.110024 $ GeV. Therefore, the corresponding step size for our TTS simulations is $\tau = \ln2/\Lambda \approx 6.3 \ \text{GeV}^{-1} $. 

We take integer number of time steps with $r=25$ in our numerical demonstrations. The total simulation time is $x^+ = r \tau = 157.5$ GeV$^{-1}$. In principle, longer simulation time and fractional simulation step (i.e., $r$ is not an integer) are also achievable. 

For benchmark purposes, we also perform the simulations via the Trotter algorithm. The total number of qubits necessary for the simulation is $6$. These are the necessary system qubits to encode the selected basis set. Recall that we save the qubits for encoding the bases of the longitudinal momentum and the helicity. Also, the ancilla qubits are not required for the Trotter implementation. 
We take the step size of the Trotter simulation to be the same as that of the TTS algorithm in this work.

\subsection{Simulation conditions and measurement}

Without loss of generality, we take the initial quark state $ \ket{\psi; x^+ }$ at $x^+ =0$ to be of $\ket{\boldsymbol{p}^{\perp}}=\ket{p^1 , p^2}= \ket{0,\ 0}$, and $\ket{c} = \ket{\text{Red} }$ in our numerical demonstrations. We fix the longitudinal momentum $p^+ =850$ GeV and the helicity $\lambda = 1/2$ in our model problem. 
Other initial configurations can also be implemented, e.g., the color neutral state of the quark.

We simulate the evolution of the quark state via the TTS and Trotter approaches with the IBM QASM simulator \cite{Qiskit} according to the procedures discussed in Sec. \ref{sec:algorithms}. Based on the simulations, we obtain the state $ \ket{\psi;x^{+}} $ of the scattered quark at $x^+ > 0 $ during the scattering. In each simulation, we set the shot number to be $10^{6}$ in order for good statistics.

The evolved state $ \ket{\psi;x^{+}} $ is projected onto various basis states [see Eq.~\eqref{eq:amplitude_basis_expansion}]. The probability $ | c_{\beta }(x^+) |^2$ of measuring each basis state $\ket{\beta } $ is recorded, which presents our knowledge of the scattered quark. Based on the knowledge of $ \ket{\psi;x^{+}} $, we can also evaluate other quantities that characterize the properties of the scattering system.

\section{Results and discussions}
\label{sec:results_discussion}

\begin{figure} 
	\includegraphics[scale=0.6]{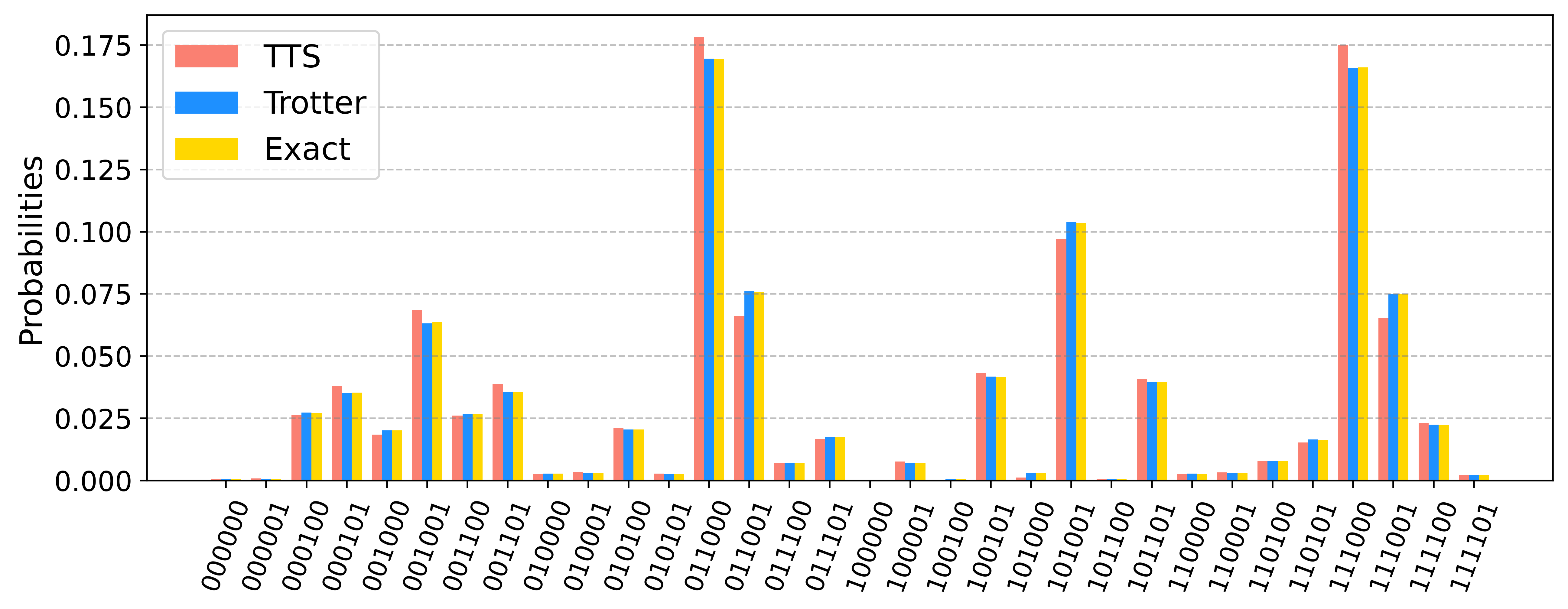}
	\caption{(color online)
		Final probability distribution of different basis states at the end of the 
        simulations via different approaches. The probability distributions obtained by the classical calculation (denoted as ``Exact") are compared with those obtained by the TTS and Trotter approaches. Each basis state is labeled by the set of quantum numbers $\{p^1,\ p^2,\ c\}$ (recall that $p^+$ and $\lambda$ are fixed for the in our model problem and are hence omitted), which is denoted by a 6-digit binary on the horizontal axis, where the first/middle/last (from bottom to top) two digits denote the value of $p^1$/$p^2$/$c$. 
        The simulation time is $ x^+ = 157.5 $ GeV$^{-1}$. 
	}
	\label{fig:Pr_distribution}
\end{figure}

We perform the dynamics simulations via the TTS algorithm, together with the first-order Trotter algorithm as a benchmark, for our model problem discussed in Sec. \ref{sec:Model_Problem}, where we also present the simulation conditions and circuit parameters for both implementations. We present the final probability distribution of different basis states at $ x^+ = 157.5\ \text{GeV}^{-1}$ in Fig. \ref{fig:Pr_distribution}, where these results are compared with those from exact classical calculations via \texttt{Mathematica} \cite{Wolfram_One}. 
We postpone the systematic investigations of the errors, noises and corresponding error corrections \cite{nielsen_chuang_2010,Devitt_2013} in the simulations, which is involving while our focus of the present work is the setup of the formal quantum algorithm for the simulating the QCD dynamics.

We note that only those basis states with $ \ket{\text{Red}} \mapsto \ket{00}$ and $ \ket{\text{Green}} \mapsto \ket{01}$ in the color space receive population at the end of the evolution. No state with $ \ket{\text{Blue}} \mapsto \ket{10}$ is populated at the end of the evolution. This is because we only include the contribution of  ${T}^{1}$ in the interaction $V$, which makes $ \ket{\text{Blue}} $ a dark state (i.e., a state that is forbidden by any transitions from or to other states). Nor is the state with $ \ket{\text{Null}} \mapsto \ket{11}$ populated as it does not map to any physical state in our encoding scheme.

We find that the probability distribution obtained by the TTS algorithm with $K_r = 3$ agrees with the classical calculation.
We expect that the precision of the simulation can be significantly improved by including the higher-order terms in the Taylor series expansion in the TTS implementation \cite{PhysRevLett.114.090502}. 
The initial quark state is prepared to have $p^1=p^2=0$ and $c=$Red, which corresponds to the binary string `101000' in the system register following our compact encoding scheme. During the scattering, we find other basis states of the quark system (i.e., the eigenbases of the asymptotic scattering system) gain populations via the transitions induced by the coupling between the quark and the external color field. These probabilities can be applied to determine the elastic and inelastic scattering cross sections of the quark system \cite{PhysRevD.88.065014,PhysRevD.101.076016,PhysRevD.104.056014}.

\begin{figure}
	\begin{subfigure}{.5\textwidth}
		\centering
		\includegraphics[scale = 0.55]{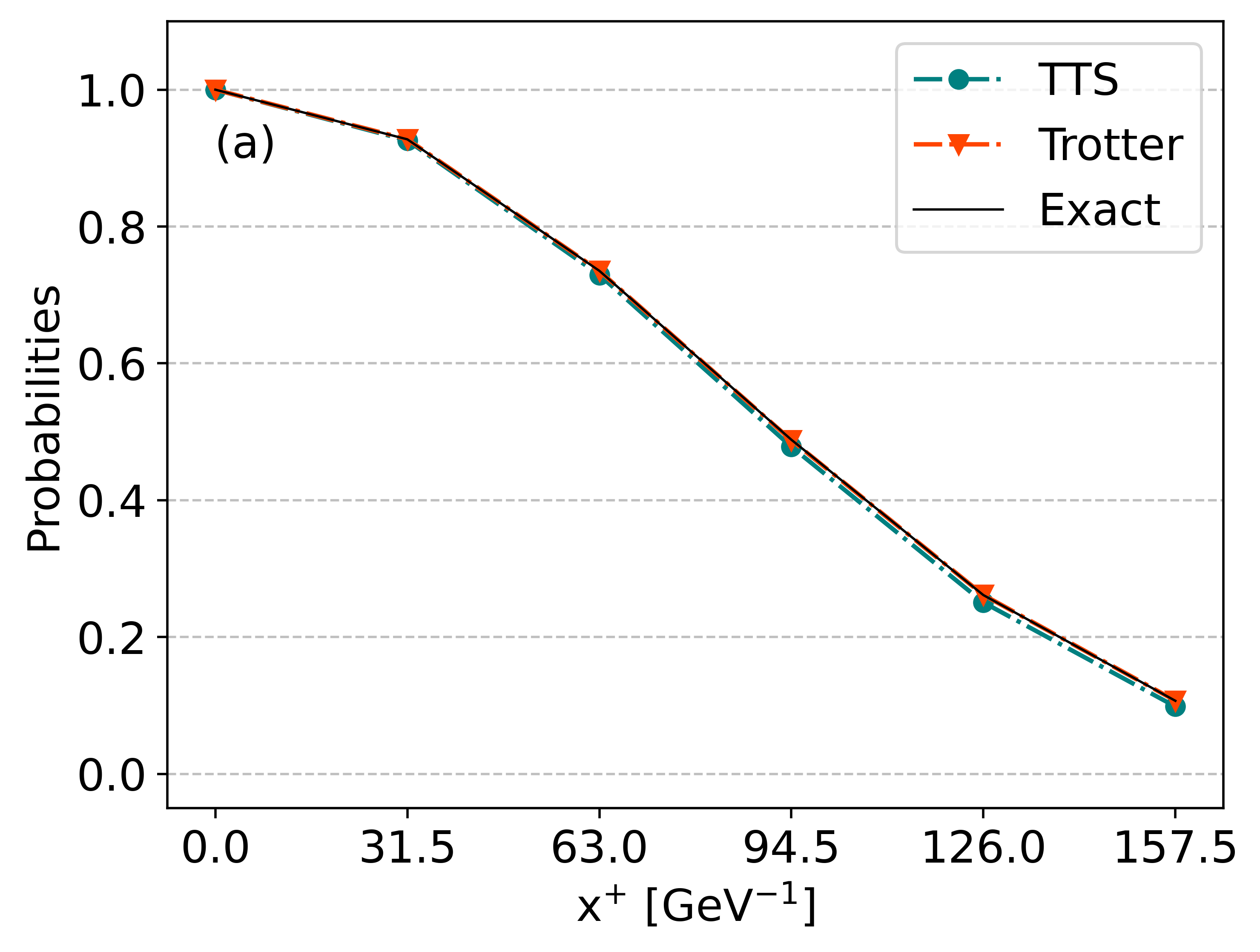}
	\end{subfigure}%
	\begin{subfigure}{.5\textwidth}
		\centering
		\includegraphics[scale = 0.55]{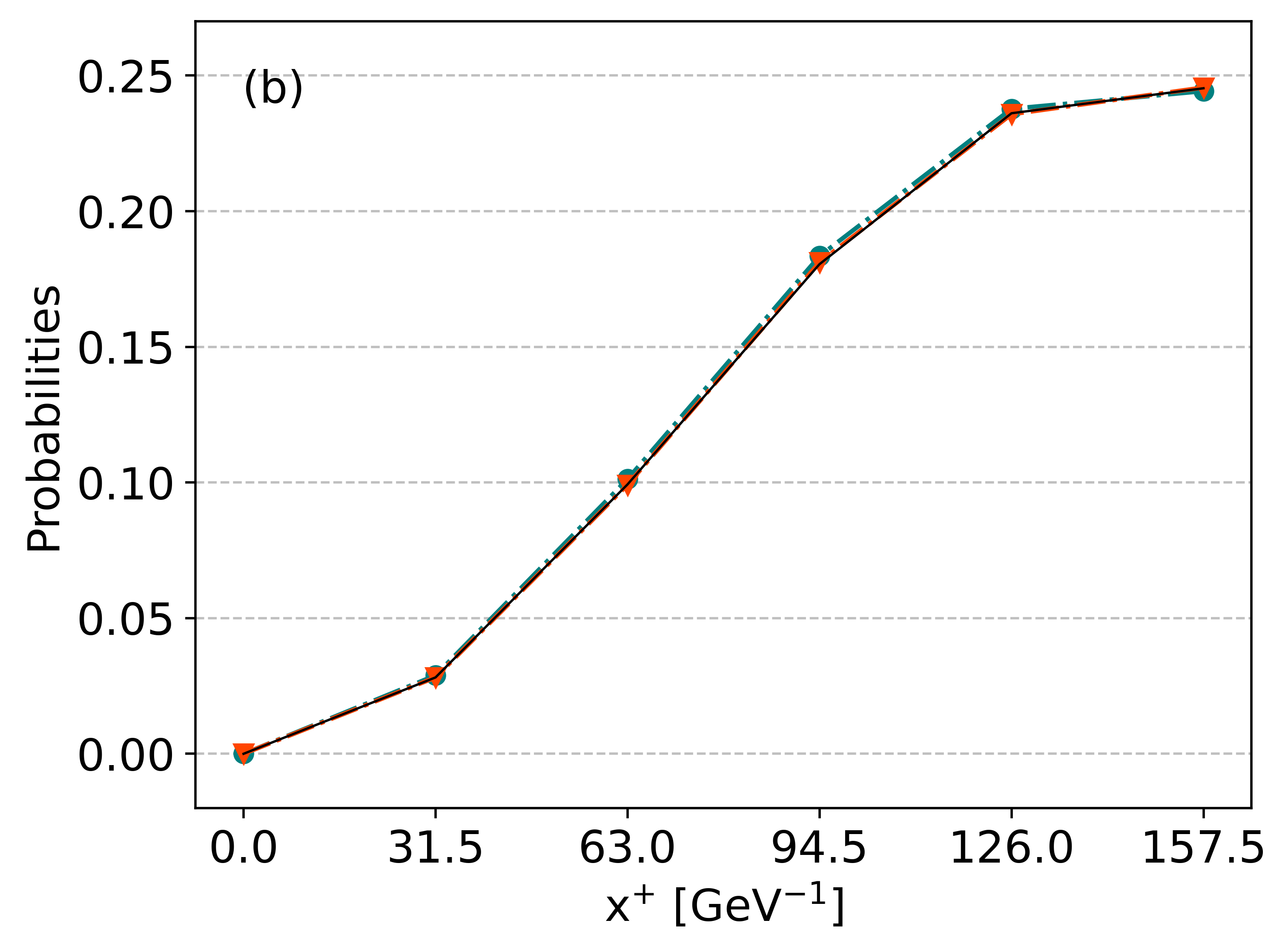}
	\end{subfigure}
	\caption{(color online)
            The probabilities of two transverse momentum modes of the quark during the scattering. Panels (a) and (b) present the probabilities of the modes with the transverse momenta $ \{ p^{1}=0,\ p^{2}=0 \}$  and $ \{  p^{1}=- a_p^{\perp} ,\ p^{2}=0 \} $ as functions of the light front time, respectively. The TTS algorithm with $K_r=3$ (green discs) and the Trotter algorithm (red triangles) are employed in the simulations. The results via different algorithms are joined by dot-dashed lines to guide the eye. The exact results from classical calculations (black solid line) are also presented for comparison. 
            }
	\label{fig:evo_of_Pr}
\end{figure}

We study the probabilities of the modes with different transverse momenta of the quark under the scattering with the external SU(3) color field. In Fig. \ref{fig:evo_of_Pr}, we present the probabilities of two elected sets of basis states, each with specific transverse momentum, as functions of the evolution time. These results are obtained from the quark wave functions solved via the TTS (with $K_r=3$) {and the Trotter approaches}, where all the degrees of freedom other than the transverse momentum are traced out.
We find that the probability of the basis states with $  p^{1}=0 $ and $p^{2}=0 $ starts from $100\%$ and drops monotonically during the evolution [Fig.~\ref{fig:evo_of_Pr}(a)]. 
Recall that we prepare the initial quark state to be of $  p^{1}=0 $ and $ p^{2}=0 $.
We also present the probability of the basis states with $  p^1 =- a_p^{ \perp } $ and $ p^2 =0 $ (where $a_p^{\perp} = 2\pi /10 $ GeV) as a function of the evolution time in Fig. \ref{fig:evo_of_Pr}(b); these states gain populations during the scattering. 
These population evolutions indicate the transfer of the transverse momentum between the external field (medium) and the quark during scattering, which induces the momentum broadening in the jet \cite{Barata:2022wim}.

\begin{figure} 
	\includegraphics[scale=0.6]{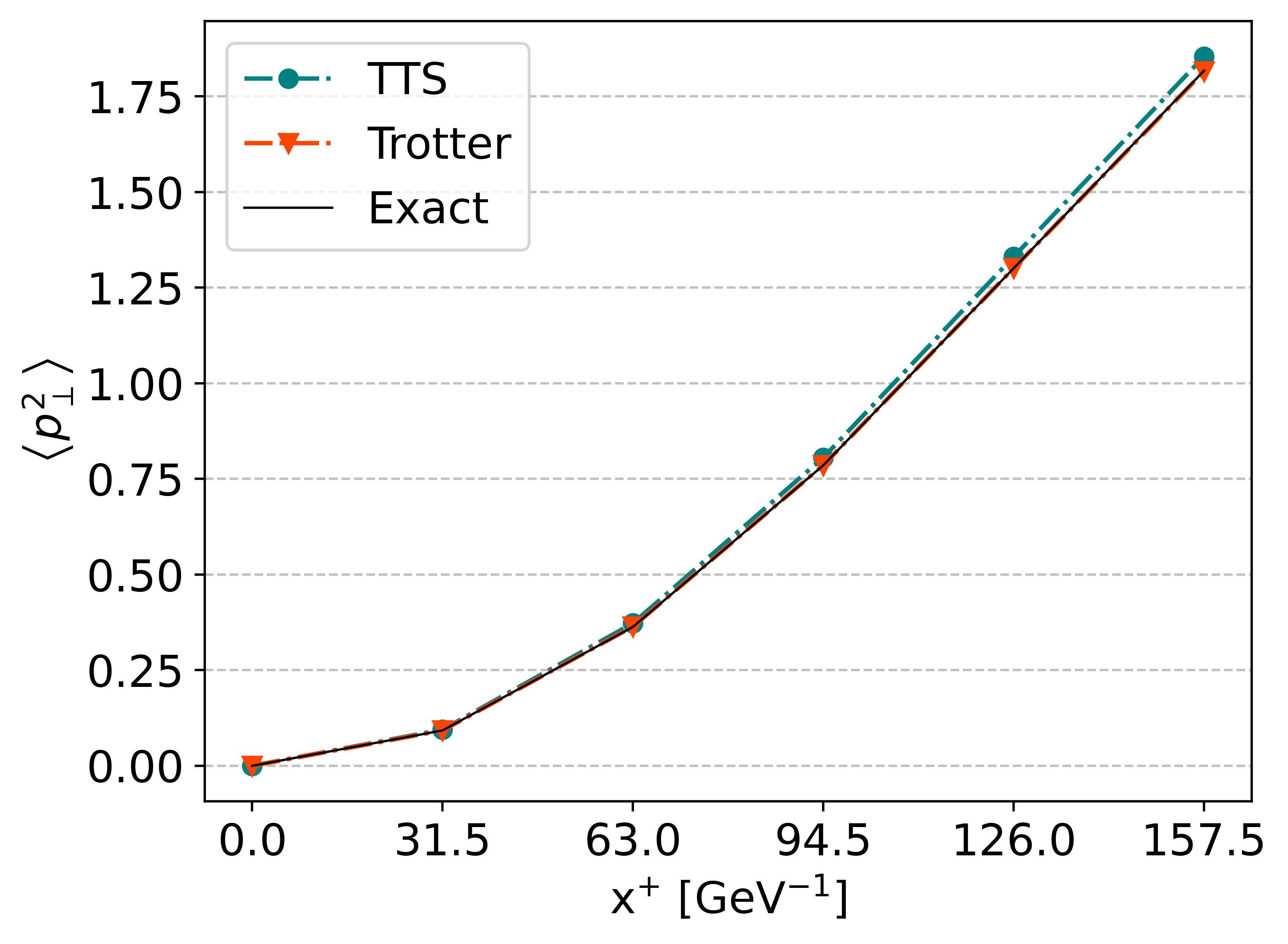}
	\caption{(color online)
		The evolution of the expectation $\braket{\boldsymbol{p}^{2}_{\perp}}$ of the quark. 
		The results marked as green discs and red triangles are obtained based on the simulations via the TTS (with $K_r=3$) and Trotter algorithms. The black solid line denotes the exact results obtained from classical calculations. 
		}
	\label{fig:evo_of_p^2}
\end{figure}

We compute the expectation of the squared transverse momentum operator $\boldsymbol{p}_{\perp}^{2}$ based on the probability of the transverse momentum modes of the quark during the scattering.
The expectation value $\braket{\boldsymbol{p}_{\perp}^{2}}$ is of vital importance in studying the jet process in heavy ion collisions \cite{Barata:2021yri,Barata:2022wim}. In Fig. \ref{fig:evo_of_p^2}, we present the results of $\braket{\boldsymbol{p}_{\perp}^{2}}$ as a function of the evolution time. We find that the results of $\braket{\boldsymbol{p}_{\perp}^{2}}$ by the TTS approach agree with the exact results. In addition, we note that $\braket{\boldsymbol{p}_{\perp}^{2}}$ increases during the scattering. While we prepare the initial quark state to be with  $\braket{\boldsymbol{p}_{\perp}^{2}}=0$, the scattered quark disperses in transverse momentum due to its interaction with the external field generated by the nucleus.

\begin{figure}
	\begin{subfigure}{.5\textwidth}
		\centering
		\includegraphics[scale = 0.55]{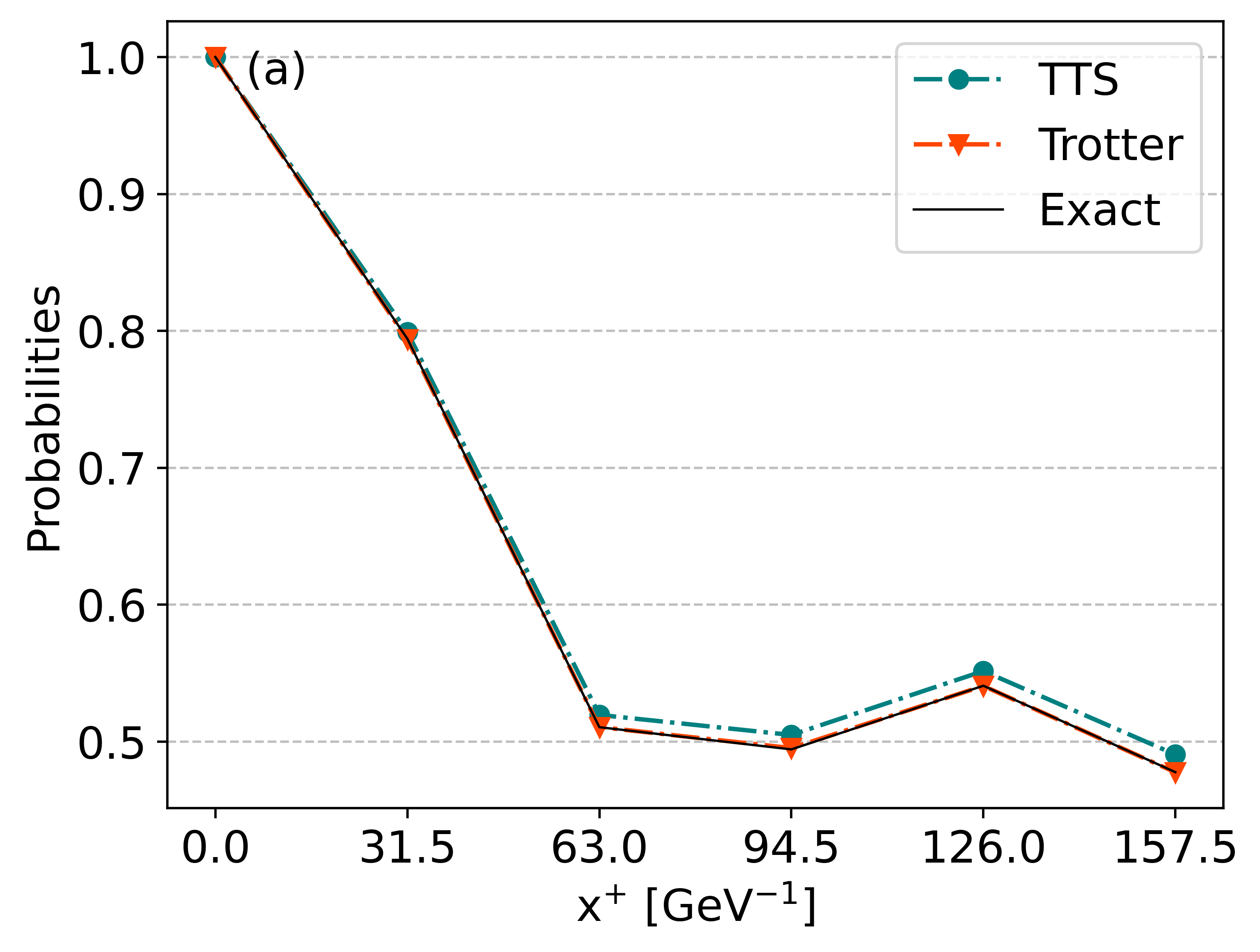}
		\label{fig:Red}
	\end{subfigure}%
	\begin{subfigure}{.5\textwidth}
		\centering
		\includegraphics[scale = 0.55]{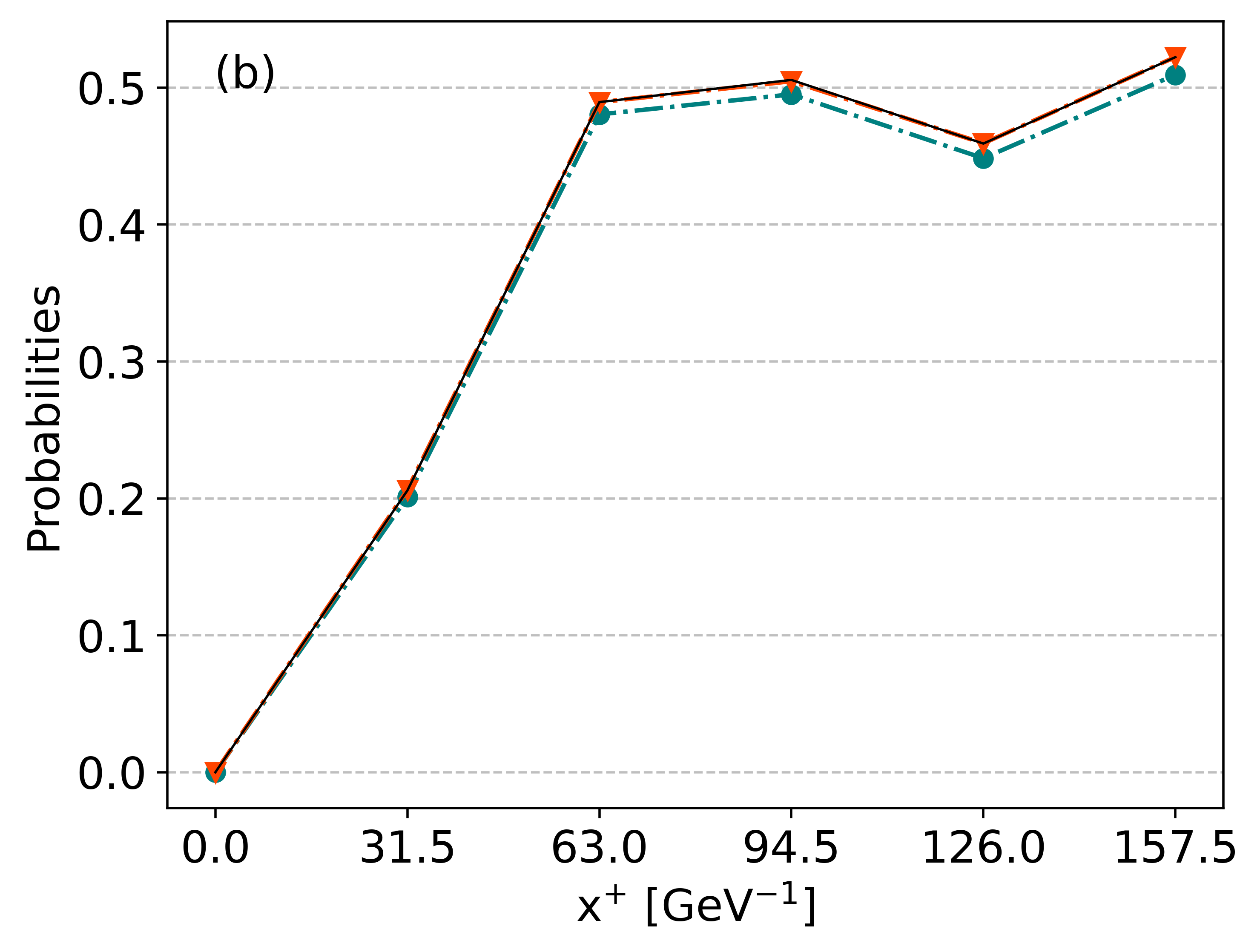}
		\label{fig:Green}
	\end{subfigure}
	\caption{(color online)
            Probabilities of the color of the quark during the scattering. Panels (a) and (b) show the evolutions of the probabilities of $\ket{\text{Red}}$ and  $\ket{\text{Green}}$, respectively.
	}
	\label{fig:evo_of_Color}
\end{figure} 

Similar to Fig. \ref{fig:evo_of_p^2}, we compute the probability of the color based on the quark wave functions obtained via the TTS (with $K_{r}=3$) and the Trotter approaches at each moment during the evolution, in which process we trace out all the degrees of freedom other than the color. These results are shown in Fig. \ref{fig:evo_of_Color}. 
We note that the probability of $\ket{\text{Red} }$ drops from $100\%$ to about $50\%$ during the evolution, while that of $\ket{\text{Green} }$ increases from $0$ to about $50\%$ . This can be understood from the facts that 1) we prepare the initial state of the quark to be in the $\ket{\text{Red} }$ state; and 2) we restrict the background field to allow transitions only between the $\ket{\text{Red} }$ and $\ket{\text{Green} }$. The $\ket{\text{Blue} } $ is a dark state in our restrict model problem, and the probabilities of $\ket{\text{Red} }$ and $\ket{\text{Green} }$ complement each other throughout the scattering. The probabilities of $\ket{\text{Red} }$ and $\ket{\text{Green} }$ become almost equal soon after the scattering begins. 
If we had included all the terms of $V$ [Eq. \eqref{eq:statistical_LF_pot}] in our simulations (recall from Eq. \eqref{eq:vPrime_part} that we do not include all the terms of $V$ due to the restrictions on our computational resources), we would expect the final probabilities of the states $\ket{\text{Red} }$, $\ket{\text{Green} }$, and $\ket{\text{Blue} } $ to be approximately $1/3$, which is an indication of the color depolarization because of the color transfer between the randomized color source (the background field) and the quark system through the transition processes.

\begin{figure} 
	\includegraphics[scale=0.6]{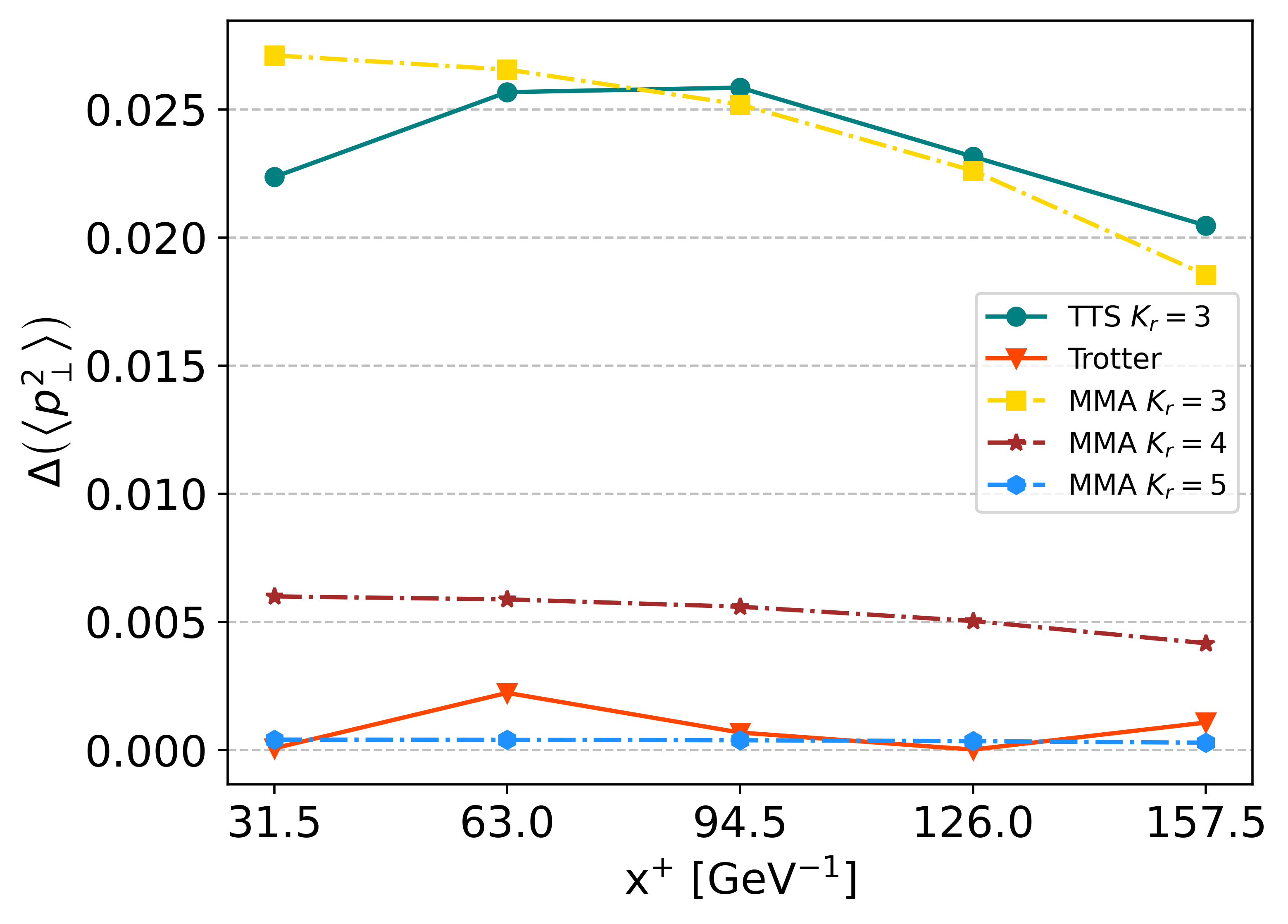}
	\caption{(color online)
		Relative difference $ \Delta\left( \braket{\boldsymbol{p}_{\perp}^{2}} \right) $ [Eq. \eqref{eq:deviation}]  of the results obtained via the TTS and Trotter approaches with respect to the exact results as a function of the simulation time. The results between $x^+ \in [31.5, \ 157.5]$ GeV$^{-1}$ are presented. The results marked as green discs and red triangles are obtained based on the TTS (with $K_r=3$) and Trotter algorithms via simulations on the IBM QASM simulator. The results marked as yellow squares, dark red stars, and blue pentagons are obtained based on the matrix multiplications according to Eq. \eqref{eq:OAA_total_eq} via \texttt{Mathematica}, which correspond to the classical predictions (denoted by ``MMA") of the TTS algorithm with the truncation orders being $K_r=3, \ 4, \ \text{and} \ 5$, respectively. These results are all compared with the exact results from classical calculations.
		Results obtained based on quantum simulations are joined by solid lines, while those obtained via pure classical calculations are joined by dash-dotted lines.
		}
	\label{fig:dev_of_p^2}
\end{figure}

We present the absolute relative deviation 
\begin{equation}
	\Delta\left( \braket{\boldsymbol{p}_{\perp}^{2}} \right) \equiv \frac{ | \braket{\boldsymbol{p}_{\perp}^{2}}_\text{sim} - \braket{\boldsymbol{p}_{\perp}^{2}}_{\text{exact}} | }{ \braket{\boldsymbol{p}_{\perp}^{2}}_\text{exact}}
	\label{eq:deviation},
\end{equation}
of the quark as a function of the simulation time and algorithm in Fig. \ref{fig:dev_of_p^2}. Here $\braket{\boldsymbol{p}_{\perp}^{2}}_\text{exact}$ denotes the exact results via classical calculations, while $ \braket{\boldsymbol{p}_{\perp}^{2}}_\text{sim} $ denotes the results via either the TTS algorithm (with $K_r =3$) or the Trotter algorithm. In addition, we also compute $\braket{\boldsymbol{p}_{\perp}^{2}}$ according to the TTS algorithm in terms of the classical matrix multiplications according to Eq. \eqref{eq:OAA_total_eq}, where the truncation orders $K_r$ are taken to be 3, 4, and 5. These results are also compared with the exact results according to Eq. \eqref{eq:deviation}, and the corresponding relative deviations are shown in Fig. \ref{fig:dev_of_p^2}.
We find that the results from the TTS approaches, either via the quantum simulation or via classical matrix multiplications, agree well with the exact results. 
We also note that the precision of the TTS simulations improves significantly with increasing truncation order $K_r$.   

In Figs. \ref{fig:Pr_distribution}--\ref{fig:dev_of_p^2}, we present the Trotter results to benchmark with respective TTS results. We find that the Trotter results are more precise than those from the TTS approach for our limited model problem, while the gate count to implement the Trotter approach is less than those to implement the TTS approach. It seems that the Trotter approach is more advantageous than the TTS approach.
However, this only holds for the problems with limited Hamiltonian norms and simulation times. 
Indeed, the gate scaling with the simulation time is a major consideration for achieving the efficiency of practical quantum simulations aiming at finite precision goal (e.g., $\epsilon \leq 10^{-4}$). While the gate cost of the TTS algorithm is linear in the simulation time $x^+$ [Eq. \eqref{eq:gate_cost_by_Pauli_strings_overall}], the gate scaling is quadratic with $x^+$ for the first-order Trotter algorithm [Eq. \eqref{eq:complexity_Trotter}].
Meanwhile, the gate cost is also sensitive to the $\mathbb{L}^1$ norm $\Lambda$ of the expansion coefficients of the Hamiltonian [Eq. \eqref{eq:Hamiltonian_decomposition}], where $ \Lambda $ is expected to increase with the basis dimension.
We find that the Trotter algorithm presents a less favorable gate scaling to the Hamiltonian norm than the TTS.
With these major considerations, one expects the TTS algorithm to be more advantageous in efficiency than the Trotter for larger-scale simulations with long simulation time, letting alone the factor that the TTS algorithm is optimal in the error scaling compared to the scaling of $1/\epsilon$ for the first-order Trotter approach.

\section{Summary and outlook}
\label{sec:conclusions_outlook}

We present a framework to quantum simulate the ultra-relativistic quark-nucleus scattering dynamics. 
This framework employs the truncated Taylor series (TTS) \cite{PhysRevLett.114.090502} algorithm that is more advantageous for precise and efficient large-scale dynamics simulations on future quantum computers than the Trotter algorithm \cite{lloyd1996universal} adopted in previous studies \cite{Barata:2021yri,Yao:2022eqm,Barata:2022wim,Barata:2023clv}.
Working with the light-front (LF) Hamiltonian formalism, we employ the eigenbasis of the asymptotic scattering system to solve the scattering. 
We lattice discretize these bases and encode the discretized bases with a compact encoding scheme. 
We also exploit the unique operator structure of the LF Hamiltonian of the scattering system, which enables an efficient Hamiltonian input scheme that takes advantage of the quantum Fourier transformation (qFT) \cite{nielsen_chuang_2010}.
This framework can be generalized for large-scale dynamics simulations of a wide range of QCD systems on future fault-tolerant quantum hardwares.

We simulate the dynamics of the scattering system utilizing the LF Hamiltonian formalism based on the idea of the time-dependent basis light-front quantization \cite{PhysRevD.88.065014,PhysRevD.101.076016,PhysRevD.104.056014}. 
For simplicity, we restrict our focus to the simulation of time-independent Hamiltonian in this work, while this approach can also be generalized to simulate time-dependent Hamiltonians \cite{PhysRevD.101.076016,PhysRevD.104.056014}.
Working in the Schr\"odinger picture, the LF Hamiltonian of the scattering system is categorized into two parts: 1) the reference part, which determines the kinematics of the quark when the external interaction is absent; and 2) the interaction part, which describes the coupling between the quark and the SU(3) color field generated by the nucleus. 
Indeed, the reference Hamiltonian determines the kinematics of the asymptotic scattering system, while the interaction Hamiltonian drives the dynamics during the scattering.
The reference Hamiltonian is a local operator in the basis of the longitudinal momentum, the transverse momentum, the helicity, and the color. By contrast, the interaction Hamiltonian is local in the longitudinal momentum space and the transverse coordinate space, while it induces transitions in the helicity and color spaces.

We elect the basis to be $  \ket{\beta}  = \ket{p^+ } \otimes \ket{p^1} \otimes \ket{p^2} \otimes \ket{\lambda} \otimes \ket{c}  $, where $ \ket{p^+ }  $,  $ \ket{p^1} \otimes \ket{p^2} $, $ \ket{\lambda }$,  and $ \ket{c} $ are the eigenstates of the longitudinal momentum, the transverse momentum, the helicity, and the color, respectively. Notably, this basis is the eigenbasis of the reference Hamiltonian.
Applying the lattice discretization and regularization schemes to the longitudinal and transverse momenta, we retain a set of discretized eigenbases in terms of the multidimensional lattice states (note that the helicity and color states are inherently discrete and they are mapped to corresponding lattice states as well). 
We encode these lattice states as binary strings on the quantum computer via the compact encoding scheme. 
The required number of qubits to encode these states is the logarithm of the total number of the discretized bases (or, equivalently, the total number of the corresponding
multidimensional lattice states).

With the choices of the basis and its encoding scheme, we input the reference Hamiltonian and the interaction Hamiltonian separately in view of their different operator structures.
We note that the reference Hamiltonian is diagonal (of sparsity 1) in the basis representation.
For the interaction Hamiltonian, it is non-sparse matrix in the chosen basis, which results in difficulties in the Hamiltonian input scheme. The major complexity results from the fact that the interaction Hamiltonian is local in the transverse coordinate basis (instead of the transverse momentum basis). This incurs inefficiency in inputting the Hamiltonian. However, we can improve the efficiency of the input scheme for the interaction Hamiltonian by implementing the qFT between the transverse momentum and the transverse coordinate bases.
As for the other degrees of freedom, the structure of the interaction matrix is simple due to the facts that the interaction is local in the longitudinal momentum basis, and that the number of helicity and color bases is limited for the scattering system.
In one word, the total LF Hamiltonian of scattering system can be expressed in terms of sparse matrices taking advantage of the efficient basis transformations via the qFT. 	
One can encode these sparse matrices in terms of the linear combination of unitaries, which are taken as the tensor products of Pauli matrices in this work. 

Based on the  qFT-assisted Hamiltonian input scheme, we solve the time-evolution operator according to the TTS algorithm and design the corresponding quantum circuit. 
We then simulate the dynamics of the scattering with the input state of the scattering system.
We obtain the transition probabilities (which in turn determine the elastic and inelastic scattering cross sections) and the other quantities of the scattering system based on the simulation results.
As for the benchmark purposes, we also perform simulations via the first-order Trotter algorithm utilizing the same qFT-assisted Hamiltonian input scheme.

We present the analysis of the gate and qubit cost based on our elected basis, the qFT-assisted Hamiltonian input scheme, and the simulation algorithms. 
We find that the asymptotic qubit costs of the TTS and Trotter approaches are the same, which scale as the logarithm of the basis dimension of the scattering system for practical simulations.
Compared to the Trotter algorithm, the gate cost of the TTS algorithm presents optimal scaling with the simulation error and near-optimal scaling with the simulation time at the cost of additional ancilla qubits. 
We can further reduce the gate cost by improving our Hamiltonian input scheme. 
For example, better options of unitaries are helpful to improve the efficiency of the Hamiltonian input scheme. It is also possible to implement other efficient oracle-based input schemes (see, e.g., Appendix \ref{sec:decompo_to_one_sparse} and Ref. \cite{berry2017exponential}) to encode the Hamiltonian. 
For future large-scale dynamics simulations beyond our restricted model problem, we expect the TTS algorithm to be more precise and efficient than the Trotter algorithm due to its optimal gate scaling with the simulation error and near optimal gate scaling with simulation time. 

We demonstrate our approach with a simple model problem.
In this problem, we model the SU(3) color field of the nucleus according to the theory of color glass condensate \cite{gelis2010color}. We take the color field to be time independent and retain only a few terms of the interaction Hamiltonian for simplicity. 
We simulate the scattering dynamics with the IBM QASM simulator \cite{Qiskit} via the TTS algorithm, where we retain the terms up to the third order of the Taylor series. A few elected quantities of the scattering system are computed based on the simulation results.
We benchmark the TTS results by those from the Trotter algorithm and from classical calculations.
We find that TTS results, even with limited truncation order, agree well with the Trotter results and the classical results. 

In the next step, it is straightforward to generalize the current approach to simulate multi-quark scattering dynamics. Including more complete Fock sector expansions of the physical quark system is also interesting.
Meanwhile,  other efficient Hamiltonian input schemes, such as the oracle-based approaches \cite{berry2017exponential,Du:2023bpw,Du:2024zvr}, and simulation algorithms \cite{PhysRevLett.118.010501,gilyen2019quantum,Berry2020timedependent} are worth explorations for optimal scalings of the gate and qubit costs.
Our ultimate goal along this research line is to develop an efficient and consistent approach for the precision dynamics simulation of the multi-quark-and-gluon systems in the ultra-relativistic collisions of heavy nuclei from first principles. 
Since the algorithmic structure of the TTS algorithm connects closely to other efficient algorithms such as the truncated Dyson series \cite{PhysRevA.99.042314}, our TTS-based simulation framework can also be adapted to simulate general time-dependent QCD dynamics in a straightforward manner.

\begin{acknowledgments}
We thank Meijian Li for sharing the data of the model problem.
W.D. and J.P.V. were partially supported by US DOE Grant DE-SC0023707 under the Office of Nuclear Physics Quantum Horizons program for the ``{\bf Nu}clei and {\bf Ha}drons with {\bf Q}uantum computers ({\bf NuHaQ})" project. 
X.Z. was supported by new faculty startup funding by the Institute of Modern Physics, Chinese Academy of Sciences, by Key Research Program of Frontier Sciences, Chinese Academy of Sciences, Grant No. ZDBS-LY-7020, by the Natural Science Foundation of Gansu Province, China, Grant No. 20JR10RA067, by the Foundation for Key Talents of Gansu Province, by the Central Funds Guiding the Local Science and Technology Development of Gansu Province, Grant No. 22ZY1QA006, by Gansu International Collaboration and Talents Recruitment Base of Particle Physics (2023-2027), by International Partnership Program of the Chinese Academy of Sciences, Grant No. 016GJHZ2022103FN, by National Natural Science Foundation of China, Grant No. 12375143, by National Key R$\&$D Program of China, Grant No. 2023YFA1606903 and by the Strategic Priority Research Program of the Chinese Academy of Sciences, Grant No. XDB34000000.
\end{acknowledgments}

\appendix

\section{CONVENTIONS}
\label{sec:APP_conventions}

We denote the light-cone coordinate by $x^{\pm}=x^{0}\pm x^{3},\ x_{\perp}=\left( x^{1}, x^{2}
\right)$, where $x^{+}$ denotes the light front time, $x^{-}$ is the longitudinal coordinate, and $x_{\perp}$ is the transverse coordinates. We have 
\begin{equation}
	x.y = \frac{1}{2}x^{+}y^{-} + \frac{1}{2}x^{-}y^{+} - x_{\perp}\cdot y_{\perp}.
\end{equation}

We adopt the Minkowski metric in light-front coordinates, which reads 
\begin{equation}
	g^{\mu\nu} = 
	\begin{pmatrix}
		0 & 0 & 0 & 2 \\
		0 & -1 & 0 & 0 \\
		0 & 0 & -1 & 0 \\
		2 & 0 & 0 & 0
	\end{pmatrix}, \
	g_{\mu\nu} = 
	\begin{pmatrix}
		0 & 0 & 0 & \frac{1}{2} \\
		0 & -1 & 0 & 0 \\
		0 & 0 & -1 & 0 \\
		\frac{1}{2} & 0 & 0 & 0
	\end{pmatrix},   
\end{equation}
for $\mu,\nu = +,1,2,-$. We have $x_{-} = \frac{1}{2}x^{+},\ x_{+} = \frac{1}{2}x^{-}$. For covariant derivatives, we have 
\begin{equation}
	\partial_{+}=\frac{\partial}{\partial x^{+}} = \frac{1}{2}\partial^{-}, \
	\partial_{-}=\frac{\partial}{\partial x^{-}} = \frac{1}{2}\partial^{+}.
\end{equation}
Thus $\partial^{+}=2\frac{\partial}{\partial x^{-}}$ and $\partial^{-}=2\frac{\partial}{\partial x^{+}}$.
We choose the following representation of $\gamma$ matrices: 
\begin{equation}
	\gamma^{0} = 
	\begin{pmatrix}
		0 & -i \\
		i & 0
	\end{pmatrix}, \
	\gamma^{3} = 
	\begin{pmatrix}
		0 & i \\
		i & 0
	\end{pmatrix}, \
	\gamma^{i} = 
	\begin{pmatrix}
		-i\hat{\sigma}^{i} & 0 \\
		0 & i\hat{\sigma}^{i}
	\end{pmatrix}, \
	\gamma^{5} = 
	\begin{pmatrix}
		\sigma^{3} & 0 \\
		0 & -\sigma^{3}
	\end{pmatrix},   
\end{equation}
where $\hat{\sigma}^{1}=\sigma^{2},\ \hat{\sigma}^{2}=-\sigma^{1}$. Then we have 
\begin{equation}
	\gamma^{+} = 
	\begin{pmatrix}
		0 & 0 \\
		2i & 0
	\end{pmatrix}, \
	\gamma^{-} = 
	\begin{pmatrix}
		0 & -2i \\
		0 & 0
	\end{pmatrix}. \  
\end{equation}

The Pauli matrices are denoted as 
\begin{align}
	& {\sigma}^{0} = 
	\begin{pmatrix}
		1 & 0 \\
		0 & 1
	\end{pmatrix} 
	= {I}, \ \ \ 
	{\sigma}^{1} = 
	\begin{pmatrix}
		0 & 1 \\
		1 & 0
	\end{pmatrix} 
	= {X}, \nonumber \\  
	& {\sigma}^{2} = 
	\begin{pmatrix}
		0 & -i \\
		i & 0
	\end{pmatrix} 
	= {Y}, \ \   
	{\sigma}^{3} = 
	\begin{pmatrix}
		1 & 0 \\
		0 & -1
	\end{pmatrix} 
	= {Z}.
\end{align}

\section{DECOMPOSITION IN TERMS OF THE 1-SPARSE UNITARIES VIA AN ORACLE MODEL}
\label{sec:decompo_to_one_sparse}

As one of the alternative options, it is possible to approximate $ {H}  $ in terms of 1-sparse unitaries for a better scaling of $L$. As $P^-_{\rm QCD}$ and $V$ are both sparse matrices, we can approximate the expansion of $H $ [Eq. \eqref{eq:Hamiltonian_decomposition}] as (Lemmas 4.3 and 4.4 in Ref. \cite{berry2017exponential}) 
\begin{equation}
	{H} = \frac{1}{2}  {P}_{\rm QCD}^{-}  + \frac{1}{2} \mathcal{F} ^{\dag} {V} \mathcal{F} = \gamma \sum _{\ell _P =0}^{L'_1-1} \bar{ {h} }_{{P},\ell _P} + \gamma \left[ \mathcal{F} ^{\dag}  \left( \sum _{\ell _V=0 }^{L'_2-1} \bar{	{h} }_{V, \ell _V}  \right)  \mathcal{F} \right] ,
	\label{eq:special_orale_access}
\end{equation}
with the error $ {O}(\gamma ) $. The 1-sparse unitaries $ \{ \bar{ {h} }_{{P},\ell _P} \}$ (with $\ell _P= 0,\ 1, \ 2, \ \cdots , \ L'_1-1$) and $ \{ \bar{ {h} }_{V,\ell _V} \}$ (with $\ell _V = 0,\ 1, \ 2, \ \cdots , \ L'_2-1$) can be obtained from the sparse matrices ${P}^{-} _{\rm QCD}  /2$ and $ {V} /2$ via oracle queries \cite{berry2017exponential,PhysRevLett.114.090502}, respectively. All the terms are of the same expansion coefficient $ \gamma >0 $. Hence, we have the $\mathbb{L}^1$ norm of the expansion coefficients to be $\Lambda = (L'_1 +L_2' ) \gamma $.

We estimate the scaling of $L'=L'_1 + L'_2$ as follows. The number of unitaries in the decomposition of $ {P}^{-} _{\rm QCD} /2 $ is $L'_1 \in {O} \big(d'_P || {P}^{-} _{\rm QCD} ||_{\rm max } /\gamma \big) $ with $ d'_P =1$ being the sparsity of the diagonal $ {P}^{-} _{\rm QCD} $ matrix and $ || {P}^{-} _{\rm QCD} ||_{\rm max } $ being the max norm of $ {P}^{-} _{\rm QCD} $. Similarly, $L'_2$ can be obtained based on the $ {V} /2 $. In particular, the upper bound of $L'_2$ is $ L'_2  \in {O} \big(d'_V || {V}||_{\rm max } /\gamma \big) $, with $  || {V}||_{\rm max } $ being the max norm of $V$. $ d'_V $ denotes the sparsity of the $ {V} $ matrix, which is of limited value. In sum, we have 
\begin{equation}
	L' \in {O} \big(  ( || {P}^{-} _{\rm QCD} ||_{\rm max } + d'_V || {V}||_{\rm max } )/\gamma  \big) .	 
\end{equation}

We estimate the scaling of $L'$ in solving realistic scattering problems that necessitates sufficiently large basis space. In such cases, $ ( || {P}^{-} _{\rm QCD} ||_{\rm max } + d'_V || {V}||_{\rm max } ) $ is dominated by $|| {P}^{-} _{\rm QCD} ||_{\rm max }  $ that is determined by the ultraviolet cutoff $\Lambda _{\rm UV}$ of the transverse momentum.\footnote{$|| {V}||_{\rm max } $ is upper bounded by the saturation scale of the nuclear medium, which is less than $\Lambda _{\rm UV}$.} Meanwhile, the magnitude of $\gamma $ is required to be $ \gamma \in \Theta (\epsilon /x^+)$ in order to obtain the overall simulation error $\epsilon $ for the total simulation time $x^+$. Therefore, we can take
\begin{equation}
	L' \in {O} \big( \Lambda _{\rm UV} \cdot x^+ / \epsilon ) ,
\end{equation}
for the dynamics simulation with time $x^+$ and error $\epsilon $.

In general, the oracle-based decomposition scheme [Eq. \eqref{eq:special_orale_access}] presents a better scaling of $ L'$ with respect to the momentum cutoff than that of $L$. However, it presents explicit dependence of the simulation error $\epsilon $ and the simulation time $x^+ $. Besides, the oracle design depends on specific applications. 
As for our numerical demonstrations in this work, we mainly work with the decomposition of $H$ in terms of the Pauli strings. Indeed, this Hamiltonian decomposition scheme is straightforward though the scaling of $L$ is not optimal. 

The other oracle-based Hamiltonian input scheme [Eq. \eqref{eq:special_orale_access}] is more promising in simulation efficiency. 
Based on the discussion in Sec. \ref{sec:decompo_to_one_sparse}, it is possible to decompose $ {H} $ in terms of 1-sparse unitaries via oracle queries. 
Following the discussion in Ref. \cite{PhysRevLett.114.090502}, the total gate cost can be estimated as
\begin{equation}
	O \left(  r N_{\rm sys} K_r \log L' + r N^2_{\rm sys} \right)  , \label{eq:cost_temp_10}
\end{equation}
where the 1-sparse self-inverse operations have the complexity $ O (N_{\rm sys})$. The second term denotes the gate cost for implementing the qFT for each single-step evolution.
With $\Lambda = L' \gamma $ and $\gamma = \Theta (\epsilon / x^+)$, we have $ L'  = \Theta ( \Lambda x^+ / \epsilon ) $. The total gate cost [Eq. \eqref{eq:cost_temp_10}] can then be rewritten as 
\begin{equation}
	O \left( 
	\Lambda x^+ N_{\rm sys} \frac{\log ^2 \left( \Lambda x^+ /\epsilon\right)}{\log\log\left( \Lambda x^+ /\epsilon\right)}   + \Lambda x^+ N^2_{\rm sys}
	\right) .
\end{equation}
As claimed in Ref. \cite{PhysRevLett.114.090502}, there is no need to perform the explicit sequence of $L'$ controlled operations that implement $  \bar{ {h} }_{{P},\ell _P} $ and $ \bar{ {h} }_{{V},\ell _V} $ in Eq. \eqref{eq:special_orale_access} when the Hamiltonian is accessed by an oracle. This relaxes the dependence of the gate cost on the number of unitaries $L'$ [cf., Eq. \eqref{eq:gate_cost_by_Pauli_strings_overall}].

\normalem

\bibliography{apssamp}

\providecommand{\noopsort}[1]{}\providecommand{\singleletter}[1]{#1}%
\begin{thebibliography}{86}%
\makeatletter
\providecommand \@ifxundefined [1]{%
 \@ifx{#1\undefined}
}%
\providecommand \@ifnum [1]{%
 \ifnum #1\expandafter \@firstoftwo
 \else \expandafter \@secondoftwo
 \fi
}%
\providecommand \@ifx [1]{%
 \ifx #1\expandafter \@firstoftwo
 \else \expandafter \@secondoftwo
 \fi
}%
\providecommand \natexlab [1]{#1}%
\providecommand \enquote  [1]{``#1''}%
\providecommand \bibnamefont  [1]{#1}%
\providecommand \bibfnamefont [1]{#1}%
\providecommand \citenamefont [1]{#1}%
\providecommand \href@noop [0]{\@secondoftwo}%
\providecommand \href [0]{\begingroup \@sanitize@url \@href}%
\providecommand \@href[1]{\@@startlink{#1}\@@href}%
\providecommand \@@href[1]{\endgroup#1\@@endlink}%
\providecommand \@sanitize@url [0]{\catcode `\\12\catcode `\$12\catcode
  `\&12\catcode `\#12\catcode `\^12\catcode `\_12\catcode `\%12\relax}%
\providecommand \@@startlink[1]{}%
\providecommand \@@endlink[0]{}%
\providecommand \url  [0]{\begingroup\@sanitize@url \@url }%
\providecommand \@url [1]{\endgroup\@href {#1}{\urlprefix }}%
\providecommand \urlprefix  [0]{URL }%
\providecommand \Eprint [0]{\href }%
\providecommand \doibase [0]{https://doi.org/}%
\providecommand \selectlanguage [0]{\@gobble}%
\providecommand \bibinfo  [0]{\@secondoftwo}%
\providecommand \bibfield  [0]{\@secondoftwo}%
\providecommand \translation [1]{[#1]}%
\providecommand \BibitemOpen [0]{}%
\providecommand \bibitemStop [0]{}%
\providecommand \bibitemNoStop [0]{.\EOS\space}%
\providecommand \EOS [0]{\spacefactor3000\relax}%
\providecommand \BibitemShut  [1]{\csname bibitem#1\endcsname}%
\let\auto@bib@innerbib\@empty
\bibitem [{\citenamefont {Feynman}(2018)}]{feynman2018simulating}%
  \BibitemOpen
  \bibfield  {author} {\bibinfo {author} {\bibfnamefont {R.~P.}\ \bibnamefont
  {Feynman}},\ }\bibfield  {title} {\bibinfo {title} {Simulating physics with
  computers},\ }in\ \href@noop {} {\emph {\bibinfo {booktitle} {Feynman and
  computation}}}\ (\bibinfo  {publisher} {CRC Press},\ \bibinfo {year} {2018})\
  pp.\ \bibinfo {pages} {133--153}\BibitemShut {NoStop}%
\bibitem [{\citenamefont {Bauer}\ \emph {et~al.}(2020)\citenamefont {Bauer},
  \citenamefont {Bravyi}, \citenamefont {Motta},\ and\ \citenamefont
  {Chan}}]{bauer2020quantum}%
  \BibitemOpen
  \bibfield  {author} {\bibinfo {author} {\bibfnamefont {B.}~\bibnamefont
  {Bauer}}, \bibinfo {author} {\bibfnamefont {S.}~\bibnamefont {Bravyi}},
  \bibinfo {author} {\bibfnamefont {M.}~\bibnamefont {Motta}},\ and\ \bibinfo
  {author} {\bibfnamefont {G.~K.-L.}\ \bibnamefont {Chan}},\ }\bibfield
  {title} {\bibinfo {title} {Quantum algorithms for quantum chemistry and
  quantum materials science},\ }\href@noop {} {\bibfield  {journal} {\bibinfo
  {journal} {Chemical Reviews}\ }\textbf {\bibinfo {volume} {120}},\ \bibinfo
  {pages} {12685} (\bibinfo {year} {2020})}\BibitemShut {NoStop}%
\bibitem [{\citenamefont {Cao}\ \emph {et~al.}(2019)\citenamefont {Cao},
  \citenamefont {Romero}, \citenamefont {Olson}, \citenamefont {Degroote},
  \citenamefont {Johnson}, \citenamefont {Kieferov{\'a}}, \citenamefont
  {Kivlichan}, \citenamefont {Menke}, \citenamefont {Peropadre}, \citenamefont
  {Sawaya} \emph {et~al.}}]{cao2019quantum}%
  \BibitemOpen
  \bibfield  {author} {\bibinfo {author} {\bibfnamefont {Y.}~\bibnamefont
  {Cao}}, \bibinfo {author} {\bibfnamefont {J.}~\bibnamefont {Romero}},
  \bibinfo {author} {\bibfnamefont {J.~P.}\ \bibnamefont {Olson}}, \bibinfo
  {author} {\bibfnamefont {M.}~\bibnamefont {Degroote}}, \bibinfo {author}
  {\bibfnamefont {P.~D.}\ \bibnamefont {Johnson}}, \bibinfo {author}
  {\bibfnamefont {M.}~\bibnamefont {Kieferov{\'a}}}, \bibinfo {author}
  {\bibfnamefont {I.~D.}\ \bibnamefont {Kivlichan}}, \bibinfo {author}
  {\bibfnamefont {T.}~\bibnamefont {Menke}}, \bibinfo {author} {\bibfnamefont
  {B.}~\bibnamefont {Peropadre}}, \bibinfo {author} {\bibfnamefont {N.~P.}\
  \bibnamefont {Sawaya}}, \emph {et~al.},\ }\bibfield  {title} {\bibinfo
  {title} {Quantum chemistry in the age of quantum computing},\ }\href@noop {}
  {\bibfield  {journal} {\bibinfo  {journal} {Chemical reviews}\ }\textbf
  {\bibinfo {volume} {119}},\ \bibinfo {pages} {10856} (\bibinfo {year}
  {2019})}\BibitemShut {NoStop}%
\bibitem [{\citenamefont {McArdle}\ \emph {et~al.}(2020)\citenamefont
  {McArdle}, \citenamefont {Endo}, \citenamefont {Aspuru-Guzik}, \citenamefont
  {Benjamin},\ and\ \citenamefont {Yuan}}]{RevModPhys.92.015003}%
  \BibitemOpen
  \bibfield  {author} {\bibinfo {author} {\bibfnamefont {S.}~\bibnamefont
  {McArdle}}, \bibinfo {author} {\bibfnamefont {S.}~\bibnamefont {Endo}},
  \bibinfo {author} {\bibfnamefont {A.}~\bibnamefont {Aspuru-Guzik}}, \bibinfo
  {author} {\bibfnamefont {S.~C.}\ \bibnamefont {Benjamin}},\ and\ \bibinfo
  {author} {\bibfnamefont {X.}~\bibnamefont {Yuan}},\ }\bibfield  {title}
  {\bibinfo {title} {Quantum computational chemistry},\ }\href
  {https://doi.org/10.1103/RevModPhys.92.015003} {\bibfield  {journal}
  {\bibinfo  {journal} {Rev. Mod. Phys.}\ }\textbf {\bibinfo {volume} {92}},\
  \bibinfo {pages} {015003} (\bibinfo {year} {2020})}\BibitemShut {NoStop}%
\bibitem [{\citenamefont {Zhang}\ \emph {et~al.}(2021)\citenamefont {Zhang},
  \citenamefont {Gomes}, \citenamefont {Yao}, \citenamefont {Orth},\ and\
  \citenamefont {Iadecola}}]{PhysRevB.104.075159}%
  \BibitemOpen
  \bibfield  {author} {\bibinfo {author} {\bibfnamefont {F.}~\bibnamefont
  {Zhang}}, \bibinfo {author} {\bibfnamefont {N.}~\bibnamefont {Gomes}},
  \bibinfo {author} {\bibfnamefont {Y.}~\bibnamefont {Yao}}, \bibinfo {author}
  {\bibfnamefont {P.~P.}\ \bibnamefont {Orth}},\ and\ \bibinfo {author}
  {\bibfnamefont {T.}~\bibnamefont {Iadecola}},\ }\bibfield  {title} {\bibinfo
  {title} {Adaptive variational quantum eigensolvers for highly excited
  states},\ }\href {https://doi.org/10.1103/PhysRevB.104.075159} {\bibfield
  {journal} {\bibinfo  {journal} {Phys. Rev. B}\ }\textbf {\bibinfo {volume}
  {104}},\ \bibinfo {pages} {075159} (\bibinfo {year} {2021})}\BibitemShut
  {NoStop}%
\bibitem [{\citenamefont {Macridin}\ \emph {et~al.}(2018)\citenamefont
  {Macridin}, \citenamefont {Spentzouris}, \citenamefont {Amundson},\ and\
  \citenamefont {Harnik}}]{PhysRevLett.121.110504}%
  \BibitemOpen
  \bibfield  {author} {\bibinfo {author} {\bibfnamefont {A.}~\bibnamefont
  {Macridin}}, \bibinfo {author} {\bibfnamefont {P.}~\bibnamefont
  {Spentzouris}}, \bibinfo {author} {\bibfnamefont {J.}~\bibnamefont
  {Amundson}},\ and\ \bibinfo {author} {\bibfnamefont {R.}~\bibnamefont
  {Harnik}},\ }\bibfield  {title} {\bibinfo {title} {Electron-phonon systems on
  a universal quantum computer},\ }\href
  {https://doi.org/10.1103/PhysRevLett.121.110504} {\bibfield  {journal}
  {\bibinfo  {journal} {Phys. Rev. Lett.}\ }\textbf {\bibinfo {volume} {121}},\
  \bibinfo {pages} {110504} (\bibinfo {year} {2018})}\BibitemShut {NoStop}%
\bibitem [{\citenamefont {Yao}\ \emph {et~al.}(2021)\citenamefont {Yao},
  \citenamefont {Gomes}, \citenamefont {Zhang}, \citenamefont {Wang},
  \citenamefont {Ho}, \citenamefont {Iadecola},\ and\ \citenamefont
  {Orth}}]{PRXQuantum.2.030307}%
  \BibitemOpen
  \bibfield  {author} {\bibinfo {author} {\bibfnamefont {Y.-X.}\ \bibnamefont
  {Yao}}, \bibinfo {author} {\bibfnamefont {N.}~\bibnamefont {Gomes}}, \bibinfo
  {author} {\bibfnamefont {F.}~\bibnamefont {Zhang}}, \bibinfo {author}
  {\bibfnamefont {C.-Z.}\ \bibnamefont {Wang}}, \bibinfo {author}
  {\bibfnamefont {K.-M.}\ \bibnamefont {Ho}}, \bibinfo {author} {\bibfnamefont
  {T.}~\bibnamefont {Iadecola}},\ and\ \bibinfo {author} {\bibfnamefont
  {P.~P.}\ \bibnamefont {Orth}},\ }\bibfield  {title} {\bibinfo {title}
  {Adaptive variational quantum dynamics simulations},\ }\href
  {https://doi.org/10.1103/PRXQuantum.2.030307} {\bibfield  {journal} {\bibinfo
   {journal} {PRX Quantum}\ }\textbf {\bibinfo {volume} {2}},\ \bibinfo {pages}
  {030307} (\bibinfo {year} {2021})}\BibitemShut {NoStop}%
\bibitem [{\citenamefont {Abrams}\ and\ \citenamefont
  {Lloyd}(1997)}]{PhysRevLett.79.2586}%
  \BibitemOpen
  \bibfield  {author} {\bibinfo {author} {\bibfnamefont {D.~S.}\ \bibnamefont
  {Abrams}}\ and\ \bibinfo {author} {\bibfnamefont {S.}~\bibnamefont {Lloyd}},\
  }\bibfield  {title} {\bibinfo {title} {Simulation of many-body fermi systems
  on a universal quantum computer},\ }\href
  {https://doi.org/10.1103/PhysRevLett.79.2586} {\bibfield  {journal} {\bibinfo
   {journal} {Phys. Rev. Lett.}\ }\textbf {\bibinfo {volume} {79}},\ \bibinfo
  {pages} {2586} (\bibinfo {year} {1997})}\BibitemShut {NoStop}%
\bibitem [{\citenamefont {Wecker}\ \emph {et~al.}(2015)\citenamefont {Wecker},
  \citenamefont {Hastings}, \citenamefont {Wiebe}, \citenamefont {Clark},
  \citenamefont {Nayak},\ and\ \citenamefont {Troyer}}]{PhysRevA.92.062318}%
  \BibitemOpen
  \bibfield  {author} {\bibinfo {author} {\bibfnamefont {D.}~\bibnamefont
  {Wecker}}, \bibinfo {author} {\bibfnamefont {M.~B.}\ \bibnamefont
  {Hastings}}, \bibinfo {author} {\bibfnamefont {N.}~\bibnamefont {Wiebe}},
  \bibinfo {author} {\bibfnamefont {B.~K.}\ \bibnamefont {Clark}}, \bibinfo
  {author} {\bibfnamefont {C.}~\bibnamefont {Nayak}},\ and\ \bibinfo {author}
  {\bibfnamefont {M.}~\bibnamefont {Troyer}},\ }\bibfield  {title} {\bibinfo
  {title} {Solving strongly correlated electron models on a quantum computer},\
  }\href {https://doi.org/10.1103/PhysRevA.92.062318} {\bibfield  {journal}
  {\bibinfo  {journal} {Phys. Rev. A}\ }\textbf {\bibinfo {volume} {92}},\
  \bibinfo {pages} {062318} (\bibinfo {year} {2015})}\BibitemShut {NoStop}%
\bibitem [{\citenamefont {Smith}\ \emph {et~al.}(2019)\citenamefont {Smith},
  \citenamefont {Kim}, \citenamefont {Pollmann},\ and\ \citenamefont
  {Knolle}}]{smith2019simulating}%
  \BibitemOpen
  \bibfield  {author} {\bibinfo {author} {\bibfnamefont {A.}~\bibnamefont
  {Smith}}, \bibinfo {author} {\bibfnamefont {M.}~\bibnamefont {Kim}}, \bibinfo
  {author} {\bibfnamefont {F.}~\bibnamefont {Pollmann}},\ and\ \bibinfo
  {author} {\bibfnamefont {J.}~\bibnamefont {Knolle}},\ }\bibfield  {title}
  {\bibinfo {title} {Simulating quantum many-body dynamics on a current digital
  quantum computer},\ }\href@noop {} {\bibfield  {journal} {\bibinfo  {journal}
  {npj Quantum Information}\ }\textbf {\bibinfo {volume} {5}},\ \bibinfo
  {pages} {106} (\bibinfo {year} {2019})}\BibitemShut {NoStop}%
\bibitem [{\citenamefont {Dumitrescu}\ \emph {et~al.}(2018)\citenamefont
  {Dumitrescu}, \citenamefont {McCaskey}, \citenamefont {Hagen}, \citenamefont
  {Jansen}, \citenamefont {Morris}, \citenamefont {Papenbrock}, \citenamefont
  {Pooser}, \citenamefont {Dean},\ and\ \citenamefont
  {Lougovski}}]{PhysRevLett.120.210501}%
  \BibitemOpen
  \bibfield  {author} {\bibinfo {author} {\bibfnamefont {E.~F.}\ \bibnamefont
  {Dumitrescu}}, \bibinfo {author} {\bibfnamefont {A.~J.}\ \bibnamefont
  {McCaskey}}, \bibinfo {author} {\bibfnamefont {G.}~\bibnamefont {Hagen}},
  \bibinfo {author} {\bibfnamefont {G.~R.}\ \bibnamefont {Jansen}}, \bibinfo
  {author} {\bibfnamefont {T.~D.}\ \bibnamefont {Morris}}, \bibinfo {author}
  {\bibfnamefont {T.}~\bibnamefont {Papenbrock}}, \bibinfo {author}
  {\bibfnamefont {R.~C.}\ \bibnamefont {Pooser}}, \bibinfo {author}
  {\bibfnamefont {D.~J.}\ \bibnamefont {Dean}},\ and\ \bibinfo {author}
  {\bibfnamefont {P.}~\bibnamefont {Lougovski}},\ }\bibfield  {title} {\bibinfo
  {title} {Cloud quantum computing of an atomic nucleus},\ }\href
  {https://doi.org/10.1103/PhysRevLett.120.210501} {\bibfield  {journal}
  {\bibinfo  {journal} {Phys. Rev. Lett.}\ }\textbf {\bibinfo {volume} {120}},\
  \bibinfo {pages} {210501} (\bibinfo {year} {2018})}\BibitemShut {NoStop}%
\bibitem [{\citenamefont {Romero}\ \emph {et~al.}(2022)\citenamefont {Romero},
  \citenamefont {Engel}, \citenamefont {Tang},\ and\ \citenamefont
  {Economou}}]{PhysRevC.105.064317}%
  \BibitemOpen
  \bibfield  {author} {\bibinfo {author} {\bibfnamefont {A.~M.}\ \bibnamefont
  {Romero}}, \bibinfo {author} {\bibfnamefont {J.}~\bibnamefont {Engel}},
  \bibinfo {author} {\bibfnamefont {H.~L.}\ \bibnamefont {Tang}},\ and\
  \bibinfo {author} {\bibfnamefont {S.~E.}\ \bibnamefont {Economou}},\
  }\bibfield  {title} {\bibinfo {title} {Solving nuclear structure problems
  with the adaptive variational quantum algorithm},\ }\href
  {https://doi.org/10.1103/PhysRevC.105.064317} {\bibfield  {journal} {\bibinfo
   {journal} {Phys. Rev. C}\ }\textbf {\bibinfo {volume} {105}},\ \bibinfo
  {pages} {064317} (\bibinfo {year} {2022})}\BibitemShut {NoStop}%
\bibitem [{\citenamefont {Kiss}\ \emph {et~al.}(2022)\citenamefont {Kiss},
  \citenamefont {Grossi}, \citenamefont {Lougovski}, \citenamefont {Sanchez},
  \citenamefont {Vallecorsa},\ and\ \citenamefont
  {Papenbrock}}]{PhysRevC.106.034325}%
  \BibitemOpen
  \bibfield  {author} {\bibinfo {author} {\bibfnamefont {O.}~\bibnamefont
  {Kiss}}, \bibinfo {author} {\bibfnamefont {M.}~\bibnamefont {Grossi}},
  \bibinfo {author} {\bibfnamefont {P.}~\bibnamefont {Lougovski}}, \bibinfo
  {author} {\bibfnamefont {F.}~\bibnamefont {Sanchez}}, \bibinfo {author}
  {\bibfnamefont {S.}~\bibnamefont {Vallecorsa}},\ and\ \bibinfo {author}
  {\bibfnamefont {T.}~\bibnamefont {Papenbrock}},\ }\bibfield  {title}
  {\bibinfo {title} {Quantum computing of the $^{6}\mathrm{Li}$ nucleus via
  ordered unitary coupled clusters},\ }\href
  {https://doi.org/10.1103/PhysRevC.106.034325} {\bibfield  {journal} {\bibinfo
   {journal} {Phys. Rev. C}\ }\textbf {\bibinfo {volume} {106}},\ \bibinfo
  {pages} {034325} (\bibinfo {year} {2022})}\BibitemShut {NoStop}%
\bibitem [{\citenamefont {Du}\ \emph {et~al.}(2021{\natexlab{a}})\citenamefont
  {Du}, \citenamefont {Vary}, \citenamefont {Zhao},\ and\ \citenamefont
  {Zuo}}]{Du:2020glq}%
  \BibitemOpen
  \bibfield  {author} {\bibinfo {author} {\bibfnamefont {W.}~\bibnamefont
  {Du}}, \bibinfo {author} {\bibfnamefont {J.~P.}\ \bibnamefont {Vary}},
  \bibinfo {author} {\bibfnamefont {X.}~\bibnamefont {Zhao}},\ and\ \bibinfo
  {author} {\bibfnamefont {W.}~\bibnamefont {Zuo}},\ }\bibfield  {title}
  {\bibinfo {title} {{Quantum simulation of nuclear inelastic scattering}},\
  }\href {https://doi.org/10.1103/PhysRevA.104.012611} {\bibfield  {journal}
  {\bibinfo  {journal} {Phys. Rev. A}\ }\textbf {\bibinfo {volume} {104}},\
  \bibinfo {pages} {012611} (\bibinfo {year} {2021}{\natexlab{a}})},\ \Eprint
  {https://arxiv.org/abs/2006.01369} {arXiv:2006.01369 [nucl-th]} \BibitemShut
  {NoStop}%
\bibitem [{\citenamefont {Turro}\ \emph
  {et~al.}(2023{\natexlab{a}})\citenamefont {Turro} \emph
  {et~al.}}]{Turro:2023xgf}%
  \BibitemOpen
  \bibfield  {author} {\bibinfo {author} {\bibfnamefont {F.}~\bibnamefont
  {Turro}} \emph {et~al.},\ }\bibfield  {title} {\bibinfo {title} {{A
  quantum-classical co-processing protocol towards simulating nuclear reactions
  on contemporary quantum hardware}},\ }\href@noop {} {\  (\bibinfo {year}
  {2023}{\natexlab{a}})},\ \Eprint {https://arxiv.org/abs/2302.06734}
  {arXiv:2302.06734 [quant-ph]} \BibitemShut {NoStop}%
\bibitem [{\citenamefont {Turro}\ \emph
  {et~al.}(2023{\natexlab{b}})\citenamefont {Turro} \emph
  {et~al.}}]{Turro:2023dhg}%
  \BibitemOpen
  \bibfield  {author} {\bibinfo {author} {\bibfnamefont {F.}~\bibnamefont
  {Turro}} \emph {et~al.},\ }\bibfield  {title} {\bibinfo {title}
  {{Demonstration of a quantum-classical coprocessing protocol for simulating
  nuclear reactions}},\ }\href {https://doi.org/10.1103/PhysRevA.108.032417}
  {\bibfield  {journal} {\bibinfo  {journal} {Phys. Rev. A}\ }\textbf {\bibinfo
  {volume} {108}},\ \bibinfo {pages} {032417} (\bibinfo {year}
  {2023}{\natexlab{b}})}\BibitemShut {NoStop}%
\bibitem [{\citenamefont {Du}\ and\ \citenamefont {Vary}(2023)}]{Du:2023bpw}%
  \BibitemOpen
  \bibfield  {author} {\bibinfo {author} {\bibfnamefont {W.}~\bibnamefont
  {Du}}\ and\ \bibinfo {author} {\bibfnamefont {J.~P.}\ \bibnamefont {Vary}},\
  }\bibfield  {title} {\bibinfo {title} {{Multinucleon structure and dynamics
  via quantum computing}},\ }\href
  {https://doi.org/10.1103/PhysRevA.108.052614} {\bibfield  {journal} {\bibinfo
   {journal} {Phys. Rev. A}\ }\textbf {\bibinfo {volume} {108}},\ \bibinfo
  {pages} {052614} (\bibinfo {year} {2023})},\ \Eprint
  {https://arxiv.org/abs/2304.04838} {arXiv:2304.04838 [nucl-th]} \BibitemShut
  {NoStop}%
\bibitem [{\citenamefont {Du}\ \emph {et~al.}(2021{\natexlab{b}})\citenamefont
  {Du}, \citenamefont {Vary}, \citenamefont {Zhao},\ and\ \citenamefont
  {Zuo}}]{Du:2021ctr}%
  \BibitemOpen
  \bibfield  {author} {\bibinfo {author} {\bibfnamefont {W.}~\bibnamefont
  {Du}}, \bibinfo {author} {\bibfnamefont {J.~P.}\ \bibnamefont {Vary}},
  \bibinfo {author} {\bibfnamefont {X.}~\bibnamefont {Zhao}},\ and\ \bibinfo
  {author} {\bibfnamefont {W.}~\bibnamefont {Zuo}},\ }\bibfield  {title}
  {\bibinfo {title} {{Ab initio nuclear structure via quantum adiabatic
  algorithm}},\ }\href@noop {} {\  (\bibinfo {year} {2021}{\natexlab{b}})},\
  \Eprint {https://arxiv.org/abs/2105.08910} {arXiv:2105.08910 [nucl-th]}
  \BibitemShut {NoStop}%
\bibitem [{\citenamefont {P{\'{e}}rez-Obiol}\ \emph {et~al.}(2023)\citenamefont
  {P{\'{e}}rez-Obiol}, \citenamefont {Romero}, \citenamefont {Men{\'{e}}ndez},
  \citenamefont {Rios}, \citenamefont {Garc{\'{\i}}a-S{\'{a}}ez},\ and\
  \citenamefont {Juli{\'{a}}-D{\'{\i}}az}}]{P_rez_Obiol_2023}%
  \BibitemOpen
  \bibfield  {author} {\bibinfo {author} {\bibfnamefont {A.}~\bibnamefont
  {P{\'{e}}rez-Obiol}}, \bibinfo {author} {\bibfnamefont {A.~M.}\ \bibnamefont
  {Romero}}, \bibinfo {author} {\bibfnamefont {J.}~\bibnamefont
  {Men{\'{e}}ndez}}, \bibinfo {author} {\bibfnamefont {A.}~\bibnamefont
  {Rios}}, \bibinfo {author} {\bibfnamefont {A.}~\bibnamefont
  {Garc{\'{\i}}a-S{\'{a}}ez}},\ and\ \bibinfo {author} {\bibfnamefont
  {B.}~\bibnamefont {Juli{\'{a}}-D{\'{\i}}az}},\ }\bibfield  {title} {\bibinfo
  {title} {Nuclear shell-model simulation in digital quantum computers},\
  }\bibfield  {journal} {\bibinfo  {journal} {Scientific Reports}\ }\textbf
  {\bibinfo {volume} {13}},\ \href {https://doi.org/10.1038/s41598-023-39263-7}
  {10.1038/s41598-023-39263-7} (\bibinfo {year} {2023})\BibitemShut {NoStop}%
\bibitem [{\citenamefont {Lv}\ \emph {et~al.}(2022)\citenamefont {Lv},
  \citenamefont {Wei}, \citenamefont {Xie},\ and\ \citenamefont
  {Long}}]{lv2022qcsh}%
  \BibitemOpen
  \bibfield  {author} {\bibinfo {author} {\bibfnamefont {P.}~\bibnamefont
  {Lv}}, \bibinfo {author} {\bibfnamefont {S.-J.}\ \bibnamefont {Wei}},
  \bibinfo {author} {\bibfnamefont {H.-N.}\ \bibnamefont {Xie}},\ and\ \bibinfo
  {author} {\bibfnamefont {G.-L.}\ \bibnamefont {Long}},\ }\href@noop {}
  {\bibinfo {title} {Qcsh: a full quantum computer nuclear shell-model
  package}} (\bibinfo {year} {2022}),\ \Eprint
  {https://arxiv.org/abs/2205.12087} {arXiv:2205.12087 [quant-ph]} \BibitemShut
  {NoStop}%
\bibitem [{\citenamefont {Stetcu}\ \emph {et~al.}(2022)\citenamefont {Stetcu},
  \citenamefont {Baroni},\ and\ \citenamefont {Carlson}}]{PhysRevC.105.064308}%
  \BibitemOpen
  \bibfield  {author} {\bibinfo {author} {\bibfnamefont {I.}~\bibnamefont
  {Stetcu}}, \bibinfo {author} {\bibfnamefont {A.}~\bibnamefont {Baroni}},\
  and\ \bibinfo {author} {\bibfnamefont {J.}~\bibnamefont {Carlson}},\
  }\bibfield  {title} {\bibinfo {title} {Variational approaches to constructing
  the many-body nuclear ground state for quantum computing},\ }\href
  {https://doi.org/10.1103/PhysRevC.105.064308} {\bibfield  {journal} {\bibinfo
   {journal} {Phys. Rev. C}\ }\textbf {\bibinfo {volume} {105}},\ \bibinfo
  {pages} {064308} (\bibinfo {year} {2022})}\BibitemShut {NoStop}%
\bibitem [{\citenamefont {Siwach}\ and\ \citenamefont
  {Arumugam}(2022)}]{PhysRevC.105.064318}%
  \BibitemOpen
  \bibfield  {author} {\bibinfo {author} {\bibfnamefont {P.}~\bibnamefont
  {Siwach}}\ and\ \bibinfo {author} {\bibfnamefont {P.}~\bibnamefont
  {Arumugam}},\ }\bibfield  {title} {\bibinfo {title} {Quantum computation of
  nuclear observables involving linear combinations of unitary operators},\
  }\href {https://doi.org/10.1103/PhysRevC.105.064318} {\bibfield  {journal}
  {\bibinfo  {journal} {Phys. Rev. C}\ }\textbf {\bibinfo {volume} {105}},\
  \bibinfo {pages} {064318} (\bibinfo {year} {2022})}\BibitemShut {NoStop}%
\bibitem [{\citenamefont {Yang}\ \emph {et~al.}(2023)\citenamefont {Yang},
  \citenamefont {Wang}, \citenamefont {Lu}, \citenamefont {Li},\ and\
  \citenamefont {Xu}}]{yang2023shadowbased}%
  \BibitemOpen
  \bibfield  {author} {\bibinfo {author} {\bibfnamefont {R.}~\bibnamefont
  {Yang}}, \bibinfo {author} {\bibfnamefont {T.}~\bibnamefont {Wang}}, \bibinfo
  {author} {\bibfnamefont {B.-N.}\ \bibnamefont {Lu}}, \bibinfo {author}
  {\bibfnamefont {Y.}~\bibnamefont {Li}},\ and\ \bibinfo {author}
  {\bibfnamefont {X.}~\bibnamefont {Xu}},\ }\href@noop {} {\bibinfo {title}
  {Shadow-based quantum subspace algorithm for the nuclear shell model}}
  (\bibinfo {year} {2023}),\ \Eprint {https://arxiv.org/abs/2306.08885}
  {arXiv:2306.08885 [quant-ph]} \BibitemShut {NoStop}%
\bibitem [{\citenamefont {Siwach}\ and\ \citenamefont
  {Arumugam}(2019)}]{PhysRevResearch.1.033176}%
  \BibitemOpen
  \bibfield  {author} {\bibinfo {author} {\bibfnamefont {P.}~\bibnamefont
  {Siwach}}\ and\ \bibinfo {author} {\bibfnamefont {P.}~\bibnamefont
  {Arumugam}},\ }\bibfield  {title} {\bibinfo {title} {Neutrino oscillations in
  a quantum processor},\ }\href
  {https://doi.org/10.1103/PhysRevResearch.1.033176} {\bibfield  {journal}
  {\bibinfo  {journal} {Phys. Rev. Research}\ }\textbf {\bibinfo {volume}
  {1}},\ \bibinfo {pages} {033176} (\bibinfo {year} {2019})}\BibitemShut
  {NoStop}%
\bibitem [{\citenamefont {Wang}\ \emph {et~al.}(2024)\citenamefont {Wang},
  \citenamefont {Du}, \citenamefont {Zuo},\ and\ \citenamefont
  {Vary}}]{Wang:2024scd}%
  \BibitemOpen
  \bibfield  {author} {\bibinfo {author} {\bibfnamefont {P.}~\bibnamefont
  {Wang}}, \bibinfo {author} {\bibfnamefont {W.}~\bibnamefont {Du}}, \bibinfo
  {author} {\bibfnamefont {W.}~\bibnamefont {Zuo}},\ and\ \bibinfo {author}
  {\bibfnamefont {J.~P.}\ \bibnamefont {Vary}},\ }\bibfield  {title} {\bibinfo
  {title} {{Nuclear scattering via quantum computing}},\ }\href@noop {} {\
  (\bibinfo {year} {2024})},\ \Eprint {https://arxiv.org/abs/2401.17138}
  {arXiv:2401.17138 [nucl-th]} \BibitemShut {NoStop}%
\bibitem [{\citenamefont {Du}\ and\ \citenamefont
  {Vary}(2024{\natexlab{a}})}]{Du:2024zvr}%
  \BibitemOpen
  \bibfield  {author} {\bibinfo {author} {\bibfnamefont {W.}~\bibnamefont
  {Du}}\ and\ \bibinfo {author} {\bibfnamefont {J.~P.}\ \bibnamefont {Vary}},\
  }\bibfield  {title} {\bibinfo {title} {{Hamiltonian input model and
  spectroscopy on quantum computers}},\ }\href@noop {} {\  (\bibinfo {year}
  {2024}{\natexlab{a}})},\ \Eprint {https://arxiv.org/abs/2402.08969}
  {arXiv:2402.08969 [quant-ph]} \BibitemShut {NoStop}%
\bibitem [{\citenamefont {Bhoy}\ and\ \citenamefont
  {Stevenson}(2024)}]{bhoy2024shellmodel}%
  \BibitemOpen
  \bibfield  {author} {\bibinfo {author} {\bibfnamefont {B.}~\bibnamefont
  {Bhoy}}\ and\ \bibinfo {author} {\bibfnamefont {P.}~\bibnamefont
  {Stevenson}},\ }\href@noop {} {\bibinfo {title} {Shell-model study of
  $^{58}$ni using quantum computing algorithm}} (\bibinfo {year} {2024}),\
  \Eprint {https://arxiv.org/abs/2402.15577} {arXiv:2402.15577 [nucl-th]}
  \BibitemShut {NoStop}%
\bibitem [{\citenamefont {Liu}\ \emph {et~al.}(2024)\citenamefont {Liu},
  \citenamefont {Du}, \citenamefont {Lin}, \citenamefont {Vary},\ and\
  \citenamefont {Yang}}]{Liu:2024hmm}%
  \BibitemOpen
  \bibfield  {author} {\bibinfo {author} {\bibfnamefont {D.}~\bibnamefont
  {Liu}}, \bibinfo {author} {\bibfnamefont {W.}~\bibnamefont {Du}}, \bibinfo
  {author} {\bibfnamefont {L.}~\bibnamefont {Lin}}, \bibinfo {author}
  {\bibfnamefont {J.~P.}\ \bibnamefont {Vary}},\ and\ \bibinfo {author}
  {\bibfnamefont {C.}~\bibnamefont {Yang}},\ }\bibfield  {title} {\bibinfo
  {title} {{An Efficient Quantum Circuit for Block Encoding a Pairing
  Hamiltonian}},\ }\href@noop {} {\  (\bibinfo {year} {2024})},\ \Eprint
  {https://arxiv.org/abs/2402.11205} {arXiv:2402.11205 [nucl-th]} \BibitemShut
  {NoStop}%
\bibitem [{\citenamefont {Jordan}\ \emph {et~al.}(2012)\citenamefont {Jordan},
  \citenamefont {Lee},\ and\ \citenamefont {Preskill}}]{Jordan_2012}%
  \BibitemOpen
  \bibfield  {author} {\bibinfo {author} {\bibfnamefont {S.~P.}\ \bibnamefont
  {Jordan}}, \bibinfo {author} {\bibfnamefont {K.~S.~M.}\ \bibnamefont {Lee}},\
  and\ \bibinfo {author} {\bibfnamefont {J.}~\bibnamefont {Preskill}},\
  }\bibfield  {title} {\bibinfo {title} {Quantum algorithms for quantum field
  theories},\ }\href {https://doi.org/10.1126/science.1217069} {\bibfield
  {journal} {\bibinfo  {journal} {Science}\ }\textbf {\bibinfo {volume}
  {336}},\ \bibinfo {pages} {1130} (\bibinfo {year} {2012})}\BibitemShut
  {NoStop}%
\bibitem [{\citenamefont {Klco}\ \emph {et~al.}(2018)\citenamefont {Klco},
  \citenamefont {Dumitrescu}, \citenamefont {McCaskey}, \citenamefont {Morris},
  \citenamefont {Pooser}, \citenamefont {Sanz}, \citenamefont {Solano},
  \citenamefont {Lougovski},\ and\ \citenamefont
  {Savage}}]{PhysRevA.98.032331}%
  \BibitemOpen
  \bibfield  {author} {\bibinfo {author} {\bibfnamefont {N.}~\bibnamefont
  {Klco}}, \bibinfo {author} {\bibfnamefont {E.~F.}\ \bibnamefont
  {Dumitrescu}}, \bibinfo {author} {\bibfnamefont {A.~J.}\ \bibnamefont
  {McCaskey}}, \bibinfo {author} {\bibfnamefont {T.~D.}\ \bibnamefont
  {Morris}}, \bibinfo {author} {\bibfnamefont {R.~C.}\ \bibnamefont {Pooser}},
  \bibinfo {author} {\bibfnamefont {M.}~\bibnamefont {Sanz}}, \bibinfo {author}
  {\bibfnamefont {E.}~\bibnamefont {Solano}}, \bibinfo {author} {\bibfnamefont
  {P.}~\bibnamefont {Lougovski}},\ and\ \bibinfo {author} {\bibfnamefont
  {M.~J.}\ \bibnamefont {Savage}},\ }\bibfield  {title} {\bibinfo {title}
  {Quantum-classical computation of schwinger model dynamics using quantum
  computers},\ }\href {https://doi.org/10.1103/PhysRevA.98.032331} {\bibfield
  {journal} {\bibinfo  {journal} {Phys. Rev. A}\ }\textbf {\bibinfo {volume}
  {98}},\ \bibinfo {pages} {032331} (\bibinfo {year} {2018})}\BibitemShut
  {NoStop}%
\bibitem [{\citenamefont {Jordan}\ \emph {et~al.}(2019)\citenamefont {Jordan},
  \citenamefont {Lee},\ and\ \citenamefont {Preskill}}]{jordan2019quantum}%
  \BibitemOpen
  \bibfield  {author} {\bibinfo {author} {\bibfnamefont {S.~P.}\ \bibnamefont
  {Jordan}}, \bibinfo {author} {\bibfnamefont {K.~S.~M.}\ \bibnamefont {Lee}},\
  and\ \bibinfo {author} {\bibfnamefont {J.}~\bibnamefont {Preskill}},\
  }\href@noop {} {\bibinfo {title} {Quantum computation of scattering in scalar
  quantum field theories}} (\bibinfo {year} {2019}),\ \Eprint
  {https://arxiv.org/abs/1112.4833} {arXiv:1112.4833 [hep-th]} \BibitemShut
  {NoStop}%
\bibitem [{\citenamefont {Klco}\ \emph {et~al.}(2020)\citenamefont {Klco},
  \citenamefont {Savage},\ and\ \citenamefont {Stryker}}]{PhysRevD.101.074512}%
  \BibitemOpen
  \bibfield  {author} {\bibinfo {author} {\bibfnamefont {N.}~\bibnamefont
  {Klco}}, \bibinfo {author} {\bibfnamefont {M.~J.}\ \bibnamefont {Savage}},\
  and\ \bibinfo {author} {\bibfnamefont {J.~R.}\ \bibnamefont {Stryker}},\
  }\bibfield  {title} {\bibinfo {title} {Su(2) non-abelian gauge field theory
  in one dimension on digital quantum computers},\ }\href
  {https://doi.org/10.1103/PhysRevD.101.074512} {\bibfield  {journal} {\bibinfo
   {journal} {Phys. Rev. D}\ }\textbf {\bibinfo {volume} {101}},\ \bibinfo
  {pages} {074512} (\bibinfo {year} {2020})}\BibitemShut {NoStop}%
\bibitem [{\citenamefont {de~Jong}\ \emph {et~al.}(2022)\citenamefont
  {de~Jong}, \citenamefont {Lee}, \citenamefont {Mulligan}, \citenamefont
  {P\l{}osko\ifmmode~\acute{n}\else \'{n}\fi{}}, \citenamefont {Ringer},\ and\
  \citenamefont {Yao}}]{PhysRevD.106.054508}%
  \BibitemOpen
  \bibfield  {author} {\bibinfo {author} {\bibfnamefont {W.~A.}\ \bibnamefont
  {de~Jong}}, \bibinfo {author} {\bibfnamefont {K.}~\bibnamefont {Lee}},
  \bibinfo {author} {\bibfnamefont {J.}~\bibnamefont {Mulligan}}, \bibinfo
  {author} {\bibfnamefont {M.}~\bibnamefont {P\l{}osko\ifmmode~\acute{n}\else
  \'{n}\fi{}}}, \bibinfo {author} {\bibfnamefont {F.}~\bibnamefont {Ringer}},\
  and\ \bibinfo {author} {\bibfnamefont {X.}~\bibnamefont {Yao}},\ }\bibfield
  {title} {\bibinfo {title} {Quantum simulation of nonequilibrium dynamics and
  thermalization in the schwinger model},\ }\href
  {https://doi.org/10.1103/PhysRevD.106.054508} {\bibfield  {journal} {\bibinfo
   {journal} {Phys. Rev. D}\ }\textbf {\bibinfo {volume} {106}},\ \bibinfo
  {pages} {054508} (\bibinfo {year} {2022})}\BibitemShut {NoStop}%
\bibitem [{\citenamefont {Bauer}\ \emph {et~al.}(2022)\citenamefont {Bauer},
  \citenamefont {Davoudi}, \citenamefont {Balantekin}, \citenamefont
  {Bhattacharya}, \citenamefont {Carena}, \citenamefont {de~Jong},
  \citenamefont {Draper}, \citenamefont {El-Khadra}, \citenamefont {Gemelke},
  \citenamefont {Hanada}, \citenamefont {Kharzeev}, \citenamefont {Lamm},
  \citenamefont {Li}, \citenamefont {Liu}, \citenamefont {Lukin}, \citenamefont
  {Meurice}, \citenamefont {Monroe}, \citenamefont {Nachman}, \citenamefont
  {Pagano}, \citenamefont {Preskill}, \citenamefont {Rinaldi}, \citenamefont
  {Roggero}, \citenamefont {Santiago}, \citenamefont {Savage}, \citenamefont
  {Siddiqi}, \citenamefont {Siopsis}, \citenamefont {Zanten}, \citenamefont
  {Wiebe}, \citenamefont {Yamauchi}, \citenamefont {Yeter-Aydeniz},\ and\
  \citenamefont {Zorzetti}}]{bauer2022quantum}%
  \BibitemOpen
  \bibfield  {author} {\bibinfo {author} {\bibfnamefont {C.~W.}\ \bibnamefont
  {Bauer}}, \bibinfo {author} {\bibfnamefont {Z.}~\bibnamefont {Davoudi}},
  \bibinfo {author} {\bibfnamefont {A.~B.}\ \bibnamefont {Balantekin}},
  \bibinfo {author} {\bibfnamefont {T.}~\bibnamefont {Bhattacharya}}, \bibinfo
  {author} {\bibfnamefont {M.}~\bibnamefont {Carena}}, \bibinfo {author}
  {\bibfnamefont {W.~A.}\ \bibnamefont {de~Jong}}, \bibinfo {author}
  {\bibfnamefont {P.}~\bibnamefont {Draper}}, \bibinfo {author} {\bibfnamefont
  {A.}~\bibnamefont {El-Khadra}}, \bibinfo {author} {\bibfnamefont
  {N.}~\bibnamefont {Gemelke}}, \bibinfo {author} {\bibfnamefont
  {M.}~\bibnamefont {Hanada}}, \bibinfo {author} {\bibfnamefont
  {D.}~\bibnamefont {Kharzeev}}, \bibinfo {author} {\bibfnamefont
  {H.}~\bibnamefont {Lamm}}, \bibinfo {author} {\bibfnamefont {Y.-Y.}\
  \bibnamefont {Li}}, \bibinfo {author} {\bibfnamefont {J.}~\bibnamefont
  {Liu}}, \bibinfo {author} {\bibfnamefont {M.}~\bibnamefont {Lukin}}, \bibinfo
  {author} {\bibfnamefont {Y.}~\bibnamefont {Meurice}}, \bibinfo {author}
  {\bibfnamefont {C.}~\bibnamefont {Monroe}}, \bibinfo {author} {\bibfnamefont
  {B.}~\bibnamefont {Nachman}}, \bibinfo {author} {\bibfnamefont
  {G.}~\bibnamefont {Pagano}}, \bibinfo {author} {\bibfnamefont
  {J.}~\bibnamefont {Preskill}}, \bibinfo {author} {\bibfnamefont
  {E.}~\bibnamefont {Rinaldi}}, \bibinfo {author} {\bibfnamefont
  {A.}~\bibnamefont {Roggero}}, \bibinfo {author} {\bibfnamefont {D.~I.}\
  \bibnamefont {Santiago}}, \bibinfo {author} {\bibfnamefont {M.~J.}\
  \bibnamefont {Savage}}, \bibinfo {author} {\bibfnamefont {I.}~\bibnamefont
  {Siddiqi}}, \bibinfo {author} {\bibfnamefont {G.}~\bibnamefont {Siopsis}},
  \bibinfo {author} {\bibfnamefont {D.~V.}\ \bibnamefont {Zanten}}, \bibinfo
  {author} {\bibfnamefont {N.}~\bibnamefont {Wiebe}}, \bibinfo {author}
  {\bibfnamefont {Y.}~\bibnamefont {Yamauchi}}, \bibinfo {author}
  {\bibfnamefont {K.}~\bibnamefont {Yeter-Aydeniz}},\ and\ \bibinfo {author}
  {\bibfnamefont {S.}~\bibnamefont {Zorzetti}},\ }\href@noop {} {\bibinfo
  {title} {Quantum simulation for high energy physics}} (\bibinfo {year}
  {2022}),\ \Eprint {https://arxiv.org/abs/2204.03381} {arXiv:2204.03381
  [quant-ph]} \BibitemShut {NoStop}%
\bibitem [{\citenamefont {Davoudi}\ \emph {et~al.}(2023)\citenamefont
  {Davoudi}, \citenamefont {Mueller},\ and\ \citenamefont
  {Powers}}]{PhysRevLett.131.081901}%
  \BibitemOpen
  \bibfield  {author} {\bibinfo {author} {\bibfnamefont {Z.}~\bibnamefont
  {Davoudi}}, \bibinfo {author} {\bibfnamefont {N.}~\bibnamefont {Mueller}},\
  and\ \bibinfo {author} {\bibfnamefont {C.}~\bibnamefont {Powers}},\
  }\bibfield  {title} {\bibinfo {title} {Towards quantum computing phase
  diagrams of gauge theories with thermal pure quantum states},\ }\href
  {https://doi.org/10.1103/PhysRevLett.131.081901} {\bibfield  {journal}
  {\bibinfo  {journal} {Phys. Rev. Lett.}\ }\textbf {\bibinfo {volume} {131}},\
  \bibinfo {pages} {081901} (\bibinfo {year} {2023})}\BibitemShut {NoStop}%
\bibitem [{\citenamefont {Mueller}\ \emph {et~al.}(2023)\citenamefont
  {Mueller}, \citenamefont {Carolan}, \citenamefont {Connelly}, \citenamefont
  {Davoudi}, \citenamefont {Dumitrescu},\ and\ \citenamefont
  {Yeter-Aydeniz}}]{PRXQuantum.4.030323}%
  \BibitemOpen
  \bibfield  {author} {\bibinfo {author} {\bibfnamefont {N.}~\bibnamefont
  {Mueller}}, \bibinfo {author} {\bibfnamefont {J.~A.}\ \bibnamefont
  {Carolan}}, \bibinfo {author} {\bibfnamefont {A.}~\bibnamefont {Connelly}},
  \bibinfo {author} {\bibfnamefont {Z.}~\bibnamefont {Davoudi}}, \bibinfo
  {author} {\bibfnamefont {E.~F.}\ \bibnamefont {Dumitrescu}},\ and\ \bibinfo
  {author} {\bibfnamefont {K.}~\bibnamefont {Yeter-Aydeniz}},\ }\bibfield
  {title} {\bibinfo {title} {Quantum computation of dynamical quantum phase
  transitions and entanglement tomography in a lattice gauge theory},\ }\href
  {https://doi.org/10.1103/PRXQuantum.4.030323} {\bibfield  {journal} {\bibinfo
   {journal} {PRX Quantum}\ }\textbf {\bibinfo {volume} {4}},\ \bibinfo {pages}
  {030323} (\bibinfo {year} {2023})}\BibitemShut {NoStop}%
\bibitem [{\citenamefont {Mueller}\ \emph {et~al.}(2020)\citenamefont
  {Mueller}, \citenamefont {Tarasov},\ and\ \citenamefont
  {Venugopalan}}]{PhysRevD.102.016007}%
  \BibitemOpen
  \bibfield  {author} {\bibinfo {author} {\bibfnamefont {N.}~\bibnamefont
  {Mueller}}, \bibinfo {author} {\bibfnamefont {A.}~\bibnamefont {Tarasov}},\
  and\ \bibinfo {author} {\bibfnamefont {R.}~\bibnamefont {Venugopalan}},\
  }\bibfield  {title} {\bibinfo {title} {Deeply inelastic scattering structure
  functions on a hybrid quantum computer},\ }\href
  {https://doi.org/10.1103/PhysRevD.102.016007} {\bibfield  {journal} {\bibinfo
   {journal} {Phys. Rev. D}\ }\textbf {\bibinfo {volume} {102}},\ \bibinfo
  {pages} {016007} (\bibinfo {year} {2020})}\BibitemShut {NoStop}%
\bibitem [{\citenamefont {Lamm}\ \emph {et~al.}(2020)\citenamefont {Lamm},
  \citenamefont {Lawrence},\ and\ \citenamefont
  {Yamauchi}}]{PhysRevResearch.2.013272}%
  \BibitemOpen
  \bibfield  {author} {\bibinfo {author} {\bibfnamefont {H.}~\bibnamefont
  {Lamm}}, \bibinfo {author} {\bibfnamefont {S.}~\bibnamefont {Lawrence}},\
  and\ \bibinfo {author} {\bibfnamefont {Y.}~\bibnamefont {Yamauchi}} (\bibinfo
  {collaboration} {NuQS Collaboration}),\ }\bibfield  {title} {\bibinfo {title}
  {Parton physics on a quantum computer},\ }\href
  {https://doi.org/10.1103/PhysRevResearch.2.013272} {\bibfield  {journal}
  {\bibinfo  {journal} {Phys. Rev. Research}\ }\textbf {\bibinfo {volume}
  {2}},\ \bibinfo {pages} {013272} (\bibinfo {year} {2020})}\BibitemShut
  {NoStop}%
\bibitem [{\citenamefont {Kreshchuk}\ \emph {et~al.}(2023)\citenamefont
  {Kreshchuk}, \citenamefont {Vary},\ and\ \citenamefont
  {Love}}]{Kreshchuk:2023btr}%
  \BibitemOpen
  \bibfield  {author} {\bibinfo {author} {\bibfnamefont {M.}~\bibnamefont
  {Kreshchuk}}, \bibinfo {author} {\bibfnamefont {J.~P.}\ \bibnamefont
  {Vary}},\ and\ \bibinfo {author} {\bibfnamefont {P.~J.}\ \bibnamefont
  {Love}},\ }\bibfield  {title} {\bibinfo {title} {{Simulating Scattering of
  Composite Particles}},\ }\href@noop {} {\  (\bibinfo {year} {2023})},\
  \Eprint {https://arxiv.org/abs/2310.13742} {arXiv:2310.13742 [quant-ph]}
  \BibitemShut {NoStop}%
\bibitem [{\citenamefont {M\"uller}\ and\ \citenamefont
  {Yao}(2023)}]{Muller:2023nnk}%
  \BibitemOpen
  \bibfield  {author} {\bibinfo {author} {\bibfnamefont {B.}~\bibnamefont
  {M\"uller}}\ and\ \bibinfo {author} {\bibfnamefont {X.}~\bibnamefont {Yao}},\
  }\bibfield  {title} {\bibinfo {title} {{Simple Hamiltonian for quantum
  simulation of strongly coupled (2+1)D SU(2) lattice gauge theory on a
  honeycomb lattice}},\ }\href {https://doi.org/10.1103/PhysRevD.108.094505}
  {\bibfield  {journal} {\bibinfo  {journal} {Phys. Rev. D}\ }\textbf {\bibinfo
  {volume} {108}},\ \bibinfo {pages} {094505} (\bibinfo {year} {2023})},\
  \Eprint {https://arxiv.org/abs/2307.00045} {arXiv:2307.00045 [quant-ph]}
  \BibitemShut {NoStop}%
\bibitem [{\citenamefont {Barata}\ and\ \citenamefont
  {Salgado}(2021)}]{Barata:2021yri}%
  \BibitemOpen
  \bibfield  {author} {\bibinfo {author} {\bibfnamefont {J.~a.}\ \bibnamefont
  {Barata}}\ and\ \bibinfo {author} {\bibfnamefont {C.~A.}\ \bibnamefont
  {Salgado}},\ }\bibfield  {title} {\bibinfo {title} {{A quantum strategy to
  compute the jet quenching parameter $\hat{q}$}},\ }\href
  {https://doi.org/10.1140/epjc/s10052-021-09674-9} {\bibfield  {journal}
  {\bibinfo  {journal} {Eur. Phys. J. C}\ }\textbf {\bibinfo {volume} {81}},\
  \bibinfo {pages} {862} (\bibinfo {year} {2021})},\ \Eprint
  {https://arxiv.org/abs/2104.04661} {arXiv:2104.04661 [hep-ph]} \BibitemShut
  {NoStop}%
\bibitem [{\citenamefont {Barata}\ \emph {et~al.}(2022)\citenamefont {Barata},
  \citenamefont {Du}, \citenamefont {Li}, \citenamefont {Qian},\ and\
  \citenamefont {Salgado}}]{Barata:2022wim}%
  \BibitemOpen
  \bibfield  {author} {\bibinfo {author} {\bibfnamefont {J.~a.}\ \bibnamefont
  {Barata}}, \bibinfo {author} {\bibfnamefont {X.}~\bibnamefont {Du}}, \bibinfo
  {author} {\bibfnamefont {M.}~\bibnamefont {Li}}, \bibinfo {author}
  {\bibfnamefont {W.}~\bibnamefont {Qian}},\ and\ \bibinfo {author}
  {\bibfnamefont {C.~A.}\ \bibnamefont {Salgado}},\ }\bibfield  {title}
  {\bibinfo {title} {{Medium induced jet broadening in a quantum computer}},\
  }\href {https://doi.org/10.1103/PhysRevD.106.074013} {\bibfield  {journal}
  {\bibinfo  {journal} {Phys. Rev. D}\ }\textbf {\bibinfo {volume} {106}},\
  \bibinfo {pages} {074013} (\bibinfo {year} {2022})},\ \Eprint
  {https://arxiv.org/abs/2208.06750} {arXiv:2208.06750 [hep-ph]} \BibitemShut
  {NoStop}%
\bibitem [{\citenamefont {Barata}\ \emph {et~al.}(2023)\citenamefont {Barata},
  \citenamefont {Du}, \citenamefont {Li}, \citenamefont {Qian},\ and\
  \citenamefont {Salgado}}]{Barata:2023clv}%
  \BibitemOpen
  \bibfield  {author} {\bibinfo {author} {\bibfnamefont {J.~a.}\ \bibnamefont
  {Barata}}, \bibinfo {author} {\bibfnamefont {X.}~\bibnamefont {Du}}, \bibinfo
  {author} {\bibfnamefont {M.}~\bibnamefont {Li}}, \bibinfo {author}
  {\bibfnamefont {W.}~\bibnamefont {Qian}},\ and\ \bibinfo {author}
  {\bibfnamefont {C.~A.}\ \bibnamefont {Salgado}},\ }\bibfield  {title}
  {\bibinfo {title} {{Quantum simulation of in-medium QCD jets: Momentum
  broadening, gluon production, and entropy growth}},\ }\href
  {https://doi.org/10.1103/PhysRevD.108.056023} {\bibfield  {journal} {\bibinfo
   {journal} {Phys. Rev. D}\ }\textbf {\bibinfo {volume} {108}},\ \bibinfo
  {pages} {056023} (\bibinfo {year} {2023})},\ \Eprint
  {https://arxiv.org/abs/2307.01792} {arXiv:2307.01792 [hep-ph]} \BibitemShut
  {NoStop}%
\bibitem [{\citenamefont {Yao}(2022)}]{Yao:2022eqm}%
  \BibitemOpen
  \bibfield  {author} {\bibinfo {author} {\bibfnamefont {X.}~\bibnamefont
  {Yao}},\ }\bibfield  {title} {\bibinfo {title} {{Quantum Simulation of
  Light-Front QCD for Jet Quenching in Nuclear Environments}},\ }\href@noop {}
  {\  (\bibinfo {year} {2022})},\ \Eprint {https://arxiv.org/abs/2205.07902}
  {arXiv:2205.07902 [hep-ph]} \BibitemShut {NoStop}%
\bibitem [{\citenamefont {Kreshchuk}\ \emph {et~al.}(2022)\citenamefont
  {Kreshchuk}, \citenamefont {Kirby}, \citenamefont {Goldstein}, \citenamefont
  {Beauchemin},\ and\ \citenamefont {Love}}]{Kreshchuk:2020dla}%
  \BibitemOpen
  \bibfield  {author} {\bibinfo {author} {\bibfnamefont {M.}~\bibnamefont
  {Kreshchuk}}, \bibinfo {author} {\bibfnamefont {W.~M.}\ \bibnamefont
  {Kirby}}, \bibinfo {author} {\bibfnamefont {G.}~\bibnamefont {Goldstein}},
  \bibinfo {author} {\bibfnamefont {H.}~\bibnamefont {Beauchemin}},\ and\
  \bibinfo {author} {\bibfnamefont {P.~J.}\ \bibnamefont {Love}},\ }\bibfield
  {title} {\bibinfo {title} {{Quantum simulation of quantum field theory in the
  light-front formulation}},\ }\href
  {https://doi.org/10.1103/PhysRevA.105.032418} {\bibfield  {journal} {\bibinfo
   {journal} {Phys. Rev. A}\ }\textbf {\bibinfo {volume} {105}},\ \bibinfo
  {pages} {032418} (\bibinfo {year} {2022})},\ \Eprint
  {https://arxiv.org/abs/2002.04016} {arXiv:2002.04016 [quant-ph]} \BibitemShut
  {NoStop}%
\bibitem [{\citenamefont {Kreshchuk}\ \emph
  {et~al.}(2021{\natexlab{a}})\citenamefont {Kreshchuk}, \citenamefont {Jia},
  \citenamefont {Kirby}, \citenamefont {Goldstein}, \citenamefont {Vary},\ and\
  \citenamefont {Love}}]{Kreshchuk:2020kcz}%
  \BibitemOpen
  \bibfield  {author} {\bibinfo {author} {\bibfnamefont {M.}~\bibnamefont
  {Kreshchuk}}, \bibinfo {author} {\bibfnamefont {S.}~\bibnamefont {Jia}},
  \bibinfo {author} {\bibfnamefont {W.~M.}\ \bibnamefont {Kirby}}, \bibinfo
  {author} {\bibfnamefont {G.}~\bibnamefont {Goldstein}}, \bibinfo {author}
  {\bibfnamefont {J.~P.}\ \bibnamefont {Vary}},\ and\ \bibinfo {author}
  {\bibfnamefont {P.~J.}\ \bibnamefont {Love}},\ }\bibfield  {title} {\bibinfo
  {title} {{Light-Front Field Theory on Current Quantum Computers}},\ }\href
  {https://doi.org/10.3390/e23050597} {\bibfield  {journal} {\bibinfo
  {journal} {Entropy}\ }\textbf {\bibinfo {volume} {23}},\ \bibinfo {pages}
  {597} (\bibinfo {year} {2021}{\natexlab{a}})},\ \Eprint
  {https://arxiv.org/abs/2009.07885} {arXiv:2009.07885 [quant-ph]} \BibitemShut
  {NoStop}%
\bibitem [{\citenamefont {Kreshchuk}\ \emph
  {et~al.}(2021{\natexlab{b}})\citenamefont {Kreshchuk}, \citenamefont {Jia},
  \citenamefont {Kirby}, \citenamefont {Goldstein}, \citenamefont {Vary},\ and\
  \citenamefont {Love}}]{Kreshchuk:2020aiq}%
  \BibitemOpen
  \bibfield  {author} {\bibinfo {author} {\bibfnamefont {M.}~\bibnamefont
  {Kreshchuk}}, \bibinfo {author} {\bibfnamefont {S.}~\bibnamefont {Jia}},
  \bibinfo {author} {\bibfnamefont {W.~M.}\ \bibnamefont {Kirby}}, \bibinfo
  {author} {\bibfnamefont {G.}~\bibnamefont {Goldstein}}, \bibinfo {author}
  {\bibfnamefont {J.~P.}\ \bibnamefont {Vary}},\ and\ \bibinfo {author}
  {\bibfnamefont {P.~J.}\ \bibnamefont {Love}},\ }\bibfield  {title} {\bibinfo
  {title} {{Simulating Hadronic Physics on NISQ devices using Basis Light-Front
  Quantization}},\ }\href {https://doi.org/10.1103/PhysRevA.103.062601}
  {\bibfield  {journal} {\bibinfo  {journal} {Phys. Rev. A}\ }\textbf {\bibinfo
  {volume} {103}},\ \bibinfo {pages} {062601} (\bibinfo {year}
  {2021}{\natexlab{b}})},\ \Eprint {https://arxiv.org/abs/2011.13443}
  {arXiv:2011.13443 [quant-ph]} \BibitemShut {NoStop}%
\bibitem [{\citenamefont {Barata}\ \emph {et~al.}(2021)\citenamefont {Barata},
  \citenamefont {Mueller}, \citenamefont {Tarasov},\ and\ \citenamefont
  {Venugopalan}}]{PhysRevA.103.042410}%
  \BibitemOpen
  \bibfield  {author} {\bibinfo {author} {\bibfnamefont {J.~a.}\ \bibnamefont
  {Barata}}, \bibinfo {author} {\bibfnamefont {N.}~\bibnamefont {Mueller}},
  \bibinfo {author} {\bibfnamefont {A.}~\bibnamefont {Tarasov}},\ and\ \bibinfo
  {author} {\bibfnamefont {R.}~\bibnamefont {Venugopalan}},\ }\bibfield
  {title} {\bibinfo {title} {Single-particle digitization strategy for quantum
  computation of a ${\ensuremath{\phi}}^{4}$ scalar field theory},\ }\href
  {https://doi.org/10.1103/PhysRevA.103.042410} {\bibfield  {journal} {\bibinfo
   {journal} {Phys. Rev. A}\ }\textbf {\bibinfo {volume} {103}},\ \bibinfo
  {pages} {042410} (\bibinfo {year} {2021})}\BibitemShut {NoStop}%
\bibitem [{\citenamefont {Meglio}\ \emph {et~al.}(2023)\citenamefont {Meglio},
  \citenamefont {Jansen}, \citenamefont {Tavernelli}, \citenamefont
  {Alexandrou}, \citenamefont {Arunachalam}, \citenamefont {Bauer},
  \citenamefont {Borras}, \citenamefont {Carrazza}, \citenamefont {Crippa},
  \citenamefont {Croft}, \citenamefont {de~Putter}, \citenamefont {Delgado},
  \citenamefont {Dunjko}, \citenamefont {Egger}, \citenamefont
  {Fernandez-Combarro}, \citenamefont {Fuchs}, \citenamefont {Funcke},
  \citenamefont {Gonzalez-Cuadra}, \citenamefont {Grossi}, \citenamefont
  {Halimeh}, \citenamefont {Holmes}, \citenamefont {Kuhn}, \citenamefont
  {Lacroix}, \citenamefont {Lewis}, \citenamefont {Lucchesi}, \citenamefont
  {Martinez}, \citenamefont {Meloni}, \citenamefont {Mezzacapo}, \citenamefont
  {Montangero}, \citenamefont {Nagano}, \citenamefont {Radescu}, \citenamefont
  {Ortega}, \citenamefont {Roggero}, \citenamefont {Schuhmacher}, \citenamefont
  {Seixas}, \citenamefont {Silvi}, \citenamefont {Spentzouris}, \citenamefont
  {Tacchino}, \citenamefont {Temme}, \citenamefont {Terashi}, \citenamefont
  {Tura}, \citenamefont {Tuysuz}, \citenamefont {Vallecorsa}, \citenamefont
  {Wiese}, \citenamefont {Yoo},\ and\ \citenamefont
  {Zhang}}]{dimeglio2023quantum}%
  \BibitemOpen
  \bibfield  {author} {\bibinfo {author} {\bibfnamefont {A.~D.}\ \bibnamefont
  {Meglio}}, \bibinfo {author} {\bibfnamefont {K.}~\bibnamefont {Jansen}},
  \bibinfo {author} {\bibfnamefont {I.}~\bibnamefont {Tavernelli}}, \bibinfo
  {author} {\bibfnamefont {C.}~\bibnamefont {Alexandrou}}, \bibinfo {author}
  {\bibfnamefont {S.}~\bibnamefont {Arunachalam}}, \bibinfo {author}
  {\bibfnamefont {C.~W.}\ \bibnamefont {Bauer}}, \bibinfo {author}
  {\bibfnamefont {K.}~\bibnamefont {Borras}}, \bibinfo {author} {\bibfnamefont
  {S.}~\bibnamefont {Carrazza}}, \bibinfo {author} {\bibfnamefont
  {A.}~\bibnamefont {Crippa}}, \bibinfo {author} {\bibfnamefont
  {V.}~\bibnamefont {Croft}}, \bibinfo {author} {\bibfnamefont
  {R.}~\bibnamefont {de~Putter}}, \bibinfo {author} {\bibfnamefont
  {A.}~\bibnamefont {Delgado}}, \bibinfo {author} {\bibfnamefont
  {V.}~\bibnamefont {Dunjko}}, \bibinfo {author} {\bibfnamefont {D.~J.}\
  \bibnamefont {Egger}}, \bibinfo {author} {\bibfnamefont {E.}~\bibnamefont
  {Fernandez-Combarro}}, \bibinfo {author} {\bibfnamefont {E.}~\bibnamefont
  {Fuchs}}, \bibinfo {author} {\bibfnamefont {L.}~\bibnamefont {Funcke}},
  \bibinfo {author} {\bibfnamefont {D.}~\bibnamefont {Gonzalez-Cuadra}},
  \bibinfo {author} {\bibfnamefont {M.}~\bibnamefont {Grossi}}, \bibinfo
  {author} {\bibfnamefont {J.~C.}\ \bibnamefont {Halimeh}}, \bibinfo {author}
  {\bibfnamefont {Z.}~\bibnamefont {Holmes}}, \bibinfo {author} {\bibfnamefont
  {S.}~\bibnamefont {Kuhn}}, \bibinfo {author} {\bibfnamefont {D.}~\bibnamefont
  {Lacroix}}, \bibinfo {author} {\bibfnamefont {R.}~\bibnamefont {Lewis}},
  \bibinfo {author} {\bibfnamefont {D.}~\bibnamefont {Lucchesi}}, \bibinfo
  {author} {\bibfnamefont {M.~L.}\ \bibnamefont {Martinez}}, \bibinfo {author}
  {\bibfnamefont {F.}~\bibnamefont {Meloni}}, \bibinfo {author} {\bibfnamefont
  {A.}~\bibnamefont {Mezzacapo}}, \bibinfo {author} {\bibfnamefont
  {S.}~\bibnamefont {Montangero}}, \bibinfo {author} {\bibfnamefont
  {L.}~\bibnamefont {Nagano}}, \bibinfo {author} {\bibfnamefont
  {V.}~\bibnamefont {Radescu}}, \bibinfo {author} {\bibfnamefont {E.~R.}\
  \bibnamefont {Ortega}}, \bibinfo {author} {\bibfnamefont {A.}~\bibnamefont
  {Roggero}}, \bibinfo {author} {\bibfnamefont {J.}~\bibnamefont
  {Schuhmacher}}, \bibinfo {author} {\bibfnamefont {J.}~\bibnamefont {Seixas}},
  \bibinfo {author} {\bibfnamefont {P.}~\bibnamefont {Silvi}}, \bibinfo
  {author} {\bibfnamefont {P.}~\bibnamefont {Spentzouris}}, \bibinfo {author}
  {\bibfnamefont {F.}~\bibnamefont {Tacchino}}, \bibinfo {author}
  {\bibfnamefont {K.}~\bibnamefont {Temme}}, \bibinfo {author} {\bibfnamefont
  {K.}~\bibnamefont {Terashi}}, \bibinfo {author} {\bibfnamefont
  {J.}~\bibnamefont {Tura}}, \bibinfo {author} {\bibfnamefont {C.}~\bibnamefont
  {Tuysuz}}, \bibinfo {author} {\bibfnamefont {S.}~\bibnamefont {Vallecorsa}},
  \bibinfo {author} {\bibfnamefont {U.-J.}\ \bibnamefont {Wiese}}, \bibinfo
  {author} {\bibfnamefont {S.}~\bibnamefont {Yoo}},\ and\ \bibinfo {author}
  {\bibfnamefont {J.}~\bibnamefont {Zhang}},\ }\href@noop {} {\bibinfo {title}
  {Quantum computing for high-energy physics: State of the art and challenges.
  summary of the qc4hep working group}} (\bibinfo {year} {2023}),\ \Eprint
  {https://arxiv.org/abs/2307.03236} {arXiv:2307.03236 [quant-ph]} \BibitemShut
  {NoStop}%
\bibitem [{\citenamefont {Su}\ \emph {et~al.}(2024)\citenamefont {Su},
  \citenamefont {Osborne},\ and\ \citenamefont {Halimeh}}]{su2024coldatom}%
  \BibitemOpen
  \bibfield  {author} {\bibinfo {author} {\bibfnamefont {G.-X.}\ \bibnamefont
  {Su}}, \bibinfo {author} {\bibfnamefont {J.}~\bibnamefont {Osborne}},\ and\
  \bibinfo {author} {\bibfnamefont {J.~C.}\ \bibnamefont {Halimeh}},\
  }\href@noop {} {\bibinfo {title} {A cold-atom particle collider}} (\bibinfo
  {year} {2024}),\ \Eprint {https://arxiv.org/abs/2401.05489} {arXiv:2401.05489
  [cond-mat.quant-gas]} \BibitemShut {NoStop}%
\bibitem [{\citenamefont {Dalzell}\ \emph {et~al.}(2023)\citenamefont
  {Dalzell}, \citenamefont {McArdle}, \citenamefont {Berta}, \citenamefont
  {Bienias}, \citenamefont {Chen}, \citenamefont {Gilyén}, \citenamefont
  {Hann}, \citenamefont {Kastoryano}, \citenamefont {Khabiboulline},
  \citenamefont {Kubica}, \citenamefont {Salton}, \citenamefont {Wang},\ and\
  \citenamefont {Brandão}}]{dalzell2023quantum}%
  \BibitemOpen
  \bibfield  {author} {\bibinfo {author} {\bibfnamefont {A.~M.}\ \bibnamefont
  {Dalzell}}, \bibinfo {author} {\bibfnamefont {S.}~\bibnamefont {McArdle}},
  \bibinfo {author} {\bibfnamefont {M.}~\bibnamefont {Berta}}, \bibinfo
  {author} {\bibfnamefont {P.}~\bibnamefont {Bienias}}, \bibinfo {author}
  {\bibfnamefont {C.-F.}\ \bibnamefont {Chen}}, \bibinfo {author}
  {\bibfnamefont {A.}~\bibnamefont {Gilyén}}, \bibinfo {author} {\bibfnamefont
  {C.~T.}\ \bibnamefont {Hann}}, \bibinfo {author} {\bibfnamefont {M.~J.}\
  \bibnamefont {Kastoryano}}, \bibinfo {author} {\bibfnamefont {E.~T.}\
  \bibnamefont {Khabiboulline}}, \bibinfo {author} {\bibfnamefont
  {A.}~\bibnamefont {Kubica}}, \bibinfo {author} {\bibfnamefont
  {G.}~\bibnamefont {Salton}}, \bibinfo {author} {\bibfnamefont
  {S.}~\bibnamefont {Wang}},\ and\ \bibinfo {author} {\bibfnamefont {F.~G.
  S.~L.}\ \bibnamefont {Brandão}},\ }\href@noop {} {\bibinfo {title} {Quantum
  algorithms: A survey of applications and end-to-end complexities}} (\bibinfo
  {year} {2023}),\ \Eprint {https://arxiv.org/abs/2310.03011} {arXiv:2310.03011
  [quant-ph]} \BibitemShut {NoStop}%
\bibitem [{\citenamefont {Lloyd}(1996)}]{lloyd1996universal}%
  \BibitemOpen
  \bibfield  {author} {\bibinfo {author} {\bibfnamefont {S.}~\bibnamefont
  {Lloyd}},\ }\bibfield  {title} {\bibinfo {title} {Universal quantum
  simulators},\ }\href@noop {} {\bibfield  {journal} {\bibinfo  {journal}
  {Science}\ }\textbf {\bibinfo {volume} {273}},\ \bibinfo {pages} {1073}
  (\bibinfo {year} {1996})}\BibitemShut {NoStop}%
\bibitem [{\citenamefont {Berry}\ \emph {et~al.}(2015)\citenamefont {Berry},
  \citenamefont {Childs}, \citenamefont {Cleve}, \citenamefont {Kothari},\ and\
  \citenamefont {Somma}}]{PhysRevLett.114.090502}%
  \BibitemOpen
  \bibfield  {author} {\bibinfo {author} {\bibfnamefont {D.~W.}\ \bibnamefont
  {Berry}}, \bibinfo {author} {\bibfnamefont {A.~M.}\ \bibnamefont {Childs}},
  \bibinfo {author} {\bibfnamefont {R.}~\bibnamefont {Cleve}}, \bibinfo
  {author} {\bibfnamefont {R.}~\bibnamefont {Kothari}},\ and\ \bibinfo {author}
  {\bibfnamefont {R.~D.}\ \bibnamefont {Somma}},\ }\bibfield  {title} {\bibinfo
  {title} {Simulating hamiltonian dynamics with a truncated taylor series},\
  }\href {https://doi.org/10.1103/PhysRevLett.114.090502} {\bibfield  {journal}
  {\bibinfo  {journal} {Phys. Rev. Lett.}\ }\textbf {\bibinfo {volume} {114}},\
  \bibinfo {pages} {090502} (\bibinfo {year} {2015})}\BibitemShut {NoStop}%
\bibitem [{\citenamefont {Low}\ and\ \citenamefont
  {Chuang}(2017)}]{PhysRevLett.118.010501}%
  \BibitemOpen
  \bibfield  {author} {\bibinfo {author} {\bibfnamefont {G.~H.}\ \bibnamefont
  {Low}}\ and\ \bibinfo {author} {\bibfnamefont {I.~L.}\ \bibnamefont
  {Chuang}},\ }\bibfield  {title} {\bibinfo {title} {Optimal hamiltonian
  simulation by quantum signal processing},\ }\href
  {https://doi.org/10.1103/PhysRevLett.118.010501} {\bibfield  {journal}
  {\bibinfo  {journal} {Phys. Rev. Lett.}\ }\textbf {\bibinfo {volume} {118}},\
  \bibinfo {pages} {010501} (\bibinfo {year} {2017})}\BibitemShut {NoStop}%
\bibitem [{\citenamefont {Gily{\'e}n}\ \emph {et~al.}(2019)\citenamefont
  {Gily{\'e}n}, \citenamefont {Su}, \citenamefont {Low},\ and\ \citenamefont
  {Wiebe}}]{gilyen2019quantum}%
  \BibitemOpen
  \bibfield  {author} {\bibinfo {author} {\bibfnamefont {A.}~\bibnamefont
  {Gily{\'e}n}}, \bibinfo {author} {\bibfnamefont {Y.}~\bibnamefont {Su}},
  \bibinfo {author} {\bibfnamefont {G.~H.}\ \bibnamefont {Low}},\ and\ \bibinfo
  {author} {\bibfnamefont {N.}~\bibnamefont {Wiebe}},\ }\bibfield  {title}
  {\bibinfo {title} {Quantum singular value transformation and beyond:
  exponential improvements for quantum matrix arithmetics},\ }in\ \href@noop {}
  {\emph {\bibinfo {booktitle} {Proceedings of the 51st Annual ACM SIGACT
  Symposium on Theory of Computing}}}\ (\bibinfo {year} {2019})\ pp.\ \bibinfo
  {pages} {193--204}\BibitemShut {NoStop}%
\bibitem [{\citenamefont {Berry}\ \emph {et~al.}(2014)\citenamefont {Berry},
  \citenamefont {Childs}, \citenamefont {Cleve}, \citenamefont {Kothari},\ and\
  \citenamefont {Somma}}]{berry2017exponential}%
  \BibitemOpen
  \bibfield  {author} {\bibinfo {author} {\bibfnamefont {D.~W.}\ \bibnamefont
  {Berry}}, \bibinfo {author} {\bibfnamefont {A.~M.}\ \bibnamefont {Childs}},
  \bibinfo {author} {\bibfnamefont {R.}~\bibnamefont {Cleve}}, \bibinfo
  {author} {\bibfnamefont {R.}~\bibnamefont {Kothari}},\ and\ \bibinfo {author}
  {\bibfnamefont {R.~D.}\ \bibnamefont {Somma}},\ }\bibfield  {title} {\bibinfo
  {title} {Exponential improvement in precision for simulating sparse
  hamiltonians},\ }in\ \href {https://doi.org/10.1145/2591796.2591854} {\emph
  {\bibinfo {booktitle} {Proceedings of the forty-sixth annual {ACM} symposium
  on Theory of computing}}}\ (\bibinfo  {publisher} {{ACM}},\ \bibinfo {year}
  {2014})\BibitemShut {NoStop}%
\bibitem [{\citenamefont {Zhao}\ \emph {et~al.}(2013)\citenamefont {Zhao},
  \citenamefont {Ilderton}, \citenamefont {Maris},\ and\ \citenamefont
  {Vary}}]{PhysRevD.88.065014}%
  \BibitemOpen
  \bibfield  {author} {\bibinfo {author} {\bibfnamefont {X.}~\bibnamefont
  {Zhao}}, \bibinfo {author} {\bibfnamefont {A.}~\bibnamefont {Ilderton}},
  \bibinfo {author} {\bibfnamefont {P.}~\bibnamefont {Maris}},\ and\ \bibinfo
  {author} {\bibfnamefont {J.~P.}\ \bibnamefont {Vary}},\ }\bibfield  {title}
  {\bibinfo {title} {Scattering in time-dependent basis light-front
  quantization},\ }\href {https://doi.org/10.1103/PhysRevD.88.065014}
  {\bibfield  {journal} {\bibinfo  {journal} {Phys. Rev. D}\ }\textbf {\bibinfo
  {volume} {88}},\ \bibinfo {pages} {065014} (\bibinfo {year}
  {2013})}\BibitemShut {NoStop}%
\bibitem [{\citenamefont {Li}\ \emph {et~al.}(2020)\citenamefont {Li},
  \citenamefont {Zhao}, \citenamefont {Maris}, \citenamefont {Chen},
  \citenamefont {Li}, \citenamefont {Tuchin},\ and\ \citenamefont
  {Vary}}]{PhysRevD.101.076016}%
  \BibitemOpen
  \bibfield  {author} {\bibinfo {author} {\bibfnamefont {M.}~\bibnamefont
  {Li}}, \bibinfo {author} {\bibfnamefont {X.}~\bibnamefont {Zhao}}, \bibinfo
  {author} {\bibfnamefont {P.}~\bibnamefont {Maris}}, \bibinfo {author}
  {\bibfnamefont {G.}~\bibnamefont {Chen}}, \bibinfo {author} {\bibfnamefont
  {Y.}~\bibnamefont {Li}}, \bibinfo {author} {\bibfnamefont {K.}~\bibnamefont
  {Tuchin}},\ and\ \bibinfo {author} {\bibfnamefont {J.~P.}\ \bibnamefont
  {Vary}},\ }\bibfield  {title} {\bibinfo {title} {Ultrarelativistic
  quark-nucleus scattering in a light-front hamiltonian approach},\ }\href
  {https://doi.org/10.1103/PhysRevD.101.076016} {\bibfield  {journal} {\bibinfo
   {journal} {Phys. Rev. D}\ }\textbf {\bibinfo {volume} {101}},\ \bibinfo
  {pages} {076016} (\bibinfo {year} {2020})}\BibitemShut {NoStop}%
\bibitem [{\citenamefont {Li}\ \emph {et~al.}(2021)\citenamefont {Li},
  \citenamefont {Lappi},\ and\ \citenamefont {Zhao}}]{PhysRevD.104.056014}%
  \BibitemOpen
  \bibfield  {author} {\bibinfo {author} {\bibfnamefont {M.}~\bibnamefont
  {Li}}, \bibinfo {author} {\bibfnamefont {T.}~\bibnamefont {Lappi}},\ and\
  \bibinfo {author} {\bibfnamefont {X.}~\bibnamefont {Zhao}},\ }\bibfield
  {title} {\bibinfo {title} {Scattering and gluon emission in a color field: A
  light-front hamiltonian approach},\ }\href
  {https://doi.org/10.1103/PhysRevD.104.056014} {\bibfield  {journal} {\bibinfo
   {journal} {Phys. Rev. D}\ }\textbf {\bibinfo {volume} {104}},\ \bibinfo
  {pages} {056014} (\bibinfo {year} {2021})}\BibitemShut {NoStop}%
\bibitem [{\citenamefont {Childs}\ and\ \citenamefont
  {Wiebe}(2012)}]{childs2012hamiltonian}%
  \BibitemOpen
  \bibfield  {author} {\bibinfo {author} {\bibfnamefont {A.~M.}\ \bibnamefont
  {Childs}}\ and\ \bibinfo {author} {\bibfnamefont {N.}~\bibnamefont {Wiebe}},\
  }\bibfield  {title} {\bibinfo {title} {Hamiltonian simulation using linear
  combinations of unitary operations},\ }\href@noop {} {\bibfield  {journal}
  {\bibinfo  {journal} {Quantum Information \& Computation}\ }\textbf {\bibinfo
  {volume} {12}},\ \bibinfo {pages} {901} (\bibinfo {year} {2012})}\BibitemShut
  {NoStop}%
\bibitem [{\citenamefont {Nielsen}\ and\ \citenamefont
  {Chuang}(2010)}]{nielsen_chuang_2010}%
  \BibitemOpen
  \bibfield  {author} {\bibinfo {author} {\bibfnamefont {M.~A.}\ \bibnamefont
  {Nielsen}}\ and\ \bibinfo {author} {\bibfnamefont {I.~L.}\ \bibnamefont
  {Chuang}},\ }\href {https://doi.org/10.1017/CBO9780511976667} {\emph
  {\bibinfo {title} {Quantum Computation and Quantum Information: 10th
  Anniversary Edition}}}\ (\bibinfo  {publisher} {Cambridge University Press},\
  \bibinfo {year} {2010})\BibitemShut {NoStop}%
\bibitem [{\citenamefont {Gelis}\ \emph {et~al.}(2010)\citenamefont {Gelis},
  \citenamefont {Iancu}, \citenamefont {Jalilian-Marian},\ and\ \citenamefont
  {Venugopalan}}]{gelis2010color}%
  \BibitemOpen
  \bibfield  {author} {\bibinfo {author} {\bibfnamefont {F.}~\bibnamefont
  {Gelis}}, \bibinfo {author} {\bibfnamefont {E.}~\bibnamefont {Iancu}},
  \bibinfo {author} {\bibfnamefont {J.}~\bibnamefont {Jalilian-Marian}},\ and\
  \bibinfo {author} {\bibfnamefont {R.}~\bibnamefont {Venugopalan}},\
  }\bibfield  {title} {\bibinfo {title} {The color glass condensate},\
  }\href@noop {} {\bibfield  {journal} {\bibinfo  {journal} {Annual Review of
  Nuclear and Particle Science}\ }\textbf {\bibinfo {volume} {60}},\ \bibinfo
  {pages} {463} (\bibinfo {year} {2010})}\BibitemShut {NoStop}%
\bibitem [{\citenamefont {{Qiskit contributors}}(2023)}]{Qiskit}%
  \BibitemOpen
  \bibfield  {author} {\bibinfo {author} {\bibnamefont {{Qiskit
  contributors}}},\ }\href {https://doi.org/10.5281/zenodo.2573505} {\bibinfo
  {title} {Qiskit: An open-source framework for quantum computing}} (\bibinfo
  {year} {2023})\BibitemShut {NoStop}%
\bibitem [{\citenamefont {Brodsky}\ \emph
  {et~al.}(1998{\natexlab{a}})\citenamefont {Brodsky}, \citenamefont {Pauli},\
  and\ \citenamefont {Pinsky}}]{Brodsky:1997de}%
  \BibitemOpen
  \bibfield  {author} {\bibinfo {author} {\bibfnamefont {S.~J.}\ \bibnamefont
  {Brodsky}}, \bibinfo {author} {\bibfnamefont {H.-C.}\ \bibnamefont {Pauli}},\
  and\ \bibinfo {author} {\bibfnamefont {S.~S.}\ \bibnamefont {Pinsky}},\
  }\bibfield  {title} {\bibinfo {title} {{Quantum chromodynamics and other
  field theories on the light cone}},\ }\href
  {https://doi.org/10.1016/S0370-1573(97)00089-6} {\bibfield  {journal}
  {\bibinfo  {journal} {Phys. Rept.}\ }\textbf {\bibinfo {volume} {301}},\
  \bibinfo {pages} {299} (\bibinfo {year} {1998}{\natexlab{a}})},\ \Eprint
  {https://arxiv.org/abs/hep-ph/9705477} {arXiv:hep-ph/9705477} \BibitemShut
  {NoStop}%
\bibitem [{\citenamefont {Brodsky}\ \emph
  {et~al.}(1998{\natexlab{b}})\citenamefont {Brodsky}, \citenamefont {Pauli},\
  and\ \citenamefont {Pinsky}}]{BRODSKY1998299}%
  \BibitemOpen
  \bibfield  {author} {\bibinfo {author} {\bibfnamefont {S.~J.}\ \bibnamefont
  {Brodsky}}, \bibinfo {author} {\bibfnamefont {H.-C.}\ \bibnamefont {Pauli}},\
  and\ \bibinfo {author} {\bibfnamefont {S.~S.}\ \bibnamefont {Pinsky}},\
  }\bibfield  {title} {\bibinfo {title} {Quantum chromodynamics and other field
  theories on the light cone},\ }\href
  {https://doi.org/https://doi.org/10.1016/S0370-1573(97)00089-6} {\bibfield
  {journal} {\bibinfo  {journal} {Physics Reports}\ }\textbf {\bibinfo {volume}
  {301}},\ \bibinfo {pages} {299} (\bibinfo {year}
  {1998}{\natexlab{b}})}\BibitemShut {NoStop}%
\bibitem [{\citenamefont {Harindranath}(2005)}]{Harindranath2005}%
  \BibitemOpen
  \bibfield  {author} {\bibinfo {author} {\bibfnamefont {A.}~\bibnamefont
  {Harindranath}},\ }\bibfield  {title} {\bibinfo {title} {An introduction to
  light front field theory and light front qcd (lecture notes)}} (\bibinfo
  {year} {2005})\BibitemShut {NoStop}%
\bibitem [{\citenamefont {Pauli}\ and\ \citenamefont
  {Brodsky}(1985)}]{Pauli:1985pv}%
  \BibitemOpen
  \bibfield  {author} {\bibinfo {author} {\bibfnamefont {H.~C.}\ \bibnamefont
  {Pauli}}\ and\ \bibinfo {author} {\bibfnamefont {S.~J.}\ \bibnamefont
  {Brodsky}},\ }\bibfield  {title} {\bibinfo {title} {{Solving Field Theory in
  One Space One Time Dimension}},\ }\href
  {https://doi.org/10.1103/PhysRevD.32.1993} {\bibfield  {journal} {\bibinfo
  {journal} {Phys. Rev. D}\ }\textbf {\bibinfo {volume} {32}},\ \bibinfo
  {pages} {1993} (\bibinfo {year} {1985})}\BibitemShut {NoStop}%
\bibitem [{\citenamefont {Eller}\ \emph {et~al.}(1987)\citenamefont {Eller},
  \citenamefont {Pauli},\ and\ \citenamefont {Brodsky}}]{PhysRevD.35.1493}%
  \BibitemOpen
  \bibfield  {author} {\bibinfo {author} {\bibfnamefont {T.}~\bibnamefont
  {Eller}}, \bibinfo {author} {\bibfnamefont {H.-C.}\ \bibnamefont {Pauli}},\
  and\ \bibinfo {author} {\bibfnamefont {S.~J.}\ \bibnamefont {Brodsky}},\
  }\bibfield  {title} {\bibinfo {title} {Discretized light-cone quantization:
  The massless and the massive schwinger model},\ }\href
  {https://doi.org/10.1103/PhysRevD.35.1493} {\bibfield  {journal} {\bibinfo
  {journal} {Phys. Rev. D}\ }\textbf {\bibinfo {volume} {35}},\ \bibinfo
  {pages} {1493} (\bibinfo {year} {1987})}\BibitemShut {NoStop}%
\bibitem [{\citenamefont {Somma}(2015)}]{somma2015quantum}%
  \BibitemOpen
  \bibfield  {author} {\bibinfo {author} {\bibfnamefont {R.~D.}\ \bibnamefont
  {Somma}},\ }\bibfield  {title} {\bibinfo {title} {Quantum simulations of one
  dimensional quantum systems},\ }\href@noop {} {\bibfield  {journal} {\bibinfo
   {journal} {arXiv preprint arXiv:1503.06319}\ } (\bibinfo {year}
  {2015})}\BibitemShut {NoStop}%
\bibitem [{\citenamefont {Low}\ and\ \citenamefont
  {Chuang}(2019)}]{Low2019hamiltonian}%
  \BibitemOpen
  \bibfield  {author} {\bibinfo {author} {\bibfnamefont {G.~H.}\ \bibnamefont
  {Low}}\ and\ \bibinfo {author} {\bibfnamefont {I.~L.}\ \bibnamefont
  {Chuang}},\ }\bibfield  {title} {\bibinfo {title} {Hamiltonian {S}imulation
  by {Q}ubitization},\ }\href {https://doi.org/10.22331/q-2019-07-12-163}
  {\bibfield  {journal} {\bibinfo  {journal} {{Quantum}}\ }\textbf {\bibinfo
  {volume} {3}},\ \bibinfo {pages} {163} (\bibinfo {year} {2019})}\BibitemShut
  {NoStop}%
\bibitem [{\citenamefont {Du}\ and\ \citenamefont
  {Vary}(2024{\natexlab{b}})}]{Du:2024ixj}%
  \BibitemOpen
  \bibfield  {author} {\bibinfo {author} {\bibfnamefont {W.}~\bibnamefont
  {Du}}\ and\ \bibinfo {author} {\bibfnamefont {J.~P.}\ \bibnamefont {Vary}},\
  }\bibfield  {title} {\bibinfo {title} {{Systematic input scheme for
  many-boson Hamiltonians via quantum walk}},\ }\href@noop {} {\  (\bibinfo
  {year} {2024}{\natexlab{b}})},\ \Eprint {https://arxiv.org/abs/2407.13672}
  {arXiv:2407.13672 [quant-ph]} \BibitemShut {NoStop}%
\bibitem [{\citenamefont {Kirby}\ \emph {et~al.}(2021)\citenamefont {Kirby},
  \citenamefont {Hadi}, \citenamefont {Kreshchuk},\ and\ \citenamefont
  {Love}}]{PhysRevA.104.042607}%
  \BibitemOpen
  \bibfield  {author} {\bibinfo {author} {\bibfnamefont {W.~M.}\ \bibnamefont
  {Kirby}}, \bibinfo {author} {\bibfnamefont {S.}~\bibnamefont {Hadi}},
  \bibinfo {author} {\bibfnamefont {M.}~\bibnamefont {Kreshchuk}},\ and\
  \bibinfo {author} {\bibfnamefont {P.~J.}\ \bibnamefont {Love}},\ }\bibfield
  {title} {\bibinfo {title} {Quantum simulation of second-quantized
  hamiltonians in compact encoding},\ }\href
  {https://doi.org/10.1103/PhysRevA.104.042607} {\bibfield  {journal} {\bibinfo
   {journal} {Phys. Rev. A}\ }\textbf {\bibinfo {volume} {104}},\ \bibinfo
  {pages} {042607} (\bibinfo {year} {2021})}\BibitemShut {NoStop}%
\bibitem [{\citenamefont {Babbush}\ \emph {et~al.}(2016)\citenamefont
  {Babbush}, \citenamefont {Berry}, \citenamefont {Kivlichan}, \citenamefont
  {Wei}, \citenamefont {Love},\ and\ \citenamefont
  {Aspuru-Guzik}}]{Babbush_2016}%
  \BibitemOpen
  \bibfield  {author} {\bibinfo {author} {\bibfnamefont {R.}~\bibnamefont
  {Babbush}}, \bibinfo {author} {\bibfnamefont {D.~W.}\ \bibnamefont {Berry}},
  \bibinfo {author} {\bibfnamefont {I.~D.}\ \bibnamefont {Kivlichan}}, \bibinfo
  {author} {\bibfnamefont {A.~Y.}\ \bibnamefont {Wei}}, \bibinfo {author}
  {\bibfnamefont {P.~J.}\ \bibnamefont {Love}},\ and\ \bibinfo {author}
  {\bibfnamefont {A.}~\bibnamefont {Aspuru-Guzik}},\ }\bibfield  {title}
  {\bibinfo {title} {Exponentially more precise quantum simulation of fermions
  in second quantization},\ }\href
  {https://doi.org/10.1088/1367-2630/18/3/033032} {\bibfield  {journal}
  {\bibinfo  {journal} {New Journal of Physics}\ }\textbf {\bibinfo {volume}
  {18}},\ \bibinfo {pages} {033032} (\bibinfo {year} {2016})}\BibitemShut
  {NoStop}%
\bibitem [{\citenamefont {Shende}\ \emph {et~al.}(2006)\citenamefont {Shende},
  \citenamefont {Bullock},\ and\ \citenamefont {Markov}}]{Shende_2006}%
  \BibitemOpen
  \bibfield  {author} {\bibinfo {author} {\bibfnamefont {V.}~\bibnamefont
  {Shende}}, \bibinfo {author} {\bibfnamefont {S.}~\bibnamefont {Bullock}},\
  and\ \bibinfo {author} {\bibfnamefont {I.}~\bibnamefont {Markov}},\
  }\bibfield  {title} {\bibinfo {title} {Synthesis of quantum-logic circuits},\
  }\href {https://doi.org/10.1109/tcad.2005.855930} {\bibfield  {journal}
  {\bibinfo  {journal} {{IEEE} Transactions on Computer-Aided Design of
  Integrated Circuits and Systems}\ }\textbf {\bibinfo {volume} {25}},\
  \bibinfo {pages} {1000} (\bibinfo {year} {2006})}\BibitemShut {NoStop}%
\bibitem [{\citenamefont {Suzuki}(1985)}]{doi:10.1063/1.526596}%
  \BibitemOpen
  \bibfield  {author} {\bibinfo {author} {\bibfnamefont {M.}~\bibnamefont
  {Suzuki}},\ }\bibfield  {title} {\bibinfo {title} {Decomposition formulas of
  exponential operators and lie exponentials with some applications to quantum
  mechanics and statistical physics},\ }\href
  {https://doi.org/10.1063/1.526596} {\bibfield  {journal} {\bibinfo  {journal}
  {Journal of Mathematical Physics}\ }\textbf {\bibinfo {volume} {26}},\
  \bibinfo {pages} {601} (\bibinfo {year} {1985})},\ \Eprint
  {https://arxiv.org/abs/https://doi.org/10.1063/1.526596}
  {https://doi.org/10.1063/1.526596} \BibitemShut {NoStop}%
\bibitem [{\citenamefont {Huyghebaert}\ and\ \citenamefont
  {Raedt}(1990)}]{Huyghebaert_1990}%
  \BibitemOpen
  \bibfield  {author} {\bibinfo {author} {\bibfnamefont {J.}~\bibnamefont
  {Huyghebaert}}\ and\ \bibinfo {author} {\bibfnamefont {H.~D.}\ \bibnamefont
  {Raedt}},\ }\bibfield  {title} {\bibinfo {title} {Product formula methods for
  time-dependent schrodinger problems},\ }\href
  {https://doi.org/10.1088/0305-4470/23/24/019} {\bibfield  {journal} {\bibinfo
   {journal} {Journal of Physics A: Mathematical and General}\ }\textbf
  {\bibinfo {volume} {23}},\ \bibinfo {pages} {5777} (\bibinfo {year}
  {1990})}\BibitemShut {NoStop}%
\bibitem [{\citenamefont {Childs}\ \emph {et~al.}(2021)\citenamefont {Childs},
  \citenamefont {Su}, \citenamefont {Tran}, \citenamefont {Wiebe},\ and\
  \citenamefont {Zhu}}]{PhysRevX.11.011020}%
  \BibitemOpen
  \bibfield  {author} {\bibinfo {author} {\bibfnamefont {A.~M.}\ \bibnamefont
  {Childs}}, \bibinfo {author} {\bibfnamefont {Y.}~\bibnamefont {Su}}, \bibinfo
  {author} {\bibfnamefont {M.~C.}\ \bibnamefont {Tran}}, \bibinfo {author}
  {\bibfnamefont {N.}~\bibnamefont {Wiebe}},\ and\ \bibinfo {author}
  {\bibfnamefont {S.}~\bibnamefont {Zhu}},\ }\bibfield  {title} {\bibinfo
  {title} {Theory of trotter error with commutator scaling},\ }\href
  {https://doi.org/10.1103/PhysRevX.11.011020} {\bibfield  {journal} {\bibinfo
  {journal} {Phys. Rev. X}\ }\textbf {\bibinfo {volume} {11}},\ \bibinfo
  {pages} {011020} (\bibinfo {year} {2021})}\BibitemShut {NoStop}%
\bibitem [{\citenamefont {Jalilian-Marian}(2017)}]{PhysRevD.96.074020}%
  \BibitemOpen
  \bibfield  {author} {\bibinfo {author} {\bibfnamefont {J.}~\bibnamefont
  {Jalilian-Marian}},\ }\bibfield  {title} {\bibinfo {title} {Elastic
  scattering of a quark from a color field: Longitudinal momentum exchange},\
  }\href {https://doi.org/10.1103/PhysRevD.96.074020} {\bibfield  {journal}
  {\bibinfo  {journal} {Phys. Rev. D}\ }\textbf {\bibinfo {volume} {96}},\
  \bibinfo {pages} {074020} (\bibinfo {year} {2017})}\BibitemShut {NoStop}%
\bibitem [{\citenamefont {Kovchegov}\ and\ \citenamefont
  {Levin}(2012)}]{kovchegov_levin_2012}%
  \BibitemOpen
  \bibfield  {author} {\bibinfo {author} {\bibfnamefont {Y.~V.}\ \bibnamefont
  {Kovchegov}}\ and\ \bibinfo {author} {\bibfnamefont {E.}~\bibnamefont
  {Levin}},\ }\href {https://doi.org/10.1017/CBO9781139022187} {\emph {\bibinfo
  {title} {Quantum Chromodynamics at High Energy}}},\ Cambridge Monographs on
  Particle Physics, Nuclear Physics and Cosmology\ (\bibinfo  {publisher}
  {Cambridge University Press},\ \bibinfo {year} {2012})\BibitemShut {NoStop}%
\bibitem [{\citenamefont {Krasnitz}\ \emph {et~al.}(2003)\citenamefont
  {Krasnitz}, \citenamefont {Nara},\ and\ \citenamefont
  {Venugopalan}}]{KRASNITZ2003268}%
  \BibitemOpen
  \bibfield  {author} {\bibinfo {author} {\bibfnamefont {A.}~\bibnamefont
  {Krasnitz}}, \bibinfo {author} {\bibfnamefont {Y.}~\bibnamefont {Nara}},\
  and\ \bibinfo {author} {\bibfnamefont {R.}~\bibnamefont {Venugopalan}},\
  }\bibfield  {title} {\bibinfo {title} {Gluon production in the color glass
  condensate model of collisions of ultrarelativistic finite nuclei},\ }\href
  {https://doi.org/https://doi.org/10.1016/S0375-9474(03)00636-5} {\bibfield
  {journal} {\bibinfo  {journal} {Nuclear Physics A}\ }\textbf {\bibinfo
  {volume} {717}},\ \bibinfo {pages} {268} (\bibinfo {year}
  {2003})}\BibitemShut {NoStop}%
\bibitem [{\citenamefont {McLerran}\ and\ \citenamefont
  {Venugopalan}(1994{\natexlab{a}})}]{PhysRevD.49.2233}%
  \BibitemOpen
  \bibfield  {author} {\bibinfo {author} {\bibfnamefont {L.}~\bibnamefont
  {McLerran}}\ and\ \bibinfo {author} {\bibfnamefont {R.}~\bibnamefont
  {Venugopalan}},\ }\bibfield  {title} {\bibinfo {title} {Computing quark and
  gluon distribution functions for very large nuclei},\ }\href
  {https://doi.org/10.1103/PhysRevD.49.2233} {\bibfield  {journal} {\bibinfo
  {journal} {Phys. Rev. D}\ }\textbf {\bibinfo {volume} {49}},\ \bibinfo
  {pages} {2233} (\bibinfo {year} {1994}{\natexlab{a}})}\BibitemShut {NoStop}%
\bibitem [{\citenamefont {McLerran}\ and\ \citenamefont
  {Venugopalan}(1994{\natexlab{b}})}]{PhysRevD.49.3352}%
  \BibitemOpen
  \bibfield  {author} {\bibinfo {author} {\bibfnamefont {L.}~\bibnamefont
  {McLerran}}\ and\ \bibinfo {author} {\bibfnamefont {R.}~\bibnamefont
  {Venugopalan}},\ }\bibfield  {title} {\bibinfo {title} {Gluon distribution
  functions for very large nuclei at small transverse momentum},\ }\href
  {https://doi.org/10.1103/PhysRevD.49.3352} {\bibfield  {journal} {\bibinfo
  {journal} {Phys. Rev. D}\ }\textbf {\bibinfo {volume} {49}},\ \bibinfo
  {pages} {3352} (\bibinfo {year} {1994}{\natexlab{b}})}\BibitemShut {NoStop}%
\bibitem [{\citenamefont {{Wolfram Research{,} Inc.}}()}]{Wolfram_One}%
  \BibitemOpen
  \bibfield  {author} {\bibinfo {author} {\bibnamefont {{Wolfram Research{,}
  Inc.}}},\ }\href {https://www.wolfram.com/mathematica} {\bibinfo {title}
  {Mathematica, {V}ersion 14.0}},\ \bibinfo {note} {champaign, IL,
  2024}\BibitemShut {NoStop}%
\bibitem [{\citenamefont {Devitt}\ \emph {et~al.}(2013)\citenamefont {Devitt},
  \citenamefont {Munro},\ and\ \citenamefont {Nemoto}}]{Devitt_2013}%
  \BibitemOpen
  \bibfield  {author} {\bibinfo {author} {\bibfnamefont {S.~J.}\ \bibnamefont
  {Devitt}}, \bibinfo {author} {\bibfnamefont {W.~J.}\ \bibnamefont {Munro}},\
  and\ \bibinfo {author} {\bibfnamefont {K.}~\bibnamefont {Nemoto}},\
  }\bibfield  {title} {\bibinfo {title} {Quantum error correction for
  beginners},\ }\href {https://doi.org/10.1088/0034-4885/76/7/076001}
  {\bibfield  {journal} {\bibinfo  {journal} {Reports on Progress in Physics}\
  }\textbf {\bibinfo {volume} {76}},\ \bibinfo {pages} {076001} (\bibinfo
  {year} {2013})}\BibitemShut {NoStop}%
\bibitem [{\citenamefont {Berry}\ \emph {et~al.}(2020)\citenamefont {Berry},
  \citenamefont {Childs}, \citenamefont {Su}, \citenamefont {Wang},\ and\
  \citenamefont {Wiebe}}]{Berry2020timedependent}%
  \BibitemOpen
  \bibfield  {author} {\bibinfo {author} {\bibfnamefont {D.~W.}\ \bibnamefont
  {Berry}}, \bibinfo {author} {\bibfnamefont {A.~M.}\ \bibnamefont {Childs}},
  \bibinfo {author} {\bibfnamefont {Y.}~\bibnamefont {Su}}, \bibinfo {author}
  {\bibfnamefont {X.}~\bibnamefont {Wang}},\ and\ \bibinfo {author}
  {\bibfnamefont {N.}~\bibnamefont {Wiebe}},\ }\bibfield  {title} {\bibinfo
  {title} {Time-dependent {H}amiltonian simulation with {$L^1$}-norm scaling},\
  }\href {https://doi.org/10.22331/q-2020-04-20-254} {\bibfield  {journal}
  {\bibinfo  {journal} {{Quantum}}\ }\textbf {\bibinfo {volume} {4}},\ \bibinfo
  {pages} {254} (\bibinfo {year} {2020})}\BibitemShut {NoStop}%
\bibitem [{\citenamefont {Kieferov\'a}\ \emph {et~al.}(2019)\citenamefont
  {Kieferov\'a}, \citenamefont {Scherer},\ and\ \citenamefont
  {Berry}}]{PhysRevA.99.042314}%
  \BibitemOpen
  \bibfield  {author} {\bibinfo {author} {\bibfnamefont {M.}~\bibnamefont
  {Kieferov\'a}}, \bibinfo {author} {\bibfnamefont {A.}~\bibnamefont
  {Scherer}},\ and\ \bibinfo {author} {\bibfnamefont {D.~W.}\ \bibnamefont
  {Berry}},\ }\bibfield  {title} {\bibinfo {title} {Simulating the dynamics of
  time-dependent hamiltonians with a truncated dyson series},\ }\href
  {https://doi.org/10.1103/PhysRevA.99.042314} {\bibfield  {journal} {\bibinfo
  {journal} {Phys. Rev. A}\ }\textbf {\bibinfo {volume} {99}},\ \bibinfo
  {pages} {042314} (\bibinfo {year} {2019})}\BibitemShut {NoStop}%
\end{thebibliography}%

\end{document}